\documentclass{aa}  

\usepackage{graphicx}
\usepackage{txfonts}
\usepackage{xcolor}

\newcommand{\oiii}{[O\,\textsc{iii}]}
\newcommand{\nii}{[N\,\textsc{ii}]}
\newcommand{\sii}{[S\,\textsc{ii}]}
\newcommand{\oi}{[O\,\textsc{i}]}

\newcommand{\siii}{[S\,\textsc{iii}]}
\newcommand{\oii}{[O\,\textsc{ii}]}
\newcommand{\hii}{H\,\textsc{ii}}
\newcommand{\ha}{H$\alpha$}
\newcommand{\hb}{H$\beta$}
\newcommand{\te}{$T_{\rm e}$}
\newcommand{\fion}{$f_{\rm ion}$}
\newcommand\galaxyname[2]{#1\,#2}
\newcommand{\revone}{}

\begin{document}

   \title{A physically motivated `charge-exchange method' for measuring electron temperatures within HII regions}

   \titlerunning{A physically motivated method for measuring Te}


\newcommand{\OSU}{\label{OSU} Department of Astronomy, The Ohio State University, 140 West 18th Avenue, Columbus, Ohio 43210, USA}

\newcommand{\Alberta}{\label{Alberta} Department of Physics, University of Alberta, Edmonton, AB T6G 2E1, Canada}

\newcommand{\ANU}{\label{ANU} Research School of Astronomy and Astrophysics, Australian National University, Canberra, ACT 2611, Australia}

\newcommand{\IPAC}{\label{IPAC} Caltech-IPAC, 1200 E. California Blvd. Pasadena, CA 91125, USA}

\newcommand{\Carnegie}{\label{Carnegi} Observatories of the Carnegie Institution for Science, 813 Santa Barbara Street, Pasadena, CA 91101, USA}

\newcommand{\CCAPP}{\label{CCAPP} Center for Cosmology and Astroparticle Physics, 191 West Woodruff Avenue, Columbus, OH 43210, USA}

\newcommand{\CfA}{\label{CfA} Harvard-Smithsonian Center for Astrophysics, 60 Garden Street, Cambridge, MA 02138, USA}

\newcommand{\CITEVA}{\label{CITEVA} Centro de Astronomía (CITEVA), Universidad de Antofagasta, Avenida Angamos 601, Antofagasta, Chile}

\newcommand{\CNRS}{\label{CNRS} CNRS, IRAP, 9 Av. du Colonel Roche, BP 44346, F-31028 Toulouse cedex 4, France}

\newcommand{\ESO}{\label{ESO} European Southern Observatory, Karl-Schwarzschild Stra{\ss}e 2, D-85748 Garching bei M\"{u}nchen, Germany}

\newcommand{\HD}{\label{HD} Astronomisches Rechen-Institut, Zentrum f\"{u}r Astronomie der Universit\"{a}t Heidelberg, M\"{o}nchhofstra\ss e 12-14, D-69120 Heidelberg, Germany}

\newcommand{\ICRAR}{\label{ICRAR} International Centre for Radio Astronomy Research, University of Western Australia, 35 Stirling Highway, Crawley, WA 6009, Australia}

\newcommand{\IRAM}{\label{IRAM} Institut de Radioastronomie Millim\'{e}trique (IRAM), 300 Rue de la Piscine, F-38406 Saint Martin d'H\`{e}res, France}

\newcommand{\ITA}{\label{ITA} Universit\"{a}t Heidelberg, Zentrum f\"{u}r Astronomie, Institut f\"{u}r Theoretische Astrophysik, Albert-Ueberle-Str 2, D-69120 Heidelberg, Germany}

\newcommand{\IWR}{\label{IWR} Universit\"{a}t Heidelberg, Interdisziplin\"{a}res Zentrum f\"{u}r Wissenschaftliches Rechnen, Im Neuenheimer Feld 205, D-69120 Heidelberg, Germany}

\newcommand{\JHU}{\label{JHU} Department of Physics and Astronomy, The Johns Hopkins University, Baltimore, MD 21218, USA}

\newcommand{\Leiden}{\label{Leiden} Leiden Observatory, Leiden University, P.O. Box 9513, 2300 RA Leiden, The Netherlands}

\newcommand{\Maryland}{\label{Maryland} Department of Astronomy, University of Maryland, College Park, MD 20742, USA}

\newcommand{\MPE}{\label{MPE} Max-Planck-Institut f\"{u}r extraterrestrische Physik, Giessenbachstra{\ss}e 1, D-85748 Garching, Germany}

\newcommand{\MPIA}{\label{MPIA} Max-Planck-Institut f\"{u}r Astronomie, K\"{o}nigstuhl 17, D-69117, Heidelberg, Germany}

\newcommand{\Nagoya}{\label{Nagoya} Department of Physics, Nagoya University, Furo-cho, Chikusa-ku, Nagoya, Aichi 464-8602, Japan}

\newcommand{\NRAO}{\label{NRAO} National Radio Astronomy Observatory, 520 Edgemont Road, Charlottesville, VA 22903-2475, USA}

\newcommand{\OAN}{\label{OAN} Observatorio Astron\'{o}mico Nacional (IGN), C/Alfonso XII, 3, E-28014 Madrid, Spain}

\newcommand{\ObsParis}{\label{ObsParis} Sorbonne Universit\'{e}, Observatoire de Paris, Universit\'{e} PSL, CNRS, LERMA, F-75014, Paris, France}

\newcommand{\Princeton}{\label{Princeton} Department of Astrophysical Sciences, Princeton University, Princeton, NJ 08544 USA}

\newcommand{\UToledo}{\label{UToledo} University of Toledo, 2801 W. Bancroft St., Mail Stop 111, Toledo, OH, 43606}

\newcommand{\Toulouse}{\label{Toulouse} Universit\'{e} de Toulouse, UPS-OMP, IRAP, F-31028 Toulouse cedex 4, France}

\newcommand{\UBonn}{\label{UBonn} Argelander-Institut f\"ur Astronomie, Universit\"at Bonn, Auf dem H\"ugel 71, 53121 Bonn, Germany}

\newcommand{\UChile}{\label{UChile} Departamento de Astronom\'{i}a, Universidad de Chile, Camino del Observatorio 1515, Las Condes, Santiago, Chile}

\newcommand{\UConn}{\label{UConn} Department of Physics, University of Connecticut, Storrs, CT, 06269, USA}

\newcommand{\UCSD}{\label{UCSD} Center for Astrophysics and Space Sciences, Department of Physics,  University of California, San Diego, 9500 Gilman Drive, La Jolla, CA 92093, USA}

\newcommand{\UGent}{\label{UGent} Sterrenkundig Observatorium, Universiteit Gent, Krijgslaan 281 S9, B-9000 Gent, Belgium}

\newcommand{\ULyon}{\label{ULyon} Univ Lyon, Univ Lyon 1, ENS de Lyon, CNRS, Centre de Recherche Astrophysique de Lyon UMR5574,\\ F-69230 Saint-Genis-Laval, France}

\newcommand{\UMass}{\label{UMass} University of Massachusetts—Amherst, 710 N. Pleasant Street, Amherst, MA 01003, USA}

\newcommand{\UWyoming}{\label{UWyoming} Department of Physics and Astronomy, University of Wyoming, Laramie, WY 82071, USA}

\newcommand{\LAM}{\label{LAM} Aix Marseille Univ, CNRS, CNES, LAM (Laboratoire d’Astrophysique de Marseille), Marseille, France}

\newcommand{\UHawaii}{\label{UHawaii} Institute for Astronomy, University of Hawaii, 2680 Woodlawn Drive, Honolulu, HI 96822, USA}

\newcommand{\UCM}{\label{UCM} Departamento de F\'{\i}sica de la Tierra y Astrof\'{\i}sica, Universidad Complutense de Madrid, E-28040, Spain}

\newcommand{\IPARC}{\label{IPARC} Instituto de F\'{\i}sica de Part\'{\i}culas y del Cosmos IPARCOS, Facultad de Ciencias F\'{\i}sicas, Universidad Complutense de Madrid, E-28040, Spain}

\newcommand{\STScI}{\label{STScI} Space Telescope Science Institute, 3700 San Martin Drive, Baltimore, MD 21218, USA}

\newcommand{\McMaster}{\label{McMaster} Department of Physics and Astronomy, McMaster University, 1280 Main Street West, Hamilton, ON L8S 4M1, Canada}

\newcommand{\INAF}{\label{INAF} INAF -- Osservatorio Astrofisico di Arcetri, Largo E. Fermi 5, I-50157, Firenze, Italy}

\newcommand{\Sydney}{\label{Sydney} Sydney Institute for Astronomy, School of Physics A28, The University of Sydney, NSW 2006, Australia}

\newcommand{\UA}{\label{UA} Centro de Astronomía (CITEVA), Universidad de Antofagasta, Avenida Angamos 601, Antofagasta, Chile}

\newcommand{\CITA}{\label{CITA} Canadian Institute for Theoretical Astrophysics (CITA), University of Toronto, 60 St George St, Toronto, ON M5S 3H8, Canada}

\newcommand{\ASIAA}{\label{ASIAA} Institute of Astronomy and Astrophysics, Academia Sinica, No. 1, Sec. 4, Roosevelt Road, Taipei 10617, Taiwan}

\newcommand{\TKU}{\label{TKU} Department of Physics, Tamkang University, No.151, Yingzhuan Rd., Tamsui Dist., New Taipei City 251301, Taiwan}

\newcommand{\PSMA}{\label{PSMA} Penn State Mont Alto, 1 Campus Drive, Mont Alto, PA  17237, USA}

\newcommand{\ILL}{\label{ILL} ILL}

\newcommand{\stromlo}{\label{stromlo} Research School of Astronomy and Astrophysics, Australian National University, Mt Stromlo Observatory, Weston Creek, ACT 2611, Australia}


\author{
        Kathryn Kreckel\inst{\ref{HD}}  \and
           Oleg Egorov\inst{\ref{HD}}  \and
           Francesco Belfiore\inst{\ref{INAF}}  \and
           Brent Groves\inst{\ref{ICRAR}}  \and
           Simon C. O. Glover\inst{\ref{ITA}}  \and
           Ralf R. Klessen\inst{\ref{ITA},\ref{IWR}} \and 
           Karin Sandstrom\inst{\ref{UCSD}}  \and
           Frank Bigiel\inst{\ref{UBonn}}  \and
           Daniel A. Dale\inst{\ref{UWyoming}} \and
           Kathryn Grasha\inst{\ref{ANU}} \and
           Fabian Scheuermann\inst{\ref{HD}} \and
           Eva Schinnerer\inst{\ref{MPIA}} \and
           Thomas G. Williams\inst{\ref{MPIA}} 
}


\institute{\HD   \\  \email{kathryn.kreckel@uni-heidelberg.de}  \and  
 \INAF \and
 \ICRAR \and 
 \ITA \and
 \IWR \and
 \UCSD \and
 \UBonn \and
 \UWyoming \and
 \ANU \and
 \MPIA
}

   \date{Received XX; accepted XX}

 
  \abstract
   {}
   {Temperature uncertainties plague our understanding of abundance variations within the ISM.  Using the PHANGS-MUSE large program, we develop and apply a new technique to model the strong emission lines arising from \hii\ regions in 19 nearby spiral galaxies at $\sim$50 pc resolution and infer electron temperatures for the nebulae.  }
   {Due to the charge-exchange coupling of the ionization fraction of the atomic oxygen to that of hydrogen, the emissivity of the observed \oi $\lambda$6300/\ha\ line ratio can be modeled as a function of gas phase oxygen abundance (O/H), ionization fraction (\fion) and electron temperature (\te).  We measure (O/H) using a strong line metallicity calibration, and identify a correlation between \fion\ and \siii $\lambda$9069/\sii $\lambda$6716,6730, tracing ionization parameter variations.  }
   {We solve for \te, and test the method by reproducing direct measurements of \te(\nii $\lambda$5755) based on auroral line detections to within $\sim$600 K.  
   We apply this \textit{charge-exchange method} of calculating \te\ to 4,129 \hii\ regions across 19 PHANGS-MUSE galaxies. 
   We uncover radial temperature gradients, increased homogeneity on small scales, and azimuthal temperature variations in the disks that correspond to established  abundance patterns. 
   This new technique for measuring electron temperatures leverages the growing availability of optical integral field unit spectroscopic maps across galaxy samples, increasing the statistics available compared to direct auroral line detections.}
   {}


   \maketitle
%

\section{Introduction}

Star formation is regulated by the complex interplay of heating and cooling processes in the interstellar medium (ISM). The gaseous clouds that provide the raw materials for stars to form must cool sufficiently so that they can condense and collapse \citep{McKee2007, Klessen2016}. However, the resulting stars proceed to ionize and heat their surroundings, with the most massive stars contributing further mechanical energy and enriched materials at the end of their life via supernovae,  feeding back into their natal environment \citep{Maiolino2019}. This cycling of baryons on small scales is reflected by the temperature transitions from the dense cold molecular gas ($\sim$10 K) to the warm ionized medium ($\sim$10,000 K) found in \hii\ regions, to the super-heated shock waves driven by supernova explosions ($\sim$10$^6$ K). 

In particular, within \hii\ regions the electron temperature reflects not just the strength and hardness of the ionizing radiation field (O stars have temperatures up to $\sim$50,000 K), but the ability of the gas to efficiently cool through collisionally excited lines arising from heavier elements (e.g., carbon, oxygen, nitrogen). In fact, the balance between heating and cooling in these nebulae reflects predominantly the metal abundance present in the surrounding ISM \citep{Osterbrock2006}, although age variations also play some role \citep{Ho+2019}. 

Direct measurement of electron temperatures is possible using faint temperature-sensitive emission lines. 
In the radio, it is possible to use radio recombination lines to derive the electron temperatures for \hii\ regions, but these lines are very weak and while such observations are possible for Milky Way targets  \citep{Balser2015, Wenger2019, Pineda2019} they remain challenging for extragalactic targets \citep{Zhao1996, Kepley2011, Luisi2018, Kewley2019}. In the optical one can observe auroral lines (e.g.  \oiii $\lambda$4363, \nii $\lambda$5755, \siii $\lambda$6312, \oii $\lambda$7320,7330) in extragalactic systems, but they are also very faint, typically around $\sim$1\% of their strong line counterparts. Extensive long-slit observational projects have attempted to detect auroral lines across large samples of \hii\ regions in nearby galaxies, but collecting statistically-significant samples of these faint lines is challenging. Observational campaigns targeting low-mass, metal-poor galaxies like M33, where the auroral lines are brighter, can accumulate relatively large numbers of detections (e.g., 61 \hii\ regions with  \oiii $\lambda$4363 detections,  \citealt{Rosolowsky2008}), but this approach remains daunting for high-mass systems. 
In the CHemical Abundances of Spirals  (CHAOS) project, initial results from \cite{Berg2020} detect auroral lines for a total of 190 \hii\ regions across 4 galaxies, where they are able to make direct measurements of the electron temperatures.

With the advent of wide-field optical integral field spectrographs on large 8-meter class telescopes (e.g. VLT/MUSE and KCWI on Keck), it now becomes feasible to systematically observe much larger samples of \hii\ regions with auroral line detections, but the telescope time requirements remain high. As part of the PHANGS-MUSE survey \citep{Emsellem2022}, four hours spent surveying NGC 1672 resulted in  $\sim$80 \hii\ regions with \nii $\lambda$5755 auroral line detections \citep{Ho+2019}. As such, direct detections of tens of auroral lines per galaxy have become feasible but expensive.

In addition to these auroral line measurements remaining extremely challenging, with only tens of \hii\ regions we cannot uniformly sample the galaxy disks. Increased statistics on temperature (and metallicity) variations provides insights into the evolutionary processes regulating gas flows and galaxy evolution \citep{Kreckel2019, Sanchez-Menguiano2019, Ho+2019, Kreckel2020, Li2021}, but remain plagued by the systematics affecting strong-line metallicities.  We explore a new method to constrain the electron temperature, based on a combination of strong lines, with the goal of increasing the number of regions within which measurements of the electron temperature are possible. We present an overview of the physical motivation for the method in Section \ref{sec:method}. We describe the data used in Section \ref{sec:data}, and develop and validate our new method in Section \ref{sec:methoddev}. We discuss results in Section \ref{sec:results} and conclude in Section \ref{sec:conclusions}.


\section{Physical Motivation for the Method}
\label{sec:method}

Nature has gifted us with a convenient coincidence. The ionization potential of oxygen (13.618 eV) is very similar to the ionization potential of hydrogen (13.598 eV).  As such, the charge-exchange reaction O$^0$+H$^+$ $\leftrightarrow$ O$^+$+H$^0$ is very efficient, and acts to couple the ionization fraction of oxygen to the ionization fraction of hydrogen \cite[see, e.g.][]{Draine2011,Tielens2010}.  
This means that the number densities (which we denote as $n$(X), for species X) of oxygen and hydrogen are related by 
\begin{equation}
    \frac{n({\rm O}^0)}{n({\rm O})} \approx  \frac{n({\rm H}^0)}{n({\rm H})},
\end{equation}
where
\begin{equation}
    n({\rm H}) \equiv n({\rm H}^0) + n({\rm H}^+)
\end{equation}
and
\begin{equation}
    n({\rm O}) \equiv n({\rm O}^0) + n({\rm O}^+)
\end{equation}
Note that we define the ionization fraction of hydrogen as
\begin{equation}
\label{eqn:fion_def}
    f_{\rm ion} \equiv \frac{n({\rm H}^+)}{n({\rm H})} = \frac{n({\rm H}^+)}{n({\rm H}^+) + n({\rm H}^0)} =  \frac{n({\rm H}^+)/n({\rm H}^0)}{n({\rm H}^+)/n({\rm H}^0)+1}. 
\end{equation}
Therefore, for line emission associated with the relevant ions,  
\begin{equation}
\frac{[{\rm O}\,\textsc{i}] \lambda6300}{{\rm H}\alpha} \propto \frac{n({\rm O}^0)}{n({\rm H}^+)} = \frac{n({\rm O})}{n({\rm H})} \frac{n({\rm H}^0)}{n({\rm H}^+)} = \frac{n({\rm O})}{n({\rm H})} \frac{1-f_{\rm ion}}{f_{\rm ion}}
\end{equation}
This is significant as the line ratio \oi/\ha\ depends only on temperature (through the proportionality pre-factor), metallicity (\textit{n}(O)/\textit{n}(H)) and the ionization fraction of hydrogen (\fion).

Using the latest calculations of the \oi\ and \ha\ emission rates \citep{Barklem2007, Dong2011}, 
we derive the emissivity of \oi $\lambda$6300 relative to \ha\ (as outlined in Appendix \ref{app:derivation}) to be
\begin{equation}
\begin{split}
    \frac{[{\rm O}\textsc{i}] \lambda6300}{{\rm H}\alpha} & = 
    8492 \frac{f_{\rm OI}(T)}{T_{4}^{-0.942 - 0.031 \ln T_{4}}} \exp \left(-\frac{2.284}{T_{4}} \right) \\
    & \times \frac{n({\rm O})}{n({\rm H})} \\
    & \times  \frac{n({\rm H}^0)}{n({\rm H}^+)} \xi
\end{split}
\label{eqn}
\end{equation}
where $T_4$ is the electron temperature (\te) in units of 10,000 K, f$_{\rm OI}$ is defined as
\begin{equation}
    f_{\rm OI}(T) = \sum_{i=0}^{6} a_{i} T_{4}^{i}
    \label{eqn:foi}
\end{equation}
where the coefficients $a_i$ are listed in Table \ref{table:coeffs}, 
and $\xi$ is a function of the hydrogen ionization ratio
\begin{equation}
    \xi = \frac{1+n({\rm H}^0)/n({\rm H}^+)}{\frac{8}{9}+n({\rm H}^0)/n({\rm H}^+)}.
\label{eqn:xi}
\end{equation}
This factor $\xi$ is just above unity and ranges from 1 to 9/8. This equation holds if the \oi\ and \ha\ emission are co-spatial, arising from the same parcel of gas.  We discuss this in more detail below, and explore the implications of this assumption in Section \ref{sec:cloudy_cospatial}.  This is an update to the relation provided in \cite{Reynolds1998}.

\begin{table}[b]
\caption{Coefficients for Equation \ref{eqn:foi}.\label{table:coeffs}}
\begin{tabular}{cl}
Coefficient & Value \\
\hline
$a_{0}$ & $-1.00$ \\
$a_{1}$ & $+6.4854011$ \\
$a_{2}$ & $-4.17358515$ \\
$a_{3}$ & $+1.81446389$ \\
$a_{4}$ & $-0.514022051$ \\
$a_{5}$ & $+8.39069326 \times 10^{-2}$ \\
$a_{6}$ & $- 5.93343677 \times 10^{-3}$ \\
\hline
\end{tabular}
\end{table}

The aim of this paper is to utilize a measurement of \oi/\ha, combined with an estimate of $f_{\rm ion}$, to infer \te. While \te\ can be directly measured through the detection of faint auroral lines, there exist no well-established prescriptions to infer \fion\ from observations of strong lines.  In this paper, we determine an empirical relation between \fion\ and strong line diagnostic ratios in order to allow us to solve for \te.
This \textit{charge-exchange method} of determining \te\ represents a novel approach which may be widely applicable to existing data sets.
We develop our method using a set of line fluxes measured arising from integrated \hii\ regions, as extragalactic \hii\ regions are generally unresolved. 

For this charge-exchange method to apply, we assume that electron collisions dominate the excitation of the \oi\ line, rather than collisions with atomic hydrogen. The critical fractional ionization above which electron collisions dominate is $\sim$10$^{-3}$--10$^{-4}$, which we expect to be true for the \hii\ regions we consider. We also assume that the emitting regions are in the low density limit, rather than the local thermodynamic equilibrium limit. Since the critical density is $\sim$ 10$^{6}$ cm$^{-3}$, this should also be valid for all of the \hii\ regions we consider.

There are limitations to this method that are worth highlighting here, and that will be addressed in more detail in Section \ref{sec:whydoesitwork}. This model compares the \oi\ and \ha\ line emission. However, while the majority of the hydrogen in an \hii\ region is ionized and emitting in \ha,  \oi\ will be emitted mainly in an outer shell near the ionization front. In applying this method, we are assuming that \te\ is uniform across both the \ha\ and \oi\ emitting regions. Ideally, we would further consider emission that is co-spatial, arising from the same parcel of gas. However, for unresolved \hii\ regions, we can consider only emission integrated across the entire nebula. In our approach, we argue that \fion\ constrains the size of the partially ionized zone, accounting for this discrepancy in emitting zones. Over larger (kpc scale) regions of the diffuse ionized gas, the radiation field is more diluted (low ionization parameter) and therefore the partially ionized zone is wider and would lead to a better correspondence between \ha\ and \oi\ emitting regions.

This technique was initially developed in relation to diffuse ionized gas in the Milky Way \citep{Reynolds1998}, and later applied to four individual \hii\ regions in the Milky Way \citep{Hausen2002}. We further develop this method, and aim to apply it to the thousands of \hii\ regions within the PHANGS-MUSE sample. \oi\ emission can be challenging to measure, as it is faint and can be blended with the strong and highly variable telluric  airglow emission. In this work we target nearby ($\sim$10--20 Mpc) galaxies with sufficiently large velocity offset from terrestrial \oi, and with sufficient depth to detect \oi\ across a wide sample of \hii\ regions.  Here, we explore the regime where an entire \hii\ region is integrated over 50--100~pc scales.  While the main goal is to infer \te, we note that we need to assume a metallicity in order to apply the charge exchange method. Therefore, our \te\ measurement is dependent on the assumed strong-line metallicity, and using it to infer metallicties using the direct method would be circular. We test the metallicity dependence of the method in Section  \ref{sec:validation}. Some iterative approach may be suitable, but this is beyond the scope of this work.


\section{Data}
\label{sec:data}

\subsection{CHAOS}
To develop our method, we explored data from the CHAOS project. 
\cite{Berg2020} summarizes the latest results (see also \citealt{Berg2015, Croxall2015, Croxall2016}), including four galaxies (\galaxyname{NGC}{628}, \galaxyname{NGC}{3184}, \galaxyname{M}{51} and \galaxyname{M}{101}) with a total of 190 \hii\ regions with auroral line detections as measured from integrated \hii\ region spectra. Of these, 121 have detections of \nii $\lambda$5755, 131 have \siii $\lambda$6312, 154 have \oii $\lambda$7320,7330, and 72 have \oiii $\lambda$4363. These auroral lines, arising from high energy levels, are used in combination with the nebular lines from lower energy levels of the same ion to compute electron temperatures for each ion. Using different ions to represent different temperatures across ionization zones in the nebulae,  \cite{Berg2020} combined these measurements to compute the oxygen abundance, 12+log(O/H), using the direct method. 
In this way, from the CHAOS data set, we have measurements of \te\ and O/H for 190 \hii\ regions, as well as a full catalog of strong emission lines (\oii $\lambda$3727,3729, \oiii $\lambda$5007, \hb, \oi $\lambda$6300, \nii $\lambda$6583, \ha, \sii $\lambda$6716,6730 and  \siii $\lambda$9069). All line fluxes from their catalog have been corrected for extinction  using the Balmer series (\ha, \hb). 

These observations were carried out using the Multi-Object Double Spectrographs (MODS; \citealt{Pogge2010}) with slit masks on the Large Binocular Telescope (LBT).  Slits placed on individual \hii\ regions were designed to be 1\arcsec\ wide and 10\arcsec\ long. This slit-width is well matched to the seeing during observations, however as the brightest \hii\ regions are typically selected this does not necessarily encompass the full \hii\ region size, which for giant \hii\ regions can be as large as $\sim$100 pc \citep{Azimlu+2011, Mannucci2021}.  At the 7--11 Mpc distances of the four targets this slit-width corresponds to 1\arcsec= 30--50 pc, such that some slit losses may be expected.  No correction is applied to account for the local diffuse ionized gas (DIG) background emission.  

\subsection{PHANGS-MUSE}
\label{sec:data-muse}
\begin{table*}[ht!]
\centering
\caption{Properties of the galaxies in the PHANGS-MUSE sample}
\begin{tabular}{
    l
    c
    c
    c
    c
    c
    c
    c
    c
    }
\hline
{Name} & {Type} & {Dist$^\mathrm{a}$} & {r$_{\rm eff}^\mathrm{b}$} & {Inclination$^\mathrm{c}$} & {Pos Angle$^\mathrm{c}$} & M$_{\rm star}^\mathrm{b}$ & {$E(B-V)^\mathrm{d}_{\rm MW}$} & v$_{\rm sys}^\mathrm{b}$ \\
 &  & [Mpc] & [arcsec] &  [deg] & [deg] & [M$_\odot$] & [mag]   & [km/s] \\
\hline
\galaxyname{NGC}{0628} & Sc &  9.8 & 82 &     8 &   20 &     2.2e+10 & 0.06 &  650 \\
\galaxyname{NGC}{1087} & Sc & 15.9 & 42 & 42 &  359 &     8.6e+09 & 0.03 & 1501 \\
\galaxyname{NGC}{1300} & Sbc & 19.0 & 71 &  31 &  278 &     4.1e+10 & 0.03 & 1545 \\
\galaxyname{NGC}{1365} & Sb & 19.6 & 195$^{e}$ &  55 &  201 &     9.8e+10 & 0.02 & 1613 \\
\galaxyname{NGC}{1385} & Sc & 17.2 & 40 &  44 &  181 &     9.5e+09 & 0.02 & 1476 \\
\galaxyname{NGC}{1433} & SBa & 18.6 & 48 &  28 &  199 &     7.3e+10 & 0.01 & 1057 \\
\galaxyname{NGC}{1512} & Sa & 18.8 & 52 &  42 &  261 &     5.2e+10 & 0.01 &  871 \\
\galaxyname{NGC}{1566} & SABb & 17.7 &  37 & 29 &  214 &     6.1e+10 & 0.01 & 1483 \\
\galaxyname{NGC}{1672} & Sb & 19.4 & 36 &  42 &  134 &     5.4e+10 & 0.02 & 1318 \\
\galaxyname{NGC}{2835} & Sc & 12.2 & 56 &  41 &    1 &     1.0e+10 & 0.09 &  867 \\
\galaxyname{NGC}{3351} & Sb & 10.0 & 63 &  45 &  193 &     2.3e+10 & 0.02 &  774 \\
\galaxyname{NGC}{3627} & Sb & 11.3 & 66 &  57 &  173 &     6.8e+10 & 0.03 &  715 \\
\galaxyname{NGC}{4254} & Sc & 13.1 &  38 & 34 &   68 &     2.7e+10 & 0.03 & 2388 \\
\galaxyname{NGC}{4303} & Sbc & 17.0 &  42 & 23 &  312 &     3.3e+10 & 0.02 & 1559 \\
\galaxyname{NGC}{4321} & SABb & 15.2 &  75 & 38 &  156 &     5.6e+10 & 0.02 & 1572 \\
\galaxyname{NGC}{4535} & Sc & 15.8 & 82 &  44 &  179 &     3.4e+10 & 0.02 & 1953 \\
\galaxyname{NGC}{5068} & Sc &  5.2 &  78 & 35 &  342 &     2.5e+09 & 0.09 &  667 \\
\galaxyname{NGC}{7496} & Sb & 18.7 &  42 & 35 &  193 &     9.9e+09 & 0.01 & 1639 \\
\galaxyname{IC}{5332} & SABc &  9.0 &  83 & 26 &   74 &     4.7e+09 & 0.01 &  699 \\
\hline
\multicolumn{5}{l}{Adopted from the PHANGS sample table (v1p6; \citealt{Leroy2021})} \\
\multicolumn{5}{l}{$^\mathrm{a}$ from \citet{Anand2021}} &
\multicolumn{4}{l}{$^\mathrm{b}$ from \citet{Leroy2021}} \\
\multicolumn{5}{l}{$^\mathrm{c}$ from \citet{Lang2020}} & 
\multicolumn{4}{l}{$^\mathrm{d}$ from \citet{Schlafly2011}} \\
\multicolumn{8}{l}{$^\mathrm{e}$ Due to AGN bias, derived from the  scale length (l$_*$) as r$_{\rm eff}$ = 1.41 l$_*$ following Equation 5 in \citet{Leroy2021}.} \\
\end{tabular}
\label{tbl:phangs-muse}
\end{table*}

After initial investigations into the charge-exchange method using the CHAOS data, we develop and test our method using \hii\ regions selected from the PHANGS-MUSE dataset (\citealt{Emsellem2022}, Groves et al. in prep). The Physics at High Angular resolution in Nearby GalaxieS (PHANGS) collaboration has targeted nearby spiral disk galaxies along the star-forming main sequence (Table \ref{tbl:phangs-muse}). Targets were selected to be nearby (D$<$19 Mpc, 1\arcsec$<$100 pc) and moderately inclined (inclination < 60$^\circ$). Observations of 19 galaxies were carried out using the Very Large Telescope/Multi Unit Spectroscopic Explorer (VLT/MUSE; \citealt{Bacon2010}) instrument primarily as part of a MUSE large program (PI: Schinnerer), which observed 172 individual MUSE pointings across these 19 galaxies over 4800--9300 \AA\ at $<$1\arcsec\ resolution.  The data reduction pipeline and data analysis pipeline used to reduce this data set and extract emission line maps are described in detail in \cite{Emsellem2022}.  This results in an optical integral field spectroscopic data cube, as well as integrated line emission maps of a wide range of both strong (\oiii $\lambda$5007, \hb, \oi $\lambda$6300, \nii $\lambda$6583, \ha, \sii $\lambda$6716,6730,  \siii $\lambda$9069) and auroral (\nii $\lambda$5755, \siii $\lambda$6312, \oii $\lambda$7320,7330) emission line fluxes. For the \oiii\ and \nii\  doublets we measure only the stronger line, and for the \siii\ doublet we measure only the bluer line, and assume fixed atomic ratios \citep{Osterbrock2006, Tayal2019} as

\begin{equation}
\begin{array}{lcl}
    \rm \oiii \lambda4958 & = & \rm 0.35 \times \oiii \lambda5007 \\
    \rm \nii \lambda6548  & = & \rm 0.34 \times \nii \lambda6584 \\
    \rm \siii \lambda9532 & = & \rm 2.5 \times \siii \lambda9069 
\label{ratios}
\end{array}
\end{equation}

Using the \ha\ emission line maps, \cite{Santoro2022} morphologically identified individual nebular regions using the \mbox{\textsc{hiiphot}} package \citep{Thilker2000}, using a technique similar to that described in \cite{Kreckel2019}. This works by identifying peaks in the \ha\ distribution, and growing those peaks until a neighboring region is encountered or a termination criterion set by the flux gradient is reached.  These nebular region masks are then applied to the original data cube, and the spectra within each nebular region are summed and re-fit using the data analysis pipeline.  The resulting nebular catalog compiles a list of objects with their associated line fluxes, extinction corrected using the Balmer decrement, with approximately 31,000 regions identified across the full sample. All galaxies are at sufficiently high systemic velocities (v$_{\rm sys} > $ 650 km s$^{-1}$) that the critical \oi\ emission line is not blended with telluric airglow (Table \ref{tbl:phangs-muse}).

We apply the standard BPT \citep{Baldwin1981} diagnostic diagrams requiring S/N $>$ 3 in all relevant lines, and construct our \hii\ region catalog by selecting those regions from the nebular catalog which are consistent with both of the photoionization demarcations given by \cite{Kauffmann2003} in the \oiii/\hb\ vs. \nii/\ha\ diagram and the \cite{Kewley2001} lines in the \oiii/\hb\ vs. \sii/\ha\ diagram. Regions that have S/N $<$ 3 in any of the relevant diagnostic lines are excluded from the sample. We further exclude \hii\ regions within one 1\arcsec\ of the field edge or a foreground star. This results in an \hii\ region catalog consisting of $\sim$24,000 objects. Given that the median size of our \hii\ regions is approximately consistent with our seeing limit, we expect that most of these \hii\ regions are unresolved. In this sense, we always contain the full spatial extent of the \hii\ region in our integrated spectrum.

The PHANGS-MUSE spectra also include detections of the faint \nii\ $\lambda$5755 auroral line \citep{Ho+2019}.  
Fitting of this faint line (typically less than 1\% of the \ha\ line flux) is carried out using the integrated \hii\ region spectra, subtracting the stellar population fit, and performing a Gaussian fit on the residual.  Given the faintness of the line, very slight offsets in the stellar population fitting can result in systematic residuals at the location of the \nii\ line. To improve the flux estimate from our Gaussian fit, we fix the velocity centroid to have the same systemic velocity as we measure for the brighter \ha\ line, but additionally allow for a local linear background to account for systematics in the stellar population fit. Errors are determined by sampling the error spectrum and repeating this analysis 100 times.  This results in $\sim$840 \hii\ regions where the \nii\ $\lambda$5755 line is detected at S/N $>$10.  
As with the other measured lines, we correct for extinction in using the Balmer decrement, combine this with the dereddened measurement of \nii\ $\lambda$6583 and use \textsc{pyneb} \citep{Luridiana2015} assuming $n_{\rm e} = 10 \, {\rm cm}^{-3}$, consistent with the measured values \citep{Barnes2021}, to compute \te(\nii).

For the fainter and unresolved regions the DIG background may introduce uncertainties in recovering the intrinsic line fluxes. In order to measure the DIG contribution at the position of each \hii\ region, we mask out all nebular regions and compute the median line flux within a 10\arcsec$\times$10\arcsec\ box around each region for the lines most strongly emitted by the DIG (\hb, \oi, \ha, \nii, \sii). We require a 3$\sigma$ detection of a DIG signal in order to include it in our calculations. 
As implementing the DIG subtraction is a fairly uncertain process, we do not attempt any DIG subtraction. Instead, we aim to minimize the impact of the DIG on our measurements by requiring that \hii\ regions have a $>$50\% contrast of all lines against the surrounding DIG background.  Approximately 35\% of all \hii\ regions do not have significant detections in all necessary emission lines against the DIG background, and are excluded from further analysis.

\subsection{Strong line O/H abundances}
As this method relies on knowing O/H, we apply a strong line abundance prescription to both the CHAOS and PHANGS \hii\ region catalogs.  Our preferred strong line abundance prescription is the S calibration from \cite{Pilyugin2016}, an empirical method that relies on the combination of three diagnostic line ratios. This provides an improved ability to remove degeneracies between metallicity  and ionization parameter  when compared to prescriptions that use only one or two diagnostic ratios (for some discussion see \citealt{Ho2019_machine} and \citealt{Kreckel2019}). It has been shown to produce qualitatively similar results to the \cite{Dopita2016} N2S2 prescription \citep{Kreckel2019}, and achieves small systematic uncertainties \citep{Metha2021}. 

The S calibration relies on the following three standard diagnostic line ratios: 
\begin{equation}
\begin{array}{l}
{\rm N}_2  = ({\rm \nii \lambda 6548+ \lambda 6584}) /{{\rm H}\beta },  \\
{\rm S}_2  = ({\rm \sii \lambda 6717+ \lambda 6731}) /{{\rm H}\beta },  \\
{\rm R}_3  = ({{\rm \oiii} \lambda 4959+ \lambda 5007}) /{{\rm H}\beta }.
\end{array}
\end{equation}
The prescription is defined separately over the upper and lower branches in log~${\rm N}_{2}$. The upper branch
(log~${\rm N}_{2} \ge -0.6$) is calculated as
\begin{eqnarray}
\footnotesize
       \begin{array}{lll}
     {\rm 12+log(O/H)}  & = & \rm  8.424 + 0.030 \, \log (R_{3}/S_{2}) + 0.751 \, \log N_{2}   \\  
                          & + & \rm (-0.349 + 0.182 \, \log (R_{3}/S_{2}) + 0.508 \log N_{2})   \\ 
                          & \rm \times & \log S_{2}   \\ 
     \end{array}
\label{equation:ohsu}
\end{eqnarray}
and the lower branch
(log~$N_{2} < -0.6$) is calculated as
\begin{eqnarray}
\footnotesize
       \begin{array}{lll}
     {\rm 12+log(O/H)}  & = &  \rm 8.072 + 0.789 \, \log (R_{3}/S_{2}) + 0.726 \, \log N_{2}   \\  
                          & + & \rm  (1.069 - 0.170 \, \log (R_{3}/S_{2}) + 0.022 \log N_{2})    \\ 
                          & \times & \rm \log S_{2}   \\ 
     \end{array}
\label{equation:ohsl}
\end{eqnarray}


\section{Method development}
\label{sec:methoddev}

\begin{figure*}
\includegraphics[width=7in]{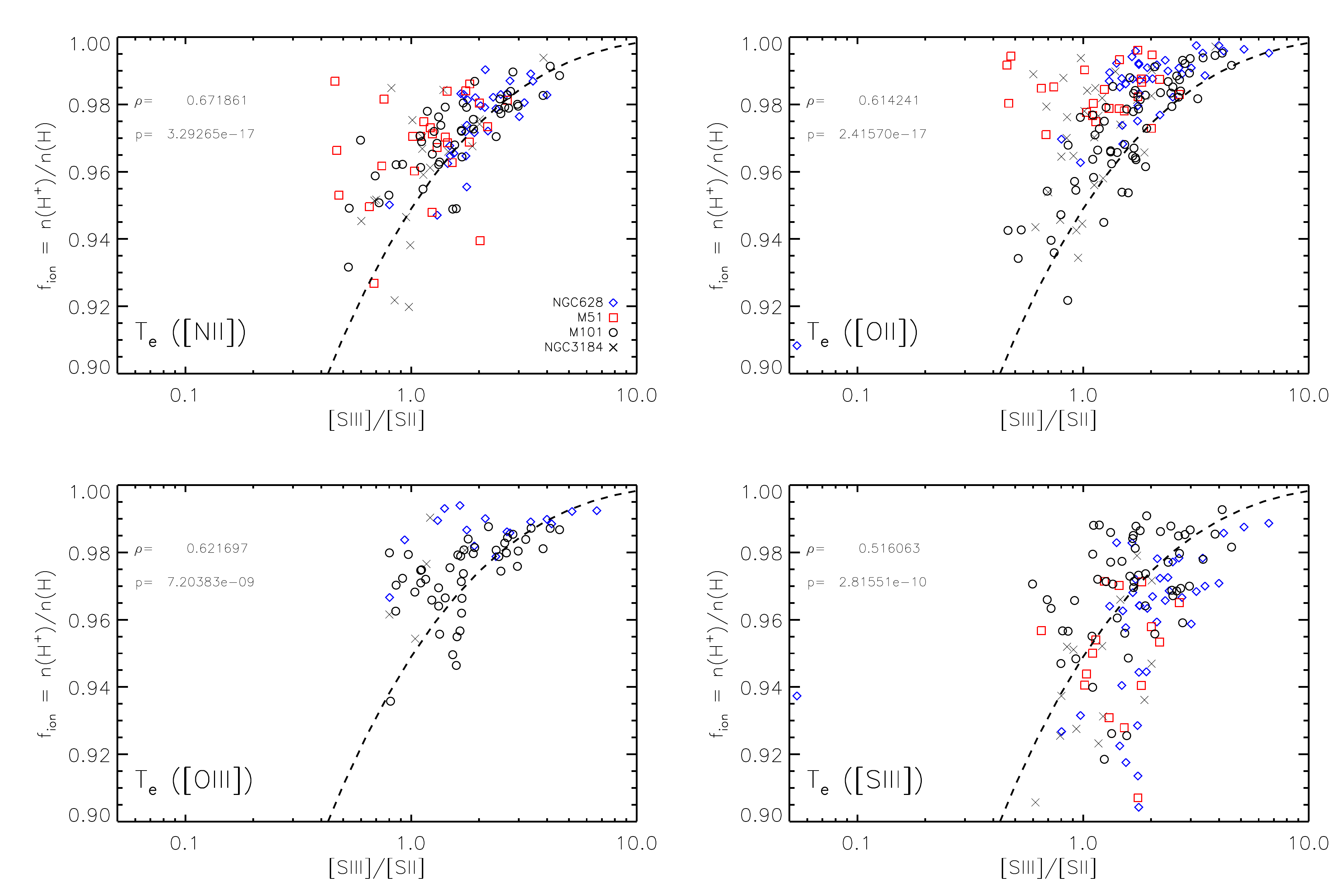}
\caption{For the CHAOS \hii\ regions, we model \fion\  based on Equation \ref{eqn}, assuming different input electron temperatures based on auroral line measurements (\te(\nii), \te(\oii), \te(\oiii), \te(\siii)) as well as their reported 12+log(O/H) and \oi/\ha. Each of the four galaxies is shown with separate symbols, and the Spearman's rank correlation coefficient ($\rho$) and its significance (p) are shown in each figure. Each value of \fion\ is plotted as a function of the \siii/\sii\ line ratio, which robustly traces changes in the ionization parameter \citep{Kewley2002}. We identify \te(\nii) as showing the strongest correlation ($\rho=0.69$) and a large number of detections, and therefore choose to adopt \te(\nii) as a reference temperature for the rest of the analysis in this paper. The dashed line shows the fit derived in Section \ref{sec:varfion}, using a combination of PHANGS-MUSE and CHAOS \hii\ regions. 
This same line is overplotted for the other panels (dashed lines) to allow comparison across the ionic temperatures.  
\label{fig:chaos_auroral}}
\end{figure*}

\begin{figure*}
\centering
	\includegraphics[width=3in]{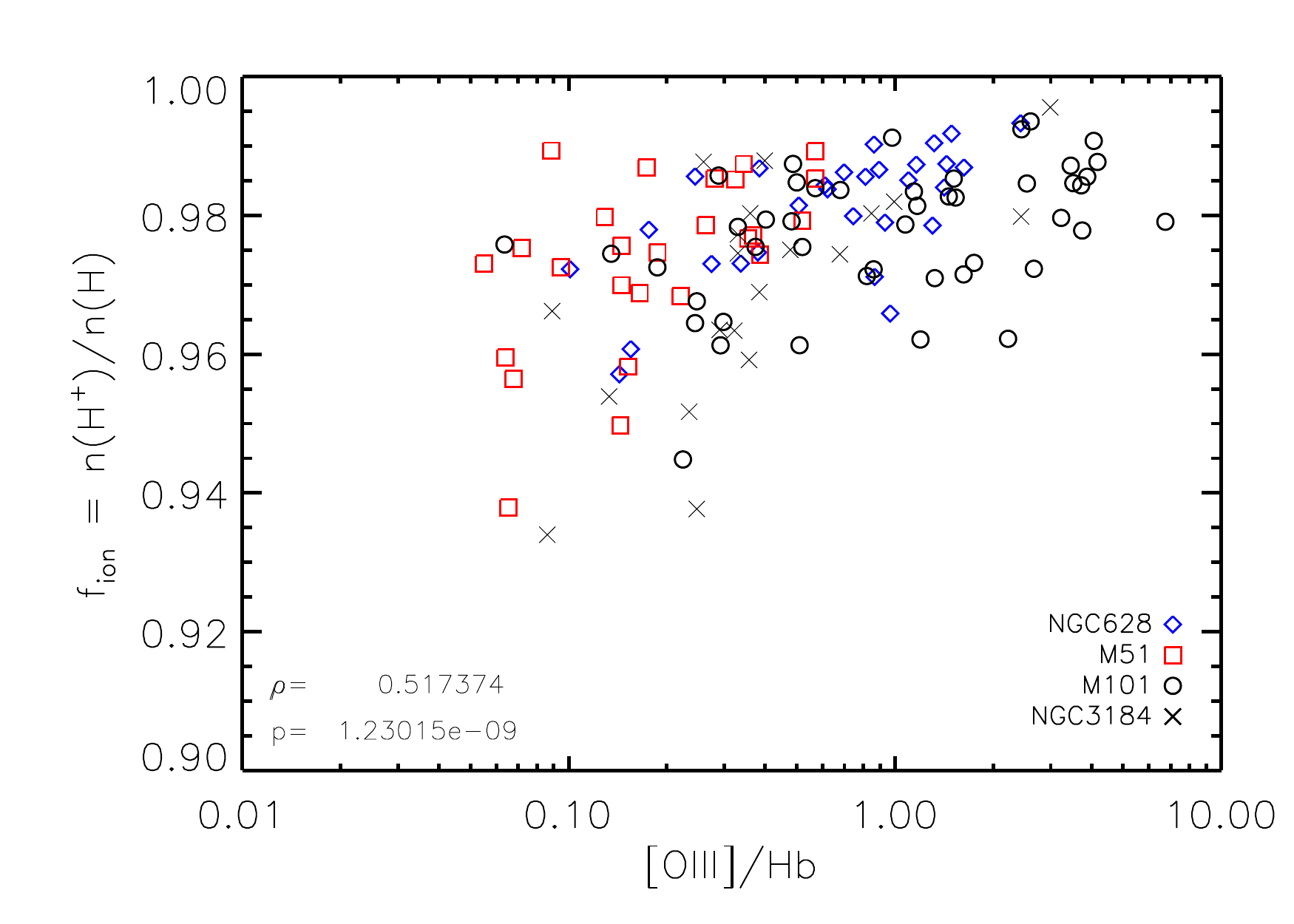}
	\includegraphics[width=3in]{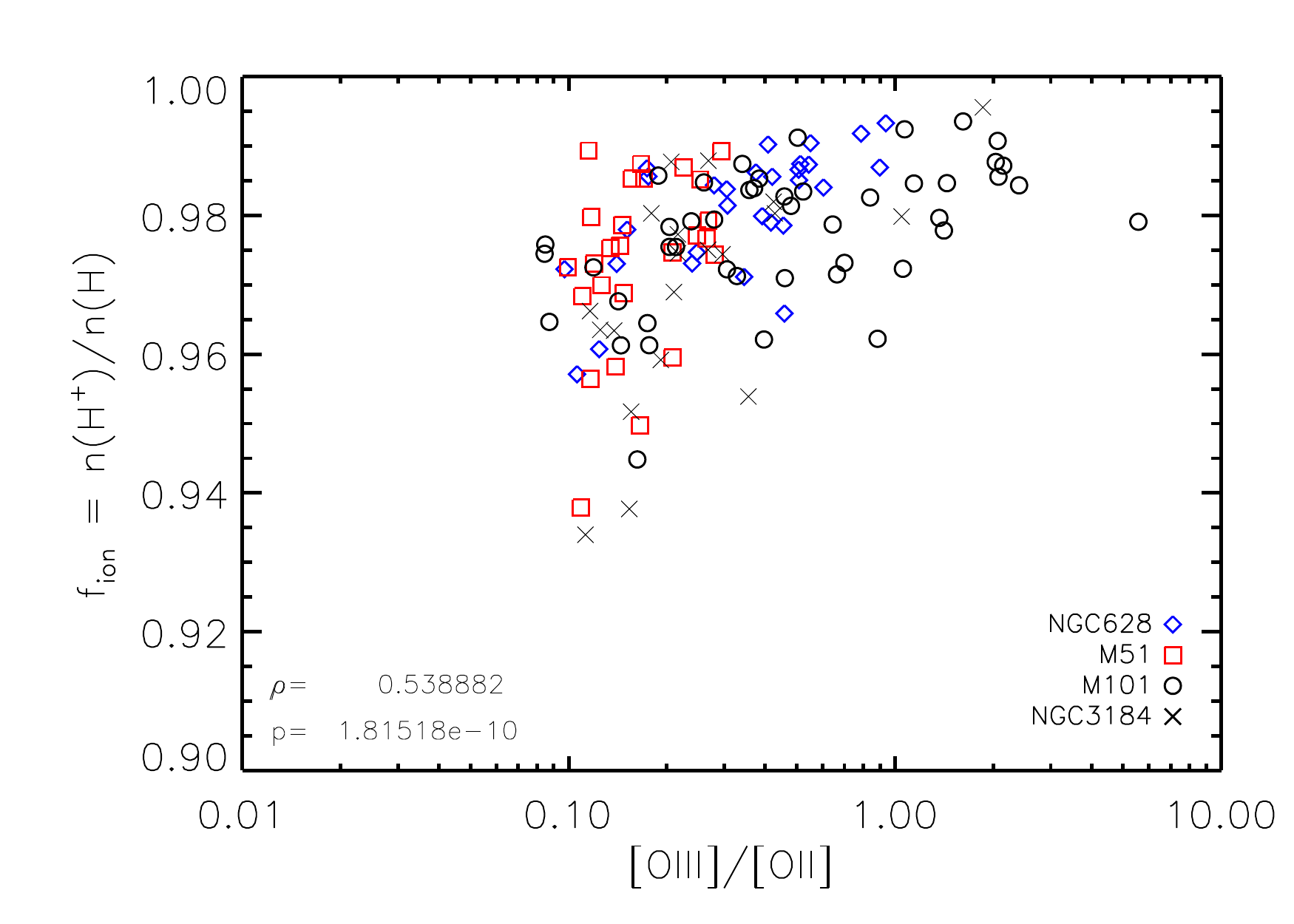}    
    \caption{We demonstrate further (weaker) correlations between \fion\ and line ratios tracing changes in ionization parameter (\oiii/\hb, left; \oiii/\oii, right) for \hii\ regions in the CHAOS sample, using \te(\nii) as our fiducial temperature measurement. Each of the four galaxies is shown with separate symbols, and the Spearman's rank correlation coefficient ($\rho$) and its significance (p) are shown in each figure. Weaker correlations are seen ($\rho$ $\sim$ 0.5) compared to the correlations with \siii/\sii\ (Figure \ref{fig:chaos_auroral}).
    \label{fig:chaos_fion}}
\end{figure*}

\begin{figure*}
\centering
    \includegraphics[width=2.3in]{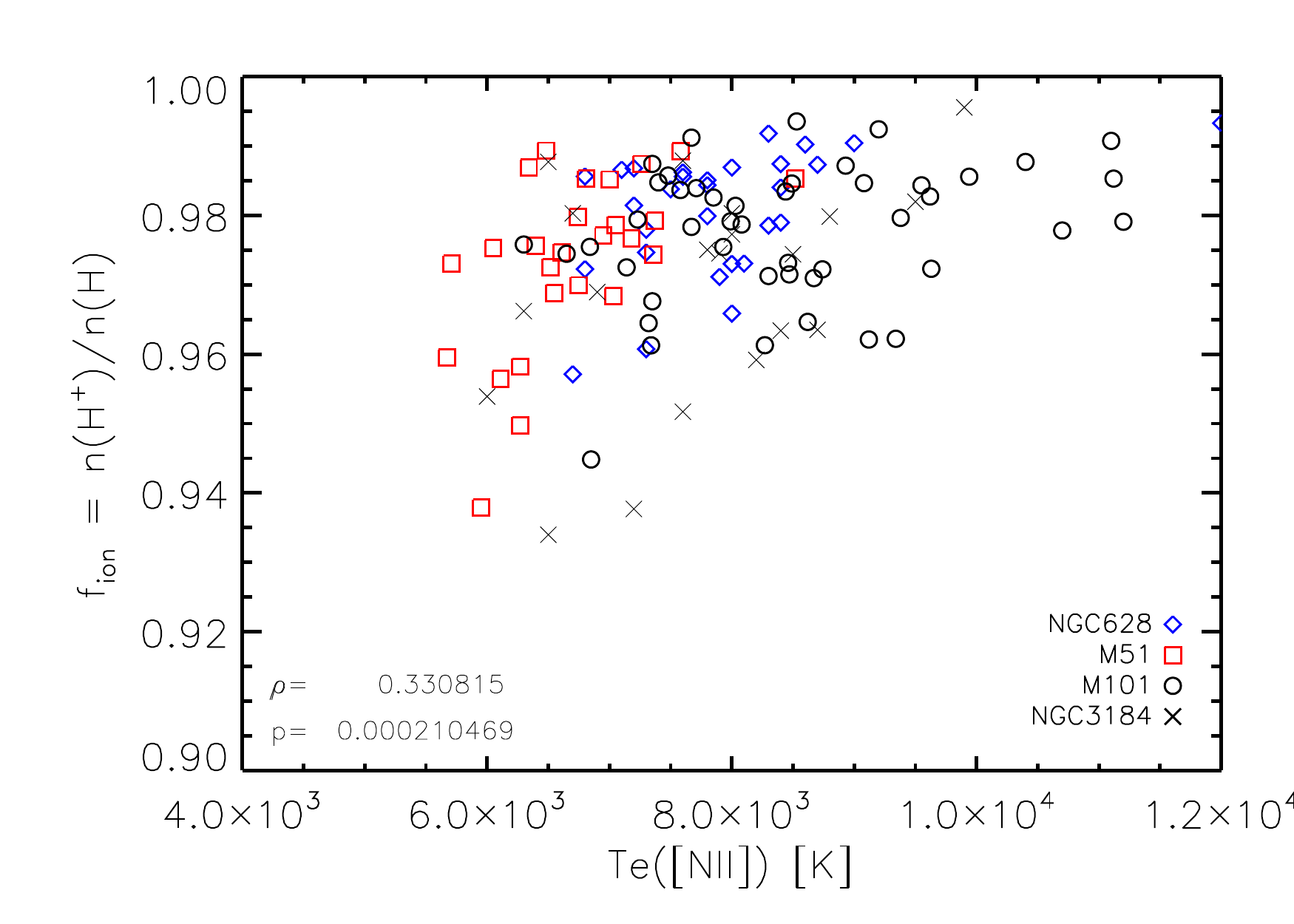}
	\includegraphics[width=2.3in]{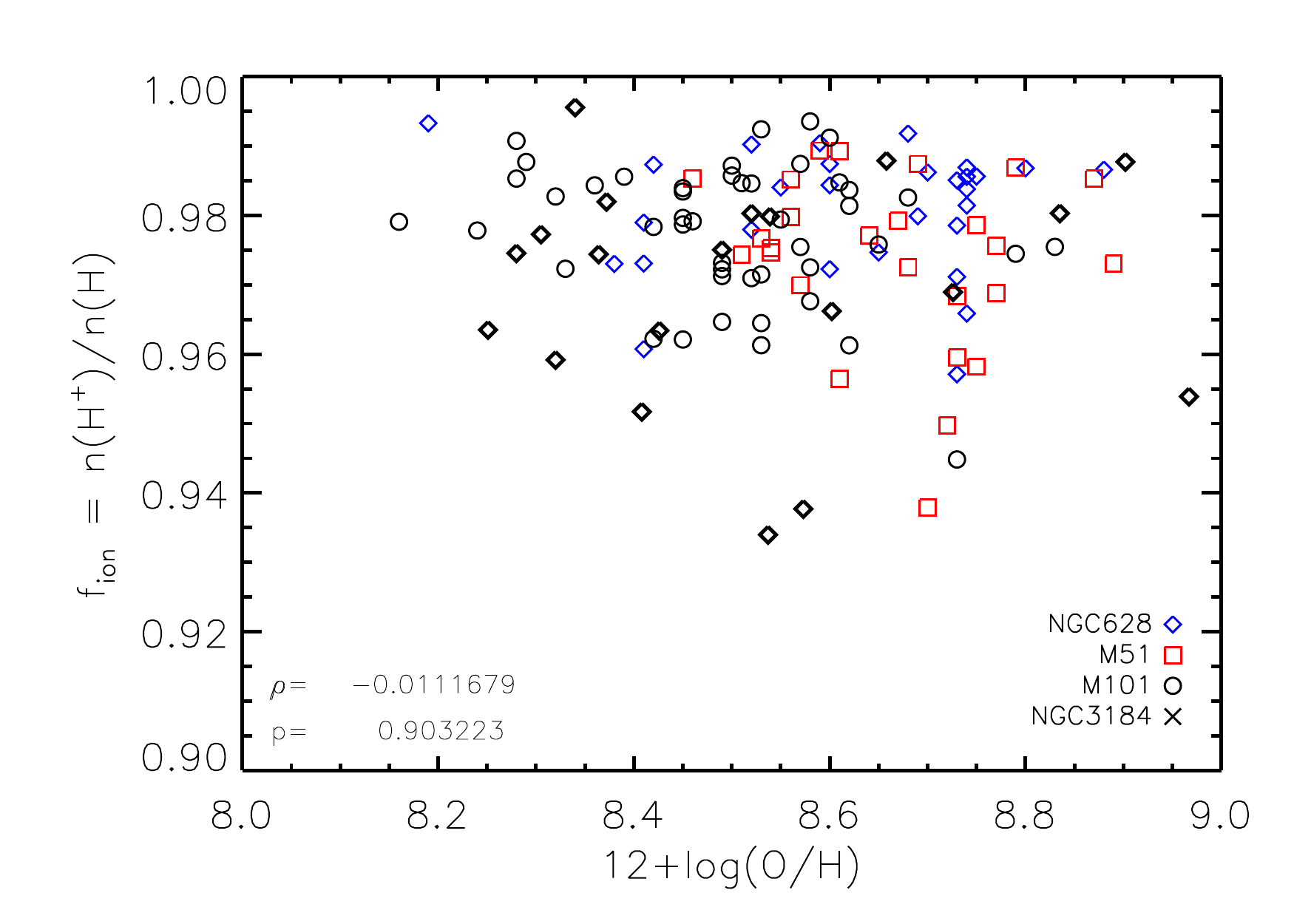}
	\includegraphics[width=2.3in]{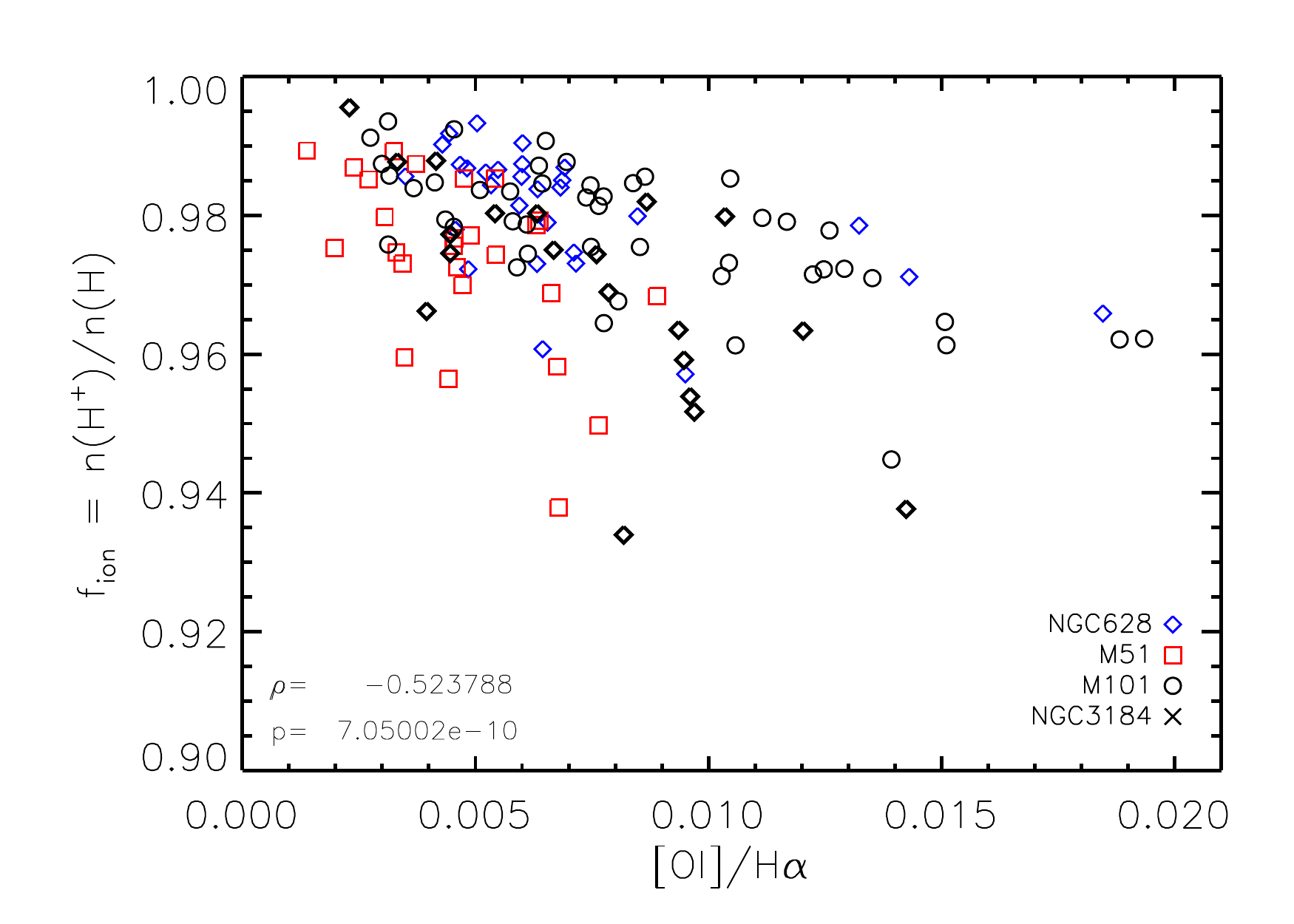}
    \caption{ Correlating \fion\ with the three input parameters (\te(\nii), 12+log(O/H),\oi/\ha) for each \hii\ region in the CHAOS sample. Each of the four galaxies is shown with separate symbols, and the Spearman's rank correlation coefficient ($\rho$) and its significance (p) are shown in each figure. The \fion\ variations correlate most strongly with changes in \oi/\ha, but show a weaker correlation than the trends with \siii/\sii\ in Figure \ref{fig:chaos_auroral}. 
    \label{fig:chaos_fion_a}}
\end{figure*}

Using the four galaxies in the CHAOS dataset, we begin by using their measured \oi/\ha\ , \te\ based on auroral line detections, and direct method metallicity in order to infer \fion\ for each of their \hii\ regions (Figure \ref{fig:chaos_auroral}). This direct method metallicity is derived by associating different \te\ values to different ionization zones in the nebula, such that the final metallicity is not independent of the \te\ measurements. Using the S calibration to derive metallicities provides similar results, although with increased scatter.   Based on the auroral line detections of \nii, \oii, \oiii, and \siii\ this approach provides four different measurements of \te, reflecting the ionic temperature in the zone where each ion is dominant. We find \fion\ values ranging from 0.93 to 0.99, and a median value of \fion=0.97 across all \hii\ regions. 

As we are integrating over the entire \hii\ region, although \oi\ is not emitted throughout the entire nebula, we expect that \fion\ will change systematically with the ionization structure of the nebulae and the size of the partially ionized region. To that end, we look for correlations with line ratios that depend on changes in the ionization parameter 
\begin{equation}
    q \equiv \frac{{\rm Q}({\rm H}^0)}{4 \pi {\rm R}^2 n({\rm H})},
\label{eqn:q}
\end{equation}
where Q(H$^0$) is the number of ionizing photons emitted per second, R is the distance between the central source and the emitting material and $n$(H) is the hydrogen density.  In photoionization models, the ionization parameter q correlates linearly with the log of \oiii/\oii, \siii/\sii, and \oiii/\hb, with both \oiii/\oii\ and \oiii/\hb\ showing additional dependences on metallicity \citep{Kewley2002, Dors2011}. Note that here we define: \\
\begin{equation}
\begin{array}{l}
{\rm \oiii/\oii} \equiv {\rm \oiii} \lambda5007/{\rm \oii} \lambda3726,3729 \\
{\rm \siii/\sii} \equiv {\rm \siii} \lambda9069,9532/{\rm \sii} \lambda6717,6731 \\
{\rm \oiii}/{\rm H}\beta \equiv {\rm \oiii} \lambda5007/{\rm H}\beta.
\end{array}    
\end{equation}

We find a strong correlation between \fion\ and \siii/\sii\ (as a proxy for ionization parameter) for all four ions (Figure \ref{fig:chaos_auroral}), with \te(\nii) producing the strongest correlation (as judged by the Spearman's rank correlation coefficient, $\rho$=0.67). This is reasonable, as the temperature of the ion tracing the low-ionization zone of the nebula is best suited to describe the \oi\ emitting region (\citealt{Berg2020}; see also Section \ref{sec:whydoesitwork}). It also shows the most uniform trend across all four galaxies in the sample. Given the strong correlation and relatively high number of detections, we choose to adopt \te(\nii) as a reference temperature for the rest of the analysis in this paper.

Weaker correlations are seen with \oiii/\hb\ or \oiii/\oii ($\rho$ $\sim$0.5, Figure \ref{fig:chaos_fion}). Some correlations are also seen with the input parameters used in calculating \fion\ (Figure \ref{fig:chaos_fion_a}), showing the strongest correlation with \oi/\ha\ ($\rho$=$-$0.57) and a weaker correlation with \te(\nii) ($\rho$=0.24). No significant correlation is seen with 12+log(O/H), suggesting \fion\ is relatively insensitive to changes in abundance.  

\subsection{Aperture biases}
\label{sec:aperture}
While the CHAOS data set presents a remarkable catalog of emission lines and derived properties, by design it is targeting only the central regions of the brightest \hii\ regions in order to improve the chances of detecting the faint auroral lines. We note that in Figure \ref{fig:chaos_auroral}, the majority of CHAOS \hii\ regions have \siii/\sii\ $>$ 1.0.  These are relatively high compared to what is reported in samples that probe fainter \hii\ regions (e.g. \citealt{Kreckel2019, Mingozzi2020}). In particular, the incomplete spatial coverage of any given \hii\ region results in an inherent bias for line ratio diagnostics that show radial structure within a \hii\ region. This is  apparent in the \siii/\sii\ line ratio, as CHAOS observations probe only the  central $\sim$50 pc ($\sim$1\arcsec) of the nebulae, where the \siii\ emission is strongest, and miss \sii\ emission in the outer nebulae (see also \citealt{Mannucci2021}).

\begin{figure}
    \centering
    \includegraphics[width=3in]{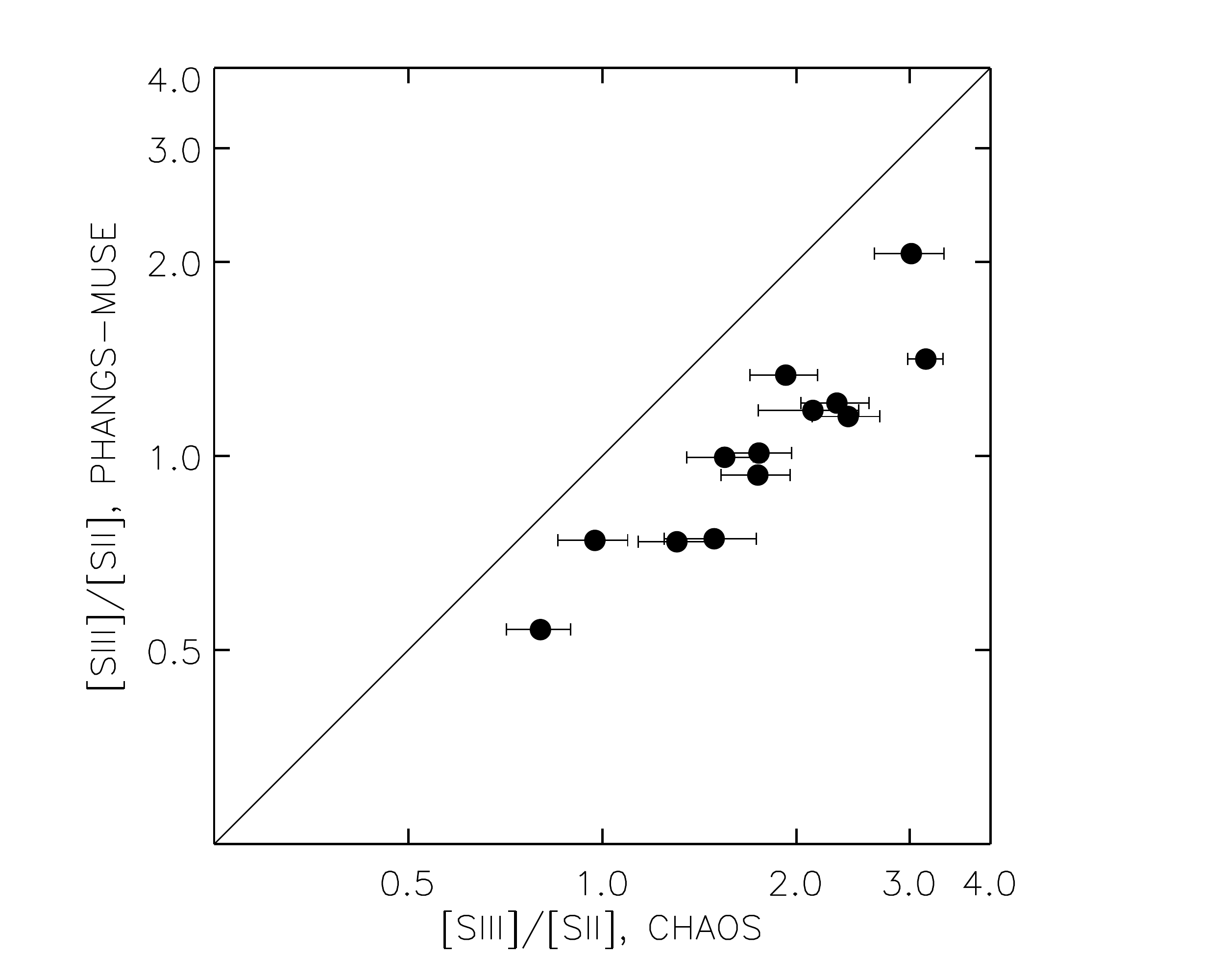}
    \caption{Comparison of the \siii/\sii\ line ratio measured across integrated \hii\ region spectra from PHANGS-MUSE with the measurement obtained within the CHAOS 1\arcsec\ slit for 13 \hii\ regions that overlap between the two surveys. Error bars are shown for both surveys, although the PHANGS errors are smaller than the symbols. The one-to-one line (black) shows that CHAOS systematically overestimates the line ratio by focusing on only the central brightest part of these extended \hii\ regions, where the \siii\ emission is preferentially located.  This demonstrates the challenges of applying such resolved measurements to broader \hii\ region samples.}
    \label{fig:chaos_siiisii_compare}
\end{figure}

We demonstrate this directly by cross matching the \hii\ regions cataloged by CHAOS and by PHANGS-MUSE in NGC~628, the only galaxy contained in both samples.  As CHAOS is predominantly targeting the outer disk while PHANGS-MUSE is limited to the inner disk, only 13 \hii\ regions overlap between the two samples. In Figure \ref{fig:chaos_siiisii_compare} we compare the \siii/\sii\ line ratios, and demonstrate how the aperture bias results in systematically higher measurements in the CHAOS sample.  This offset persists even when accounting for differences in the assumed reddening E(B$-$V) (due to both the foreground Milky Way and local reddening within the galaxy) adopted in the two surveys.  Challenges such as this motivate our use of integrated \hii\ region spectra for this paper, but similarly prevent us from relying solely on the CHAOS sample for the determination of an empirical relation between \fion\ and \siii/\sii. 

\begin{figure}
    \centering
    \includegraphics[width=3in]{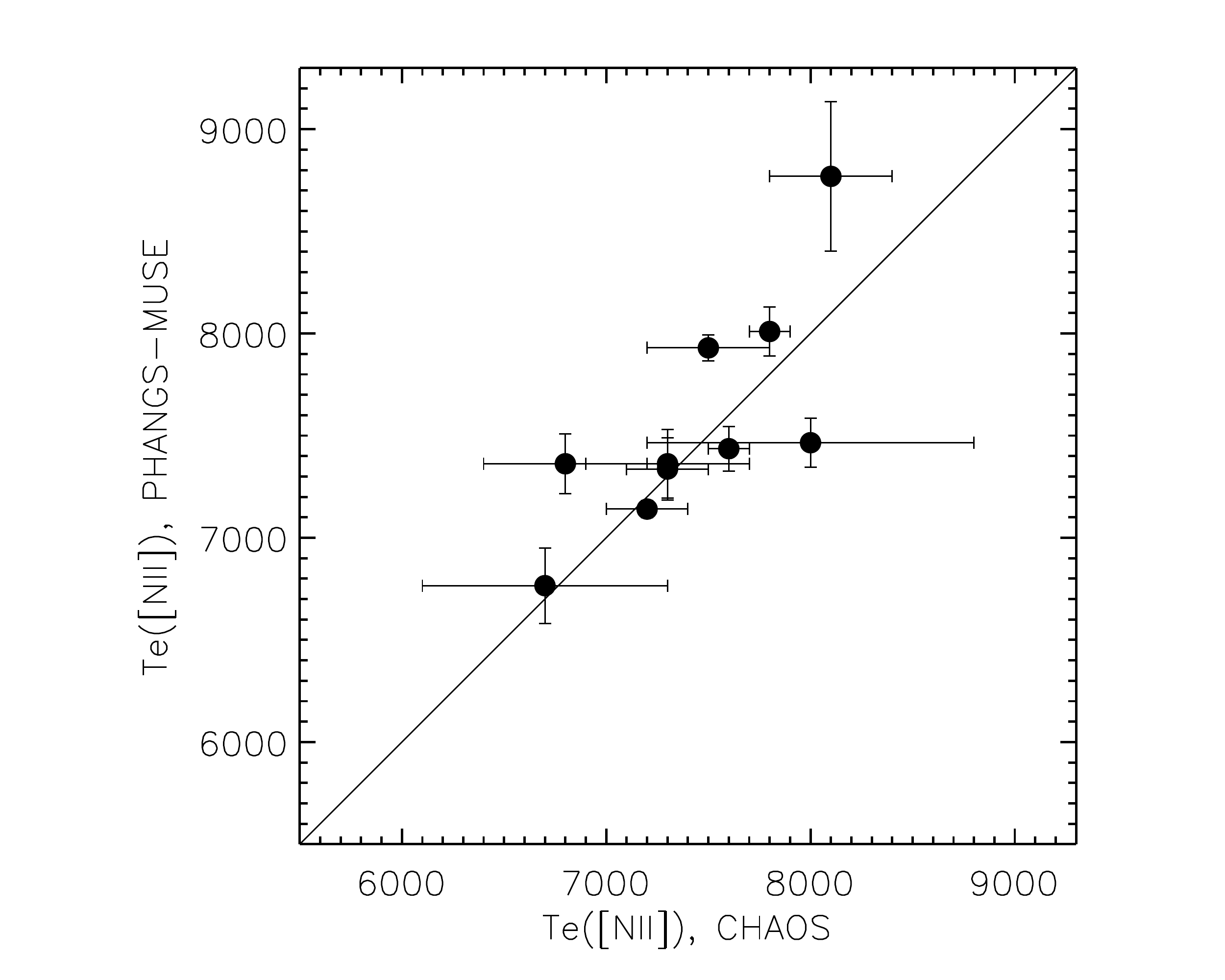}
    \caption{Comparison of \te(\nii) measured in integrated \hii\ region spectra from PHANGS-MUSE with equivalent measurements from the CHAOS 1\arcsec\ slit for 10 \hii\ regions that have reported \nii5755 line detection in both surveys. We observe  good systematic agreement between the two measurements. }
    \label{fig:comp_CHAOS_MUSE}
\end{figure}

Of the 13 \hii\ regions, only 11 have detections of \nii$\lambda$5755,  and subsequent calculations of \te(\nii), in both samples.  In Figure \ref{fig:comp_CHAOS_MUSE} we show a comparison of the derived \te(\nii), which follows very well the one-to-one line within the uncertainties. This agreement speaks to a smooth and roughly constant temperature structure within the \nii\ emitting region, as expected, as well as the high fidelity in both data sets.

\subsection{Parameterizing \fion}
\label{sec:varfion}

From the CHAOS sample, it is clear that changing the size of the partially ionized zone within the \hii\ region corresponds to a change in ionization parameter, with the strongest correlation identified between \fion\, when measured using \te(\nii), and \siii/\sii ~(as a proxy for ionization parameter). As expected from models, the ionisation fraction (i.e. basically the extent of the partially ionised zone) correlates with ionisation parameter. 
However, given the aperture bias, we cannot employ the measurements of \siii/\sii\  from CHAOS when determining an empirical relation. We therefore use the 840 \hii\ regions from PHANGS-MUSE which have \te(\nii) measurements to determine an empirical relation, such that we can use the observed \siii\ and \sii\ emission to constrain \fion. With this prescription we will then be able to measure all three necessary parameters (\fion, 12+log(O/H) and \oi/\ha) using strong line methods and thus solve for \te.

\begin{figure}
    \centering
    \includegraphics[width=3.5in]{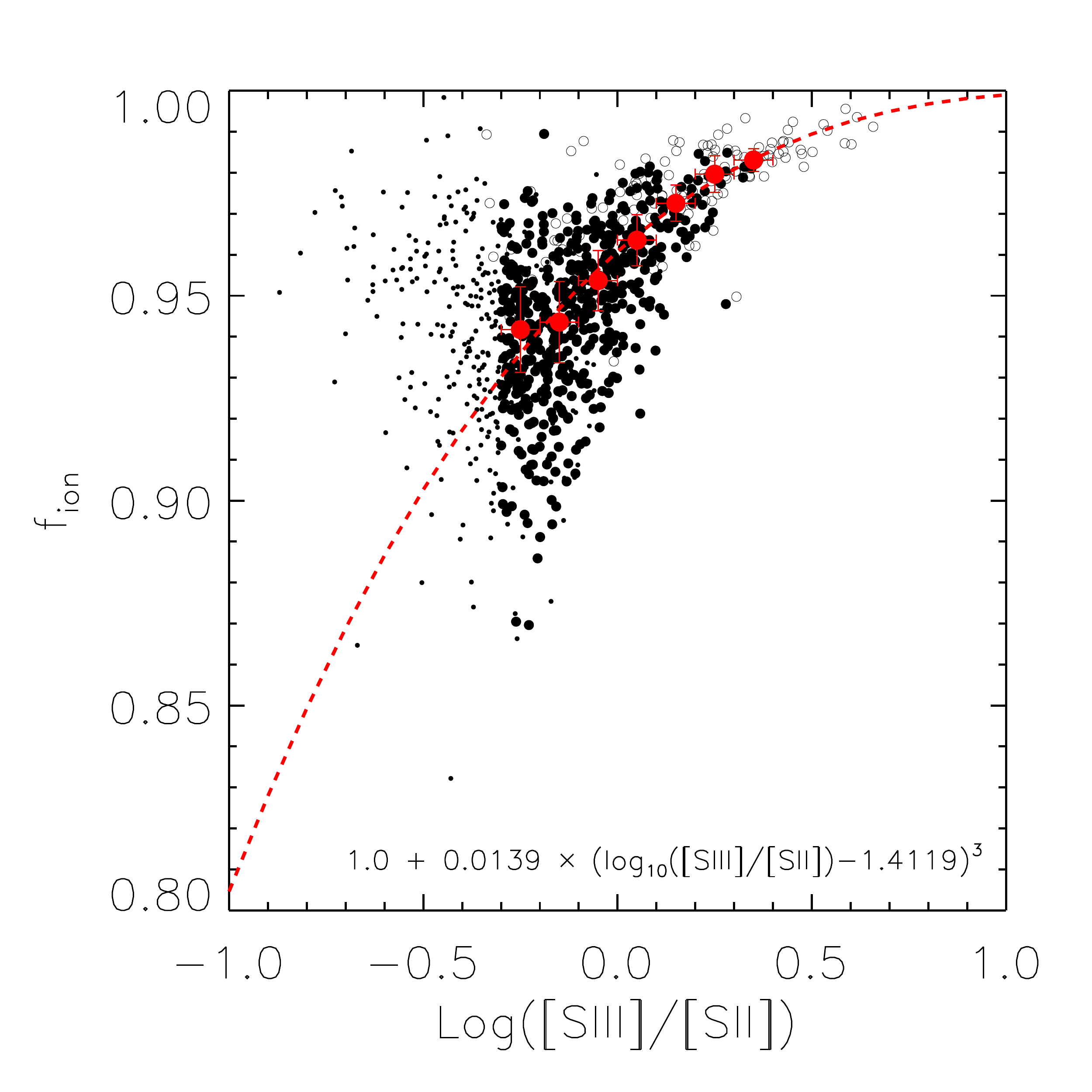}
    \caption{As in Figure \ref{fig:chaos_auroral}, we model \fion\ assuming the values of \te(\nii) measured from direct detection of auroral lines in the PHANGS-MUSE \hii\ region sample (points). This forms a continuous sequence with the \hii\ regions in the CHAOS sample (open circles), but with a significantly more pronounced non-linear trend. From the PHANGS-MUSE sample, we select only regions with more than  50\% contrast against the DIG background and with \siii/\sii\ $>$ 0.5 (filled circles). We then combine the PHANGS-MUSE and CHAOS samples to construct a binned  median (red circles).  We fit this with the functional form described in Equation \ref{eqn:fion} (red dashed line) to determine our empirical relation for \fion. }
    \label{fig:fion_relation}
\end{figure}

In Figure \ref{fig:fion_relation} we show \fion\ as a function of \siii/\sii\ for the full PHANGS-MUSE \hii\ region sample (points). We impose the requirement that regions be detected against the DIG background by at least a 50\% contrast, and consider only regions with \siii/\sii $>$ 0.5, where our sample appears more complete (filled circles).  We also show the \hii\ regions from all four galaxies from the CHAOS sample (open circles), which due to aperture biases have systematically overestimated \siii/\sii\ line ratios compared to the MUSE data (see Figure \ref{fig:chaos_siiisii_compare}), but nonetheless form a nearly continuous sequence with the PHANGS-MUSE \hii\ regions. Given the non-linear shape in this relation, it is apparent that a linear extrapolation from the CHAOS measurements (preferentially biased to high \siii/\sii) would not be suitable for the bulk of the PHANGS-MUSE sample (seen also in Figure \ref{fig:chaos_auroral}). However, given the smooth transition across samples, this also suggests that the correlation holds even when considering changes in \fion\ and changes in ionization parameter within the nebulae. To determine the most widely-applicable fit, we combine the PHANGS-MUSE and CHAOS samples and construct bins in \siii/\sii\, requiring a minimum of 10 data points, to demonstrate the median trend across the sample (red line, Figure \ref{fig:fion_relation}). To this median trend, we fit a third order power law where we have fixed the asymptote to \fion=1.0. This shows very good agreement with the binned median, and is parameterized by
\begin{equation}
    f_{\rm ion} =   1.0 + c_1 \times (\log_{10}([{\rm S}\textsc{iii}]/[{\rm S}\textsc{ii}]) - c_2)^3,
    \label{eqn:fion}
\end{equation}
where c$_1$ = 0.0139 $\pm$ 0.0060 and c$_2$ = 1.4119 $\pm$ 0.1890. 


We further overplot this fit on the relations in Figure \ref{fig:chaos_auroral} for the CHAOS data, and see reasonable agreement with many of the different ionic temperatures. The trend is reasonably consistent with \te(\nii), \te(\siii), and \te(\oiii), but clearly offset from \te(\oii). This reflects some of the previously reported discrepancies between different ionization zones \citep{Berg2020}, though it is somewhat surprising that \oiii\ (a very high ionsiation tracer) would work better than e.g. \oii\ (a low-ionisation tracer).

Given that we have imposed a minimum value of \siii/\sii$>$0.5, we cannot extrapolate this relation to lower \siii/\sii\ without additional data or modeling.  We revisit this decision in Section \ref{sec:cloudy}.

\subsection{Method summary}
\label{sec:methodsummary}
We summarize here the conditions that must be met in order to apply the charge-exchange method. Motivation for these choices is detailed at the beginning of Section \ref{sec:methoddev}.   When modeling \te\ based on integrated line fluxes within \hii\ regions:

\begin{itemize}
    \item We require S/N $>$ 10 in all of the following emission lines: \hb, \oiii, \ha, \nii, \sii, \siii. We use a fairly high threshold as the errors are underestimated by an estimated $\sim$40\% \citep{Emsellem2022}.
    \item We require a contrast of $>$50\% against the DIG background for each line.
    \item We require  \siii/\sii $>$ 0.5.
    \item We correct all lines for attenuation using the Balmer decrement.
    \item We calculate the \oi/\ha\ line ratio.
    \item We calculate 12+log(O/H) following the \cite{Pilyugin2016} S-calibration, as in Equations \ref{equation:ohsu} and \ref{equation:ohsl}.
    \item We calculate \fion\ from the \siii/\sii\ line ratio, as in Equation \ref{eqn:fion}.
    \item We use these three values to solve Equation \ref{eqn} for \te.
\end{itemize}


\subsection{Validation}
\label{sec:validation}

\begin{figure*}
    \centering
    \includegraphics[width=7in]{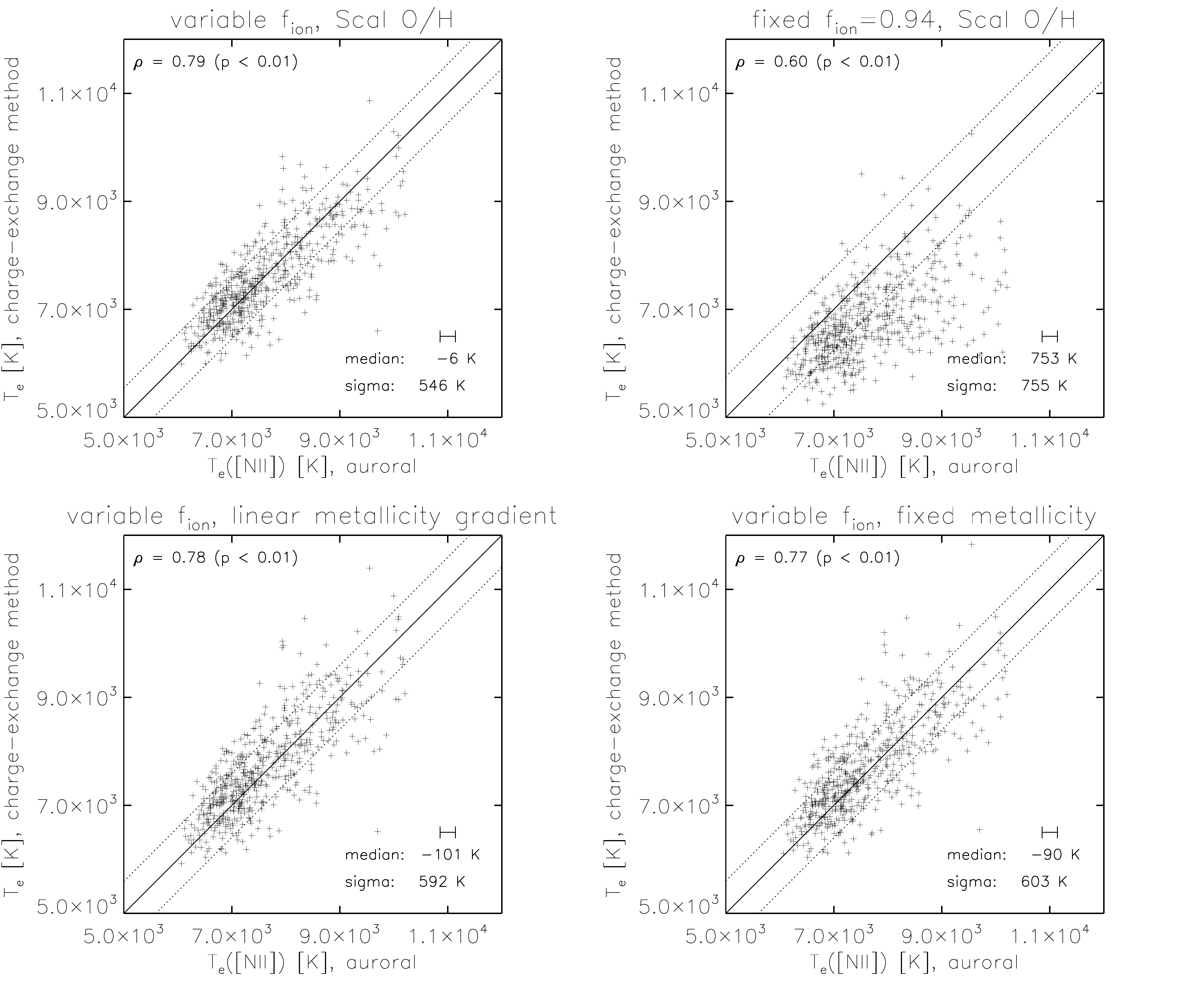}
    \caption{Comparison of the modeled \te\ as a function of the \te\ measured within PHANGS-MUSE using the \nii\ $\lambda$5755 auroral line. Each panel makes different assumptions about how the value of \fion\ or 12+log(O/H) are calculated. For each panel we show a one-to-one relation (solid line) and 1$\sigma$ relative scatter (dotted lines), as well as the Spearman's rank correlation coefficient ($\rho$). Top left: we demonstrate our adopted method (Section \ref{sec:methodsummary}), where we use our empirical prescription for \fion\ (Equation \ref{eqn:fion}) and metallicities calculated from the strong line Scal method. We recover high correlations ($\rho$=0.79) and systematic agreement within 6~K, with a scatter of 546~K. Relaxing various assumptions in our model increases the scatter. Top right: Assuming a fixed \fion=0.94 (the median for the PHANGS-MUSE sample) produces significantly higher scatter ($\sim$755~K) and clear deviations from the one-to-one relation. Bottom panels: We apply a variable \fion\ but relax our derivation of 12+log(O/H), adopting either a linear radial gradient (left) or a fixed metallicity for each galaxy (right). Both recover high correlation coefficients with moderately increased scatter ($\sim$600~K). This demonstrates the relative insensitivity to the input metallicity for this charge-exchange method of deriving \te. }
    \label{fig:predict_obs}
\end{figure*}

In Figure \ref{fig:predict_obs} we validate our charge-exchange method by applying it to the 531 PHANGS-MUSE \hii\ regions that meet the criteria in Section \ref{sec:methodsummary} and have \te(\nii) derived from auroral line methods.  We find a high correlation (Spearman's rank correlation coefficient $\rho$=0.79), good systematic agreement ($\sim$6~K) and relatively small scatter ($\sim$550~K) between the direct-method and ``charge-exchange method'' measurements of \te. Here, the scatter is calculated as the dispersion relative to the auroral line measurement, neglecting any absolute offset. Requiring a higher contrast of 300\% against the DIG background reduces the \hii\ region sample by a factor of $\sim$2 but does not significantly change these statistics. On the other hand, using a fixed value for \fion=0.94 (the median for the PHANGS-MUSE sample) introduces a significant scatter ($\sim$800~K) and non-linearity to the relation. Employing the variable \fion\ but relaxing our assumption on how the metallicity is determined (either via a fixed linear radial gradient or a fixed global value for each galaxy) retains a high degree of correlation ($\rho\sim$0.78) and only slightly higher scatter ($\sim$600~K). This demonstrates the relative insensitivity of this method to the input metallicity, though we note that our galaxies cover a fairly limited metallicity range of 8.3 $<$ 12+log(O/H) $<$ 8.7 \citep{Santoro2022}. 

\begin{figure*}
	\includegraphics[width=7in]{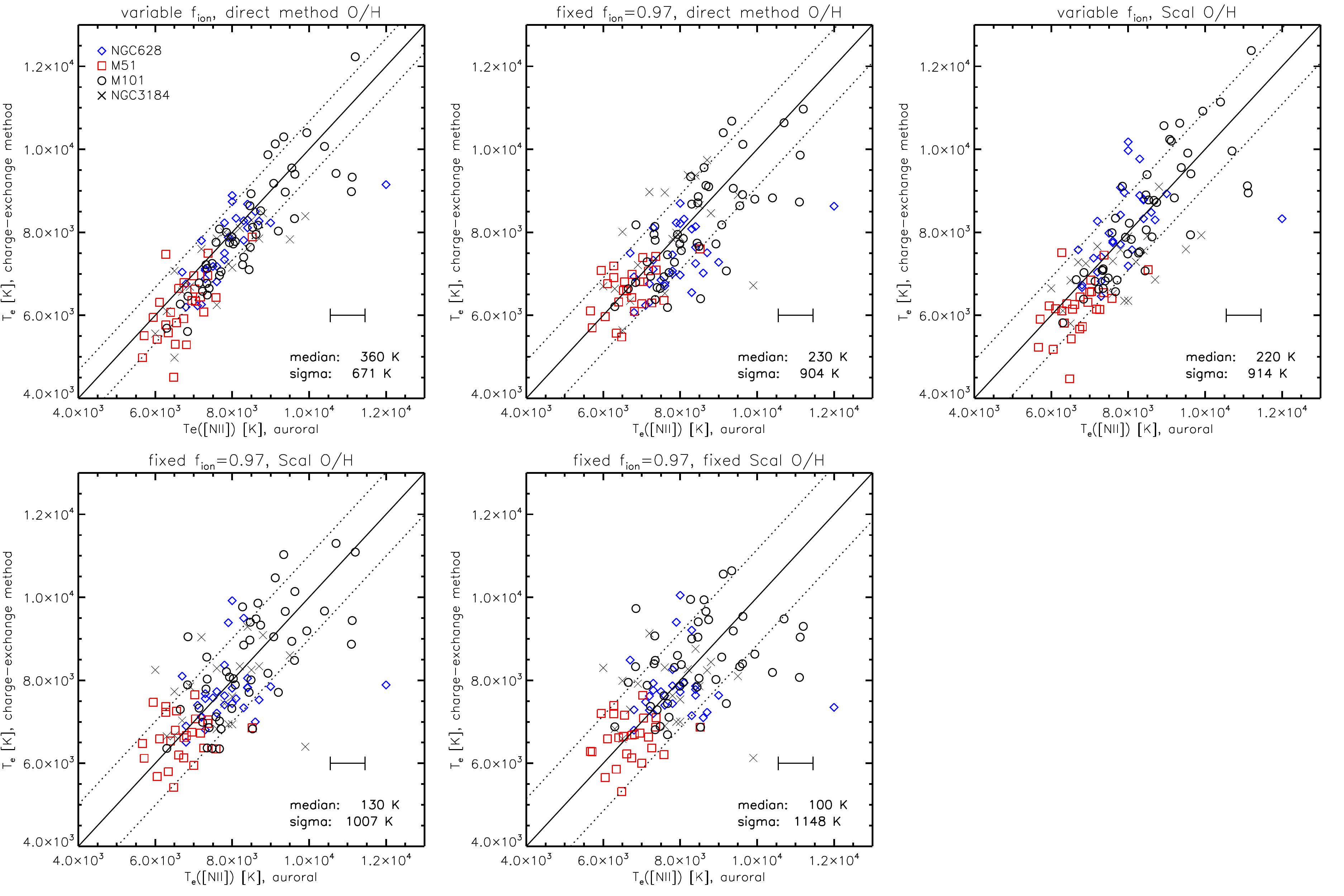}
    \caption{ Comparison of the modeled \te\ as a function of \te\ measured by CHAOS using the \nii\ $\lambda$5755 auroral line. Each panel makes different assumption about how the value of \fion\ and 12+log(O/H) are calculated. The top left plot represents the best case scenario, where we vary \fion\ as a function of \siii/\sii\ and use metallicities calculated from the direct method. This shows the lowest systematic scatter between \te\ values of 662~K. The other two plots on the top row each relax one of the assumptions, either assuming a fixed \fion=0.97 (top center) or a metallicity calculated from strong line techniques (top right). Both have increased scatter of $\sim$900~K. The two plots on the bottom row further relax the assumptions, resulting in further inaccuracies. This test includes assuming both fixed \fion=0.97 and a strong line metallicity ($\sim$950 K; bottom right) and a fixed \fion=0.97 and a fixed global metallicty per galaxy ($\sim$1100~K; bottom center).  A one-to-one relation (solid line) and 1$\sigma$ scatter (dotted lines) are shown in each plot, and the median and standard deviation are listed. While relaxing our assumptions does result in increased scatter, the overall agreement remains good. We adopt the assumptions that go into the top right figure (variable \fion\ and strong line 12+log(O/H)) for the model that we apply in this work, as summarized in Section \ref{sec:methodsummary}.  }
    \label{fig:chaos_te_comparison}
\end{figure*}

We perform an additional validation of our various assumptions by applying this charge-exchange method next to the CHAOS data set, adjusting the different assumptions we make. In Figure \ref{fig:chaos_te_comparison}, we plot our modeled \te\ against the value of \te(\nii) measured by CHAOS directly from the auroral line detections. 
For the most stringent case, where we allow a variable \fion\ that depends on \siii/\sii\ and employ direct method techniques to measure 12+log(O/H), we measure a scatter between the two \te\ values of $\sim$660~K. This scatter is only a factor of 1.5 times larger than the median uncertainty (450~K) reported for \te(\nii). Assuming a fixed \fion=0.97 (the median for the CHAOS sample) increases the scatter by almost a factor of two (860~K).  Applying a variable \fion\ prescription and instead using the strong line S calibration metallicity prescription results in a scatter of $\sim$900~K. As might be expected, adopting both of these changes, to assume a fixed \fion=0.97 and using the S calibration prescription, results in even larger scatter ($\sim$950~K). Our final test case is to assume a fixed \fion=0.97 and metallicity for each galaxy. This worst case scenario has a scatter of $\sim$1100K, which is smaller than the uncertainty quoted for 20\% of the CHAOS measurements, and still clearly retains the correlation between modeled and directly measured \te.  All of these approaches demonstrate a relatively large systematic offset of 100--300~K between the direct method and ``charge-exchange'' measurements of \te, presumably reflecting apertures biases that affect lines with significant radial extent and ionization structure  (e.g. \sii, \siii\ and \oi, \oii, \oiii).

This set of tests demonstrate that for this technique the physical motivation is quite robust, able to reproduce reliable \te values in the range 6,000-10,000K with an estimated uncertainty of $\sim$600~K for integrated \hii\ regions under reasonable assumptions (as outlined in Section \ref{sec:methodsummary}). These also justify our choice of adopting the variable prescription for \fion\ in Equation \ref{eqn:fion} that relies on changes in \siii/\sii.  The worst-case scenario test, with a fixed 12+log(O/H) for each galaxy, demonstrates that this technique is not strongly dependent on small changes in metallicity ($\sim$<0.1 dex) for each region. Our result is specifically calibrated against strong-line metallicities measured using the \cite{Pilyugin2016} S calibration, which itself is empirically calibrated against direct method metallicities, and can not account for larger (up to $\sim$0.3 dex) offsets sometimes found between different prescriptions. This discrepancy in the absolute abundance scale presents a long standing challenge facing strong line abundance prescriptions that has been much discussed in the literature \citep{Kewley2008, Kewley2019}. 


\clearpage

\clearpage


\section{Results}
\label{sec:results}
We now apply the charge-exchange method to model \te\ within the full sample of 24,173 PHANGS-MUSE \hii\ regions. 


\subsection{Electron temperatures across thousands of \hii\ regions}

We find that 4,129 \hii\ regions meet the criteria outlined in Section \ref{sec:methodsummary}, allowing us to compute \te\ for nearly a fifth  of the \hii\ region sample.  This is a factor of $\sim$4 times more \hii\ regions than have any auroral line detection, and a factor of $\sim$8 times more \hii\ regions than have direct detection of the \nii\ $\lambda$5755 line. The limiting factor for $\sim$2000 of the regions is the requirement that the emission lines be detected with sufficient contrast against the surrounding DIG background, and for $\sim$5000 \hii\ regions  \siii/\sii\ is too low. Another $\sim$5000 \hii\ regions meet neither criteria.  The remaining $\sim$7000 targets do not have sufficient S/N in all lines necessary.

\begin{figure*}
\includegraphics[width=7in]{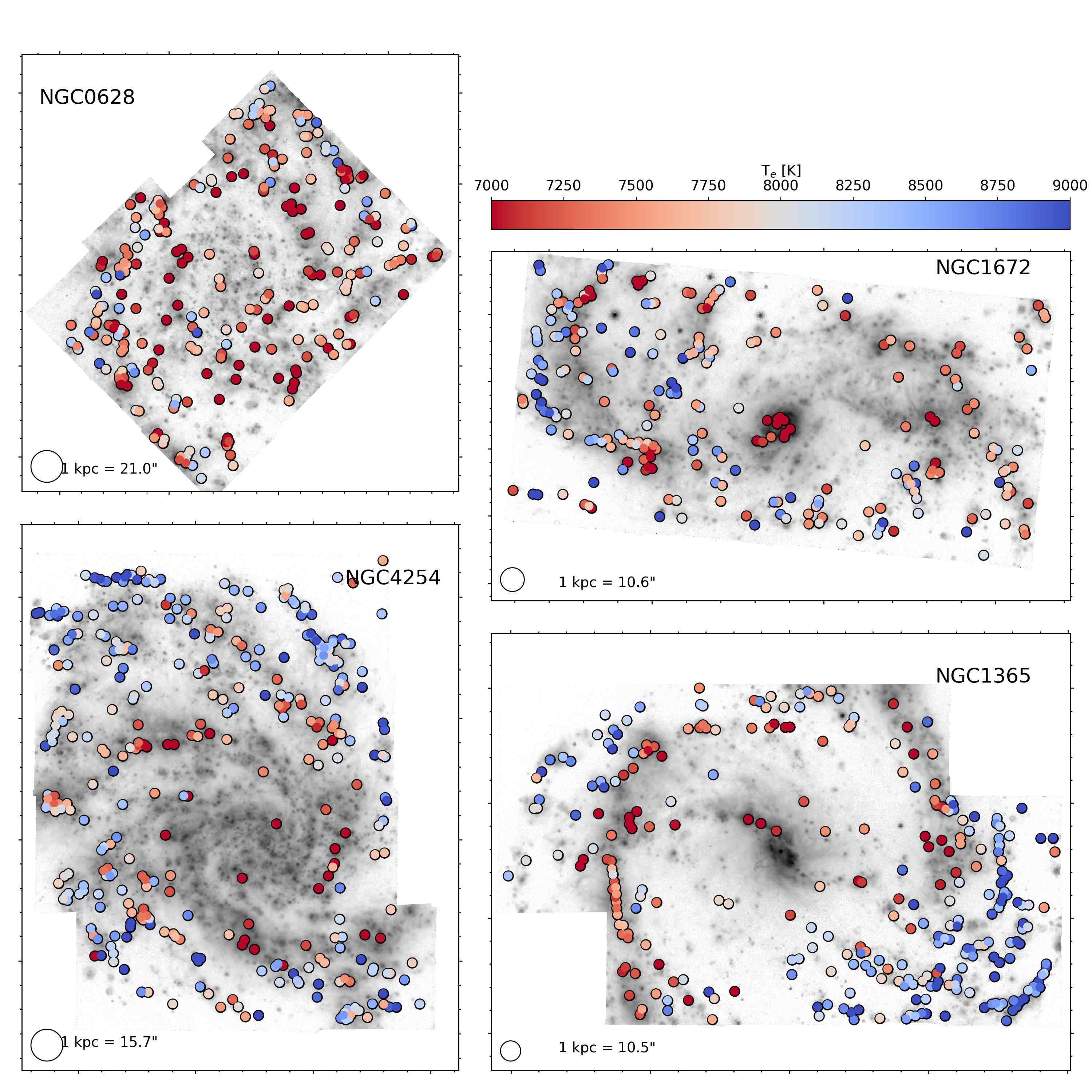}
\caption{Mapping electron temperatures within \hii\ regions across four of the PHANGS-MUSE galaxies (NGC 1672, NGC 1365, NGC 4254, NGC 628). Coverage is not just limited to spiral arms, but samples the full inner star-forming disks in our sample of galaxies. 
NGC 1365 \citep{Ho2017} and NGC 1672 \citep{Kreckel2019, Ho+2019} have both reported azimuthal abundance variations, with enhanced abundances along the spiral arms, which is supported by decreased \te\ along the eastern arms in both galaxies.  The more flocculent galaxy NGC 4254 and the bar-free grand design spiral galaxy NGC 628 both show some tentative azimuthal trends but the connection with spiral structure is more tenuous.  Small scale $<1$~kpc homogeneity in the temperature distribution is reminiscent of the trends for more uniform metallicity on small scales reported in \cite{Kreckel2020} for these targets. 
\label{fig:azimuth}}
\end{figure*}

With hundreds of measurements per galaxy, we show two dimensional maps of \te\ across four example galaxies (Figure \ref{fig:azimuth}). \galaxyname{NGC}{1672} and \galaxyname{NGC}{1365} both have previously identified azimuthal variations in their abundance distribution \citep{Ho2017, Kreckel2019}, as identified using strong line methods, with increased metallicities along the spiral arms.  We similarly observe decreased \te\ in the \hii\ regions located along the spiral arm ridges, supporting that result (see also \citealt{Ho+2019}), although the effect is less pronounced. \galaxyname{NGC}{4254} shows a much more flocculent spiral pattern, and the \te\ measurements more uniformly sample the disk.  \galaxyname{NGC}{628} hosts a two arm spiral pattern without a bar or strong bulge component, and exhibits a clear gradient in \te\ with galactocentric radius. 

\begin{figure}
    \centering
    \includegraphics[width=3.5in]{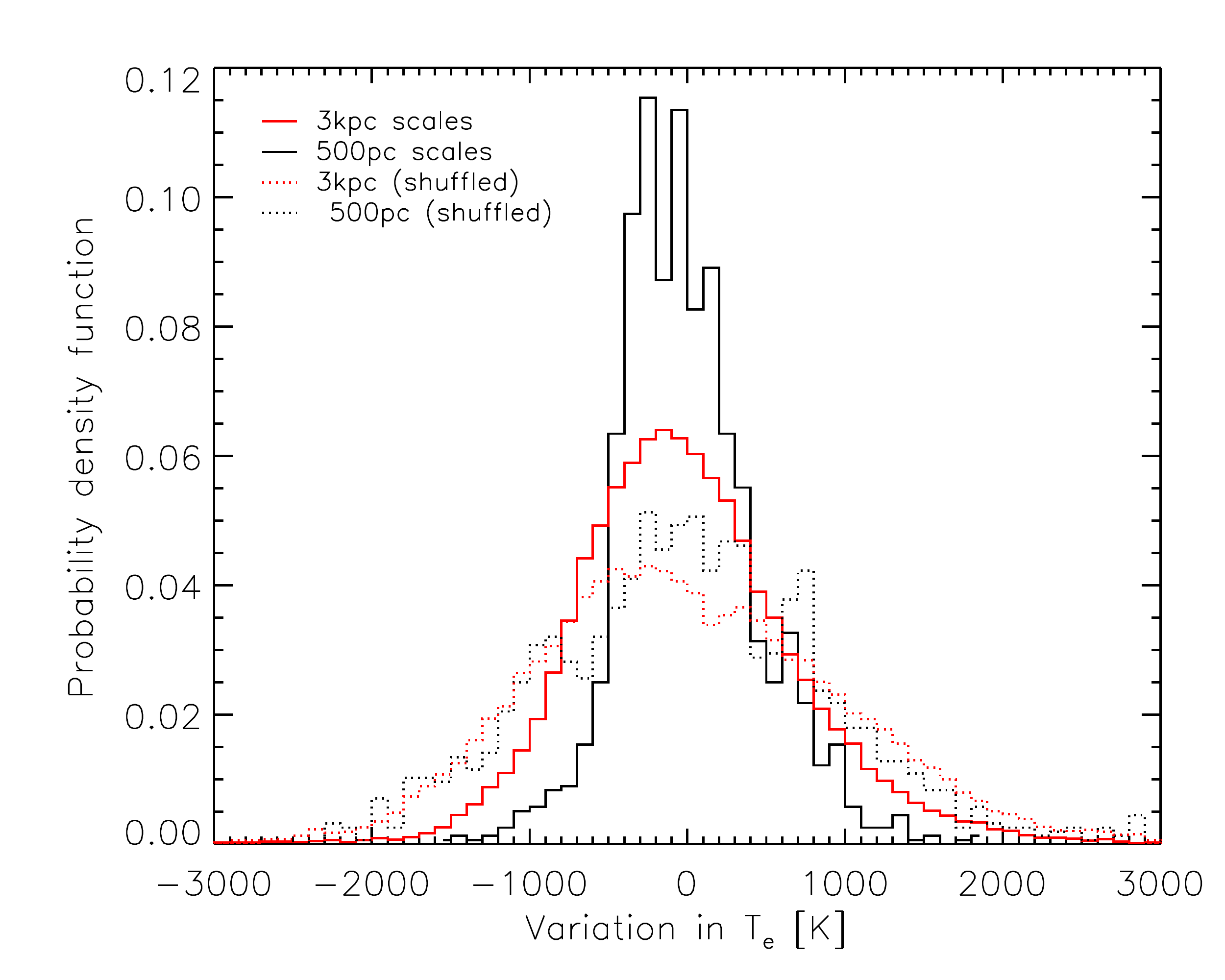}
    \caption{The probability density function of the variations in \te\ at 500~pc (black) and 3~kpc (red) sampling scales across all galaxies in the sample. We perform this calculation around each \hii\ region, relative to the mean local value at each position. We require at least five \hii\ regions be located within the relevant sampling scale length, to minimize biases due to low number statistics. Across all galaxies, we measure a standard deviation of $\sim$500~K at 500~pc scales and $\sim$800~K at 3~kpc scales. To test the null hypothesis, that all region temperatures are uncorrelated, we shuffle all values of \te\ and repeat this calculation (dotted lines).}
    \label{fig:te_scatter}
\end{figure}

Figure \ref{fig:te_scatter} quantifies the scatter in \te\ as a function of scale within all of our maps. Measuring the variation in \te\ across 500~pc or 3~kpc scales around each \hii\ region, relative to the mean local value at each position, we find that the probability density function reveals systematically smaller scatter on small physical scales. To perform this calculation, we consider the variation between neighboring regions only when 5 or more \hii\ regions are clustered on 500~pc or 3~kpc scales.  We measure a standard deviation of $\sim$500~K at 500~pc scales, and $\sim$800~K at 3~kpc scales. To test the null hypothesis, that all regions temperatures are uncorrelated, we shuffle all values of \te\ and repeat this calculation (dotted lines). This test erases the trend with physical scale and results in a broader distribution. This small scale homogeneity in the \te\ distribution reflects the spatial scales relevant to mixing, as also recently determined from strong line methods \citep{Kreckel2020, Li2021, Metha2021, Williams2022}. 

\begin{figure*}
	\includegraphics[width=7in]{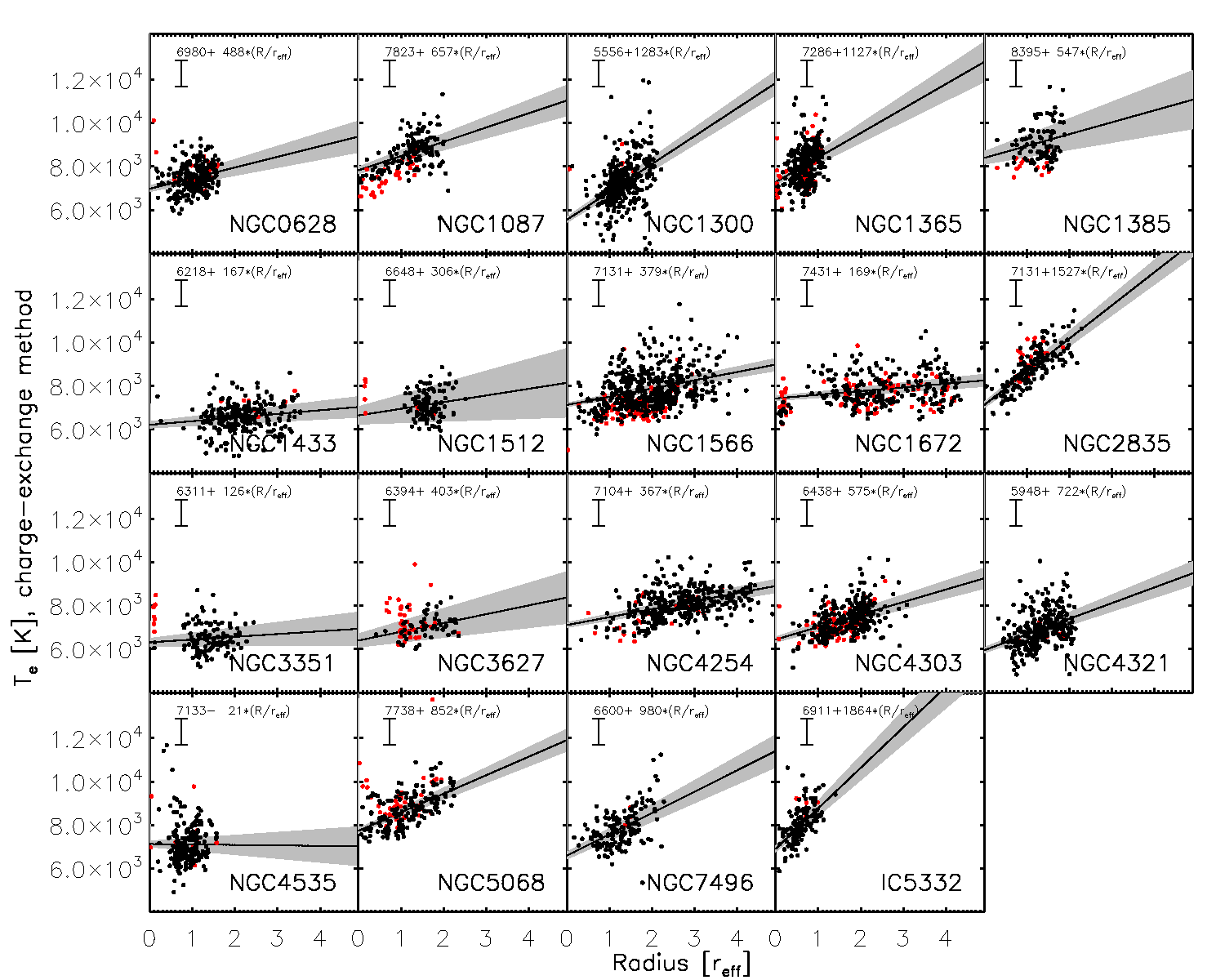}
	    \caption{The radial gradient in \te\ for each of the PHANGS-MUSE galaxies. Red points show \te\ determined from auroral line measurements, while black points show \te\ measured using the charge exchange method. Positive slopes show general agreement with the expected flat to negative metallicity gradients reported in these and other nearby galaxies \citep{Pilyugin2014, Kreckel2019}.  A linear fit is shown for each galaxy, with the fit parameters listed in the top of each plot. The grey band indicates the uncertainty in the linear fit, accounting for 600~K systematic uncertainty in the \te\ measurements. }
    \label{fig:te_grad0}
\end{figure*}

\begin{figure}
	\includegraphics[width=3in]{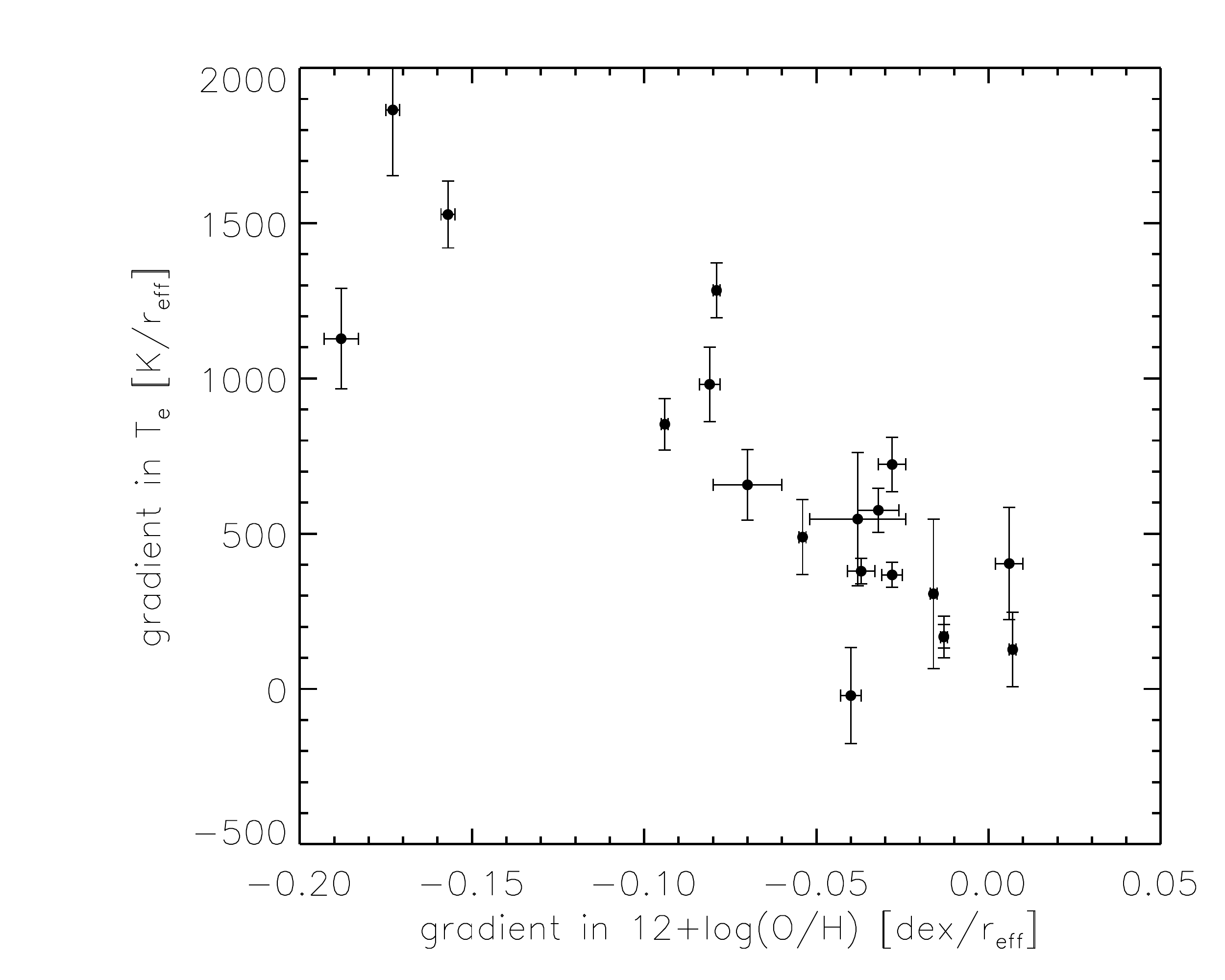}
	    \caption{The slope of the radial gradient in \te\ as a function of the slope of the radial gradient in 12+log(O/H) \citep{Santoro2022} for each PHANGS-MUSE galaxy. Positive slopes in temperature correspond (as expected) to negative slopes in metallicity, as metal-rich gas is more efficient at cooling.}
    \label{fig:te_grad_z}
\end{figure}

In Figure \ref{fig:te_grad0} we show the radial gradient in \te\ for each of the 19 galaxies.  All galaxies show positive gradients, reflecting an expected negative metallicity gradient (Figure \ref{fig:te_grad_z}), such as is typically observed across nearby galaxy samples \citep{Pilyugin2014}. In four galaxies (NGC~1300, NGC~1365, NGC~1566, NGC~4535), significant ($\sim$900--1000 K) scatter is seen in these radial trends, in excess of the uncertainty intrinsic to the method ($\sim$600 K).  This could be indicative of particularly large azimuthal abundance variations in these galaxies, which was also conclusively shown for two of these galaxies in interpolated metallicity maps \citep{Williams2022}.

 
\subsection{Why does this even work? }
\label{sec:whydoesitwork}
\begin{figure}
    \centering
    \includegraphics[width=3in]{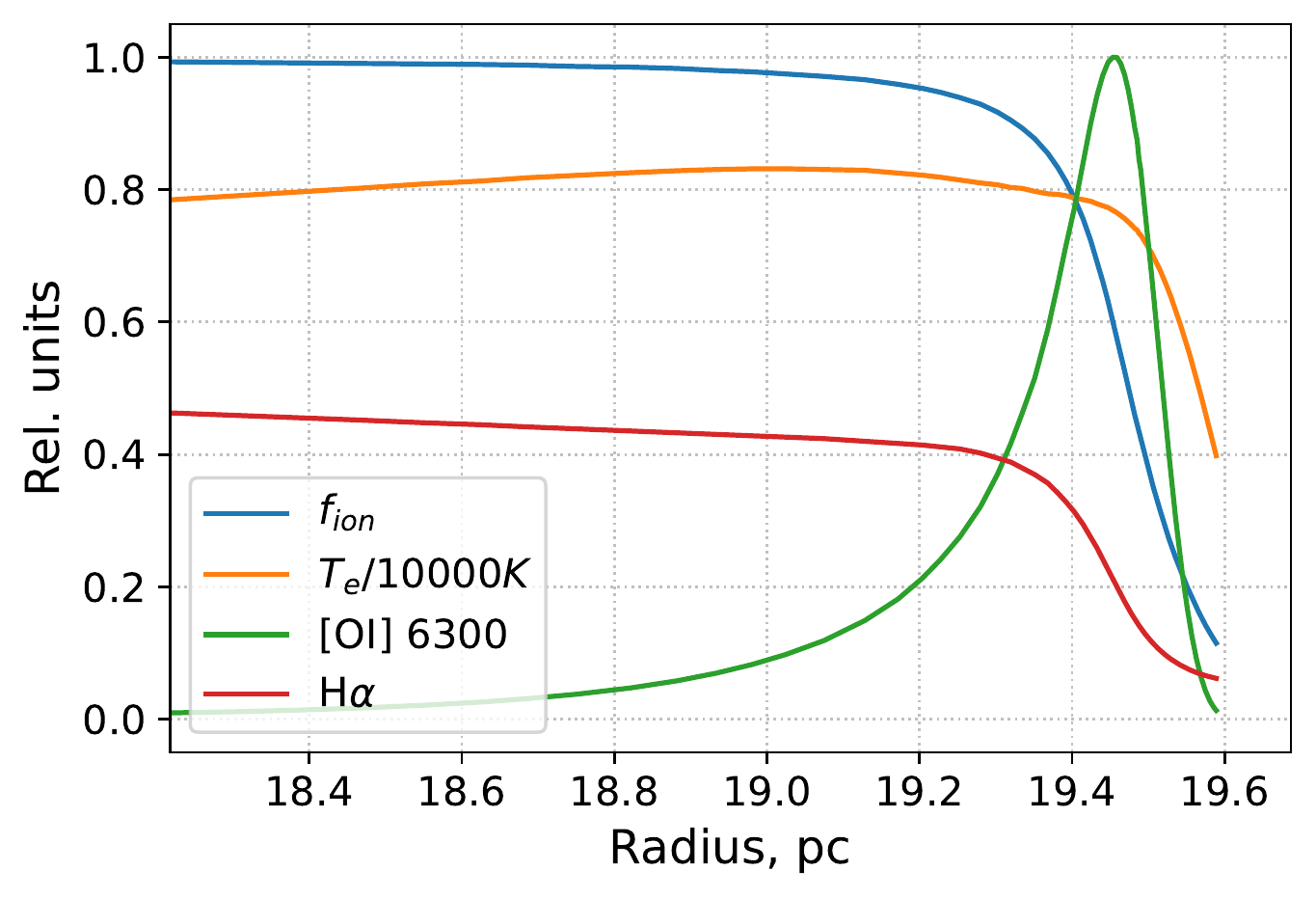}
    \caption{Cloudy modeling of the interior structure of a typical \hii\ region. The \ha\ emission (red) traces fairly closely the ionization fraction (\fion, blue) as a function of radius, while the \oi\, (green) is predominantly emitted in the outer shell of the nebula. In this way, it is apparent that very little of the \ha\ emission is co-spatial with the \oi, a key assumption in our model. At the outer edge of the nebula, the intensity of \oi\, rapidly drops together with the electron temperature \te\ (orange).}
    \label{fig:OI_extent}
\end{figure}

There are a few crucial assumptions that go into this model (see also Section \ref{sec:method}). 
This model compares the \oi\ and \ha\ line emission, however considering the ionization structure within a nebula (Figure \ref{fig:OI_extent}) it is clear that these emission lines arise from different regions, with \ha\ emitted predominantly throughout the interior while \oi\ is emitted mainly in an outer shell. For this work, we are assuming that \te\ is uniform across both the \ha\ and \oi\ emitting regions, as is shown in our Cloudy modeling (orange line, Figure \ref{fig:OI_extent}). 
At sufficiently high spatial resolution, it would be possible to resolve the internal nebular structure and compare \oi\ and \ha\ emission arising from the same parcel of gas, more directly ensuring that this assumption is true, but this is beyond what is possible given our $\sim$50~pc physical resolution. This condition of uniform \te\ and co-spatial \oi\ and \ha\ emission is also likely true in the diffuse ionized gas (e.g. \citealt{Reynolds1998}).  

This method also assumes that the gas is photoionized, and contributions from shocks are negligible. The \oi/\ha\ ratio is quite sensitive to shock excitation \citep{Allen2008}, spanning two orders of magnitude at constant metallicity and different shock velocities. Therefore, we expect the observed ratio of \oi/\ha\ to change significantly in the presence of shocks. This is directly observed in supernova remnants \citep{Kopsacheili2020}, and in general we note that the observational locus of ``star-forming regions" is not well-represented by models in the \oi\ BPT diagram (e.g.\ \citealt{Law2021}). How much of this discrepancy is due to the diffuse ionized gas (DIG), and the source of excitation for that diffuse emission, remains an active field of research \citep{Belfiore2022}. 
Our PHANGS-MUSE nebular catalog has attempted to minimize the impact of other ionizing sources by selecting peaks in \ha, using the BPT diagnostic cuts,  and having a high contrast against the DIG. 
We further require our emission lines to be detected with 50\% contrast against the DIG.  Finally, we attempt to  minimize the possibility of shock contributions by employing BPT diagnostic cuts, as described in Section \ref{sec:data-muse}. 

Finally, an essential step in our method is our ability to infer \fion\ in an independent way from the available strong-lines. We find the strongest correlation with \siii/\sii\ (Figure \ref{fig:chaos_auroral}), a robust tracer of the ionization parameter. This begs the question, why is the ionization fraction related to the ionization parameter ($q$)? Here we consider principally the incident ionization parameter, as seen by the inner edge of the nebula. 
By definition, $q=\frac{Q(H^0)}{4\pi R^2n_H}$ (Equation \ref{eqn:q}), and \fion\ = $n_{\rm H^{+}}/n_{\rm H}$ (Equation \ref{eqn:fion_def}). Assuming case B, $Q(H^0)$ could be expressed as $Q(H^0) \simeq \alpha_B \int n_e n^+ \mathrm{d}V$, where $\alpha_B$ is the recombination rate coefficient. 
Assuming a \hii\ region as a thin sphere (a common approximation), we can write that $\mathrm{d}V=4\pi R^2 \mathrm{d}l$, where $l$ is the thickness of the shell. Thus, $q \sim \alpha_B \frac{n_e n^+ l}{n_H} \sim \alpha_B f_{\rm ion} n_e l$, and we find that $q \propto f_{\rm ion}$ at constant $l$ and $n_e$. 
However, we must keep in mind for this argument that this is \fion\ measured over the full nebula, and not just over the \oi\ region in the outer shell. 

The remarkable success of this charge-exchange method for deriving \te\ in comparison with $\sim$500 \hii\ regions in the PHANGS-MUSE sample that have direct auroral line constraints (Figure \ref{fig:predict_obs}) demonstrates that the spatial averaging over the entire \hii\ region  does not invalidate this method.  Crucially, as we show through more detailed Cloudy modeling in Section \ref{sec:cloudy}, this simplistic assumption does not seem to invalidate our derived dependency of \fion\ on \siii/\sii, but in fact appears to reduce the scatter in this relation.

\subsection{Comparison of our \fion\ relation with models}
\label{sec:cloudy}

Crucial to the success of this technique is our ability to infer \fion\ directly from the \siii/\sii\ line ratio.  While we establish this empirically through comparison with the PHANGS-MUSE and CHAOS \hii\ regions (Section \ref{sec:varfion} and Figure \ref{fig:fion_relation}), we here explore how such a relation might emerge from theoretical models.

\begin{figure}
    \centering
    \includegraphics[width=\linewidth]{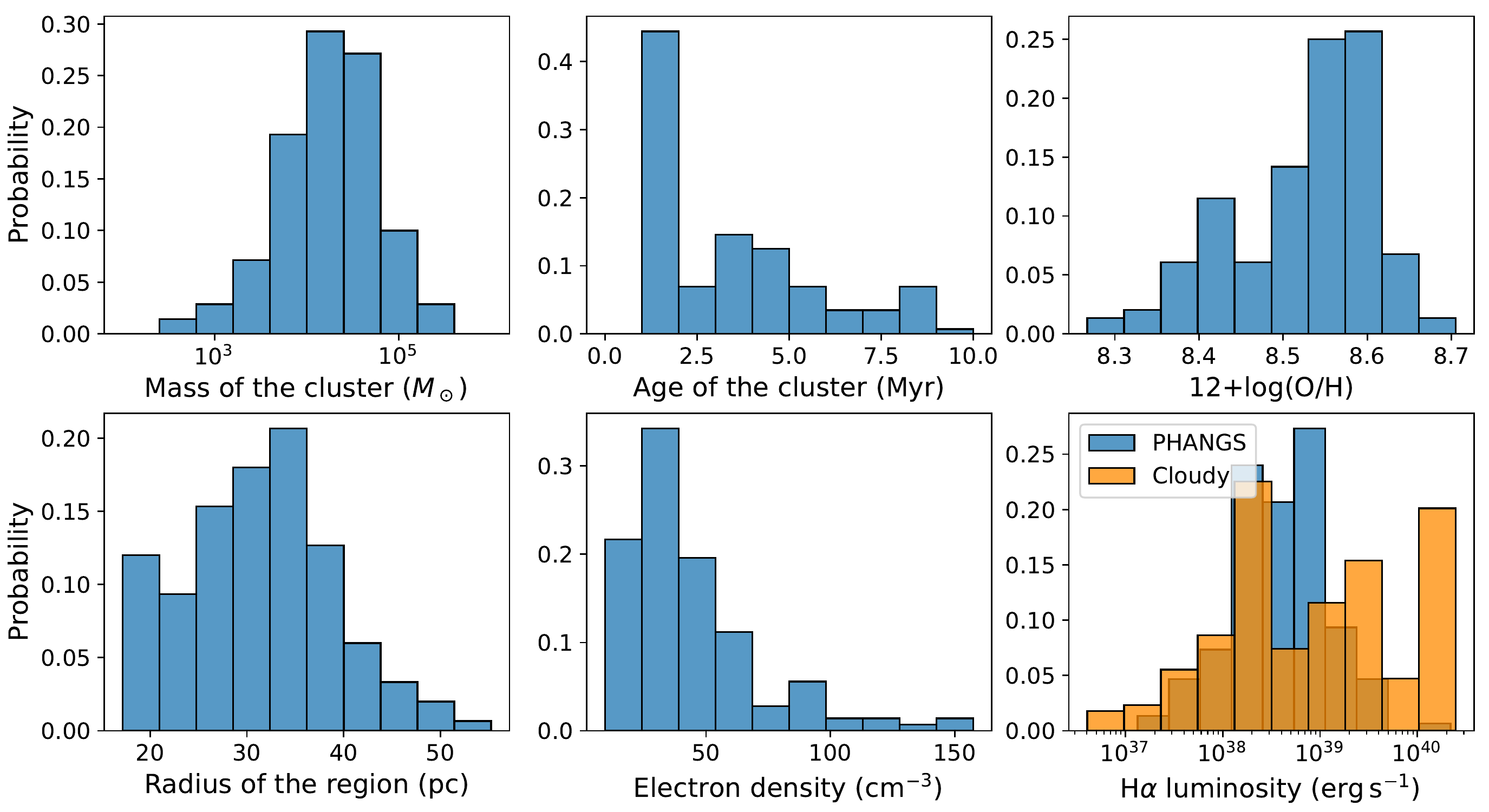}
    \caption{Distributions of the observed properties for 150 \hii\ regions with measured auroral lines and a 1-to-1 correspondence with a young stellar association (Scheuermann et al. in prep). These ranges in cluster mass, stellar cluster age, gas-phase metallicity, size and electron density are used to establish the parameter range used in our Cloudy models. The bottom-right panel shows the distribution of the reddening-corrected \ha\ luminosity in both the observational sample and computed Cloudy models.}
    \label{fig:phangs_1to1_stats}
\end{figure}

We compute a set of Cloudy models \citep{Ferland2017} for the properties of ionizing clusters and nebulae covering the same parameter space as in the observed sample. To start, we cross-match our catalog of PHANGS-MUSE \hii\, regions with measured auroral lines to a catalog of the stellar associations detected within PHANGS-HST imaging (\citealt{Lee2022}, Scheuermann et al. in prep).  These stellar associations have been identified using a watershed method applied to the NUV and V-band images, and their key parameters (stellar mass, age, reddening) have been inferred from SED modeling (Larson et al. in prep). We identify 150 objects with a one-to-one correspondence between an \hii\ region and a stellar association, where we can be fairly certain that the identified stellar association is powering the ionization of the \hii\ region. These 150 matched objects have a median age of $\sim$3 Myr and a median stellar mass of $\sim$10$^4$ M$_\odot$, and we can further constrain properties such as size ($R$), gas-phase metallicity (12+log(O/H)) and electron density ($n_{\rm e}$) as shown in Figure~\ref{fig:phangs_1to1_stats}. This cross-matched sample was used to establish the range of model parameters that we explore. 

We run Cloudy v17.02 models (using the \textsc{pycloudy} package, \citealt{Morisset2013}) consisting of a spherical gas cloud surrounding an ionizing source and take the following range in parameters based on the observations:  constant density $n_{\rm H} \simeq n_{\rm e} = 20, 90$ and 300 $\rm cm^{-3}$; metallicity $\mathrm{12+\log(O/H)} = 7.99,\, 8.47$ and 8.69 (that corresponds to $Z/\mathrm{Z}_\odot=0.2$, 0.6 and 1.0 assuming the Solar relative abundance of elements from \citealt{Grevesse2010}); inner radius of the shell $R=0$ (cloud), 15, 28, 40~pc. We also varied the volume filling factor in the range of $0.1-1.0$, but it has a neglectable effect on the output models. As an ionizing source we used the spectra produced by Starburst99 \citep{Leitherer2014} for instantaneous star formation, a Kroupa initial mass function \citep{Kroupa2001}, and Geneva stellar evolution tracks which include rotation (metallicity $Z=0.014$, \citealt{Ekstrom2012}, and $Z=0.002$, \citealt{Georgy2013})\footnote{We used used stellar evolution tracks for $Z=0.002\, (\sim0.15Z_\odot)$ only for the nebulae with $\mathrm{12+\log(O/H)=7.99}$; for the other nebulae we used $Z=0.014\, (\sim \mathrm{Z}_\odot)$.} for clusters with total stellar mass $\log(M_*/\mathrm{M}_\odot)=(3.5, 4.5, 5.5)$ and age of 0.6 Myr plus 1 to 9 Myr with steps of 1 Myr. Each Cloudy model has been iterated until convergence, and the termination criterium was reaching the lowest temperature \te$=4000$~K.  We can then compute integrated properties across the full model, including line fluxes, line ratios, and \fion. As follows from the bottom-right panel of Figure~\ref{fig:phangs_1to1_stats}, the resulting grid of Cloudy and Starburst99 models produces the same range of \ha\ luminosities as is seen for the reference \hii\ regions selected from PHANGS-MUSE, however the exact distributions do not match each other because the model parameter space is regularly sampled (in contrast to the observational data).

We derive the integrated value of \fion\ from the Cloudy models as the ratio of number densities averaged over a nebula volume:
\begin{equation}
    f_\mathrm{ion} = \frac{<n({\rm H}^+)>}{<n({\rm H})>} = \frac{\int n_{\rm H^+}(r) r^2 \mathrm{d}r}{\int n_{\rm H}(r) r^2 \mathrm{d}r},
    \label{eq:cloudy_fion}
\end{equation}
where $n_{\rm H^+}(r)$ and $n_{\rm H}(r)$ are the ionized and total hydrogen number density in each zone of the modelled nebula, and $r$ is the radius of the corresponding zone. We can easily calculate these integrals from the computed Cloudy models and thus obtain \fion\ characterizing the whole nebula.

\begin{figure*}
    \centering
    \includegraphics[width=\linewidth]{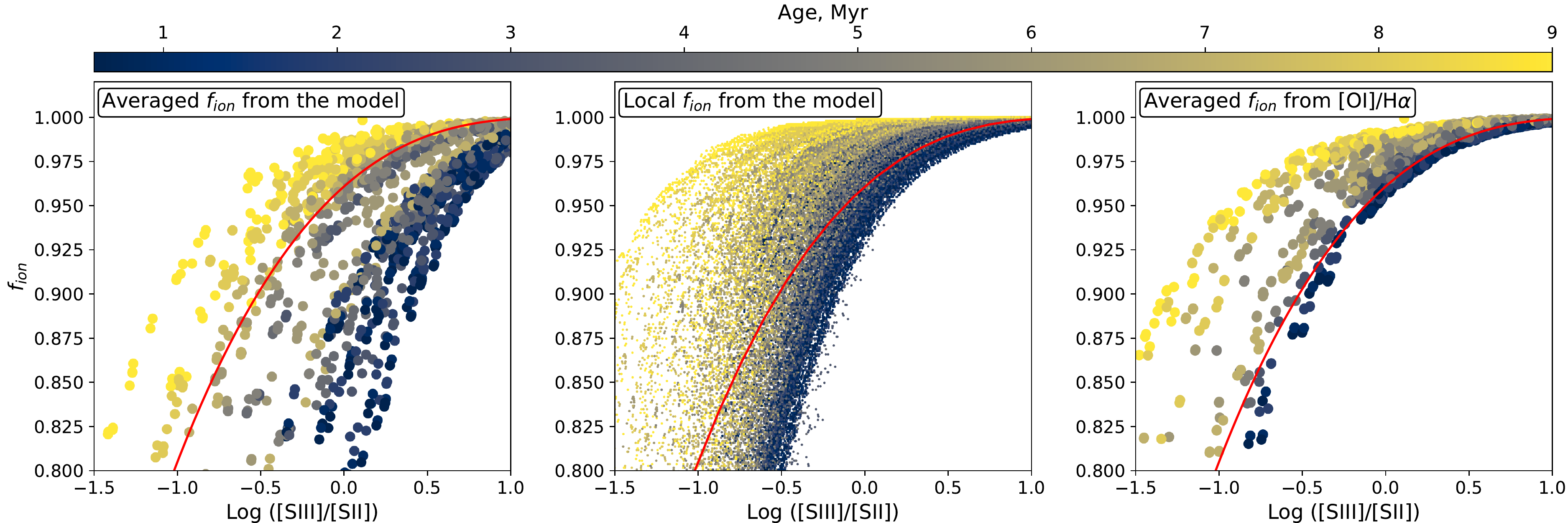}
    \caption{The relation between \fion\ and \siii/\sii\, based on the Cloudy models. On the left-hand panel the value of \fion\ is obtained directly from the model (as $<n({\rm H}^+)>/<n({\rm H})>$), while on the right-hand panel it was derived as in the observations (Equation~(\ref{eqn})) based on the integrated values of \oi/\ha\, metallicity and electron temperature \te(\nii). The central panel shows  $f_{\rm ion}=n({\rm H}^+)/n({\rm H})$ in individual zones of the models. The colors on all panels correspond to the age of the ionizing cluster, and the red curve is the empirical relation derived from the observational data (Equation~\ref{eqn:fion}). Spatially integrating across the entire line emitting region clearly reduces the scatter in this relation, and brings the model results into better agreement with the empirical relation.}
    \label{fig:cloudy_s3s2_fion}
\end{figure*}

As seen in the left-hand panel of Figure~\ref{fig:cloudy_s3s2_fion}, the range of values for \fion\ calculated in the models agrees very well with the measurements obtained from observations, but with large scatter. This scatter in the \fion\ relation as a function of \siii/\sii\ is significantly larger than in the observational measurements, driven mostly by differences in the age of the ionizing cluster. To some extent, the reduced scatter for the observational data could be a result of an observational bias, as many of the stellar associations have very young ages ($\sim$1 Myr, Figure~\ref{fig:phangs_1to1_stats}). However, we see clear offsets between the youngest Cloudy models and the empirical relation derived in Figure \ref{fig:fion_relation}. Note also that similar correlation between \fion\ and \siii/\sii\ exists also at the scales of individual zones (radii) of the models (central panel) where \fion\ is not affected by the truncation criteria of a model or by its geometry.

\subsubsection{Assumption of co-spatial \oi\ and \ha}
\label{sec:cloudy_cospatial}

One of the limitation of the charge-exchange method (as described in Section \ref{sec:whydoesitwork}) is that Equation~\ref{eqn} should only work for \oi\, and \ha\, emitted co-spatially within a nebula, but in unresolved observations we collect line emission from the whole nebula. Thus, our measurement of \oi/\ha\, differs from the value corresponding to the area where the charge-exchange process occurs. This, in turn, should lead to deviations between the \fion\ empirically calibrated for this work and \fion\ when considering only the \oi\ emitting region. 
From our Cloudy models, we directly estimate how significant this deviation is. For that we calculate  \fion\ following the method used with our observational data -- we measure the integrated value of \oi/\ha, adopt the input metallicity of the model, and derive \fion\ using Equation~\ref{eqn} assuming \te=\te(\nii) as obtained directly from the model. 
In Figure \ref{fig:cloudy_s3s2_fion} (right) we show that the discrepancy between the empirical correlation and the Cloudy models is significantly reduced when we calculate \fion\ in the same way as we did with the observational data. We note in particular that the scatter in \fion\ vs. \siii/\sii\ is very similar to that for observational data, and now the empirical relation correlates with the Cloudy models for youngest age (see right-hand panel of Figure~\ref{fig:cloudy_s3s2_fion}),  consistent with the expected observational bias.

\begin{figure*}
    \centering
    \includegraphics[width=\linewidth]{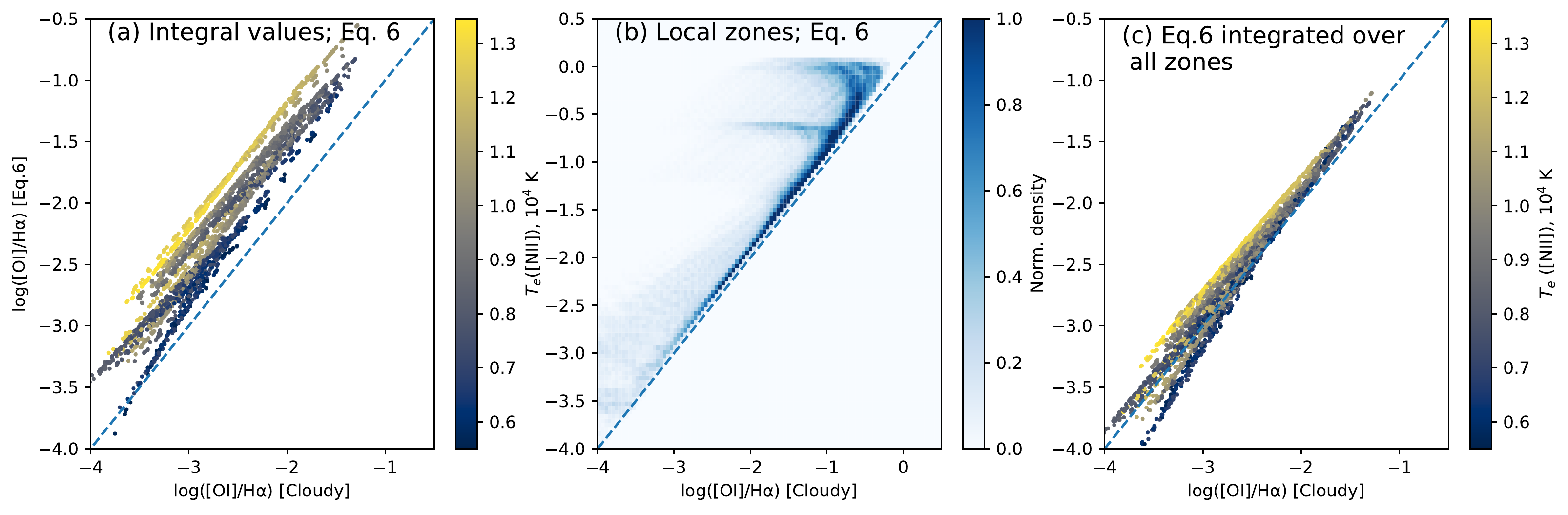}
    \caption{\revone{Validation of Equation~\ref{eqn} using our Cloudy models. Panel (a): Comparison of [O~{\sc i}]/H$\alpha$ calculated with this relation assuming the mean $f_{ion}$ (Equation~\ref{eq:cloudy_fion}) to the Cloudy output. Panel (b): The same comparison, but applied to the individual local zones in the Cloudy models. Here $f_{ion}$ corresponds to the local values of $n(\mathrm{H}^+)/n(\mathrm{H})$. Panel (c): The same test, but applied to results using Equation~\ref{eq:app_o1ha_obs} that are accounting for variations of $f_{ion}$ and the spatial extent of the \oi\, and \ha\, emitting zones.}}
    \label{fig:cloudy_eq6_test}
\end{figure*}

\begin{figure}
    \centering
    \includegraphics[width=\linewidth]{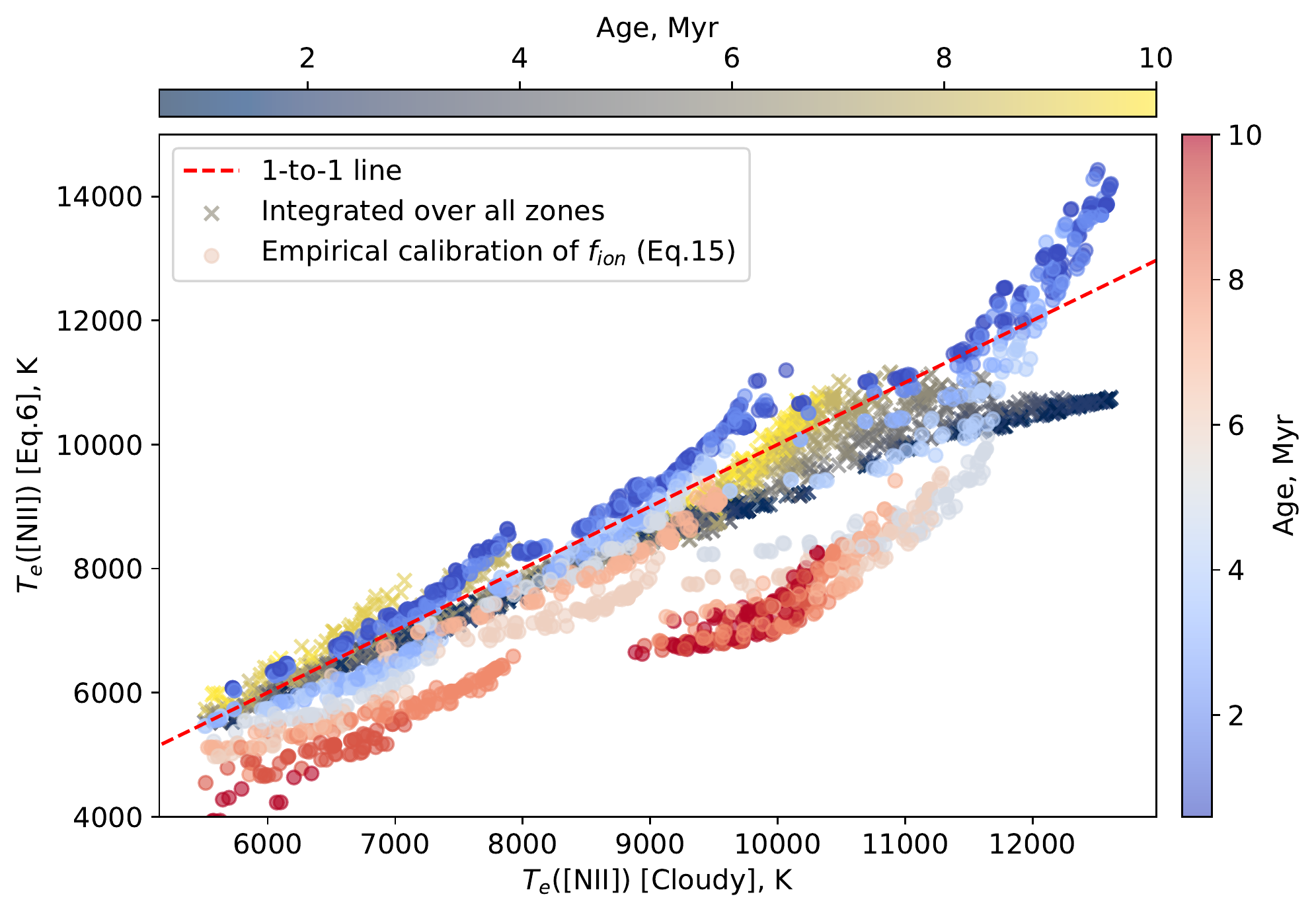}
    \caption{\revone{A validation of the resulting values for $T_e$ based on the Cloudy models. Crosses shows the values of $T_e$([N~{\sc ii}]) obtained from the volumetric version of Equation~\ref{eqn}, which accounts for variations in $f_{ion}$ and the spatial extent of the \oi\, and \ha\, emitting zones (Equation~\ref{eq:app_o1ha_obs}). Circles represent the $T_e$([N~{\sc ii}]) values yielded by Equation~\ref{eqn} assuming the empirical calibration of $f_{ion}$ (by Equation~\ref{eqn:fion}). The calculated values are compared with the `true' $T_e$([N~{\sc ii}]) from the Cloudy models. The symbol colour encodes the age of the ionizing cluster. Both data sets show generally good agreement with the 1-to-1 relation, but the empirical calibration significantly underestimate $T_e$ for old regions. Note that for the empirically-calibrated results we show only the models with $\log(\mathrm{[OI]/H\alpha}) \ge -3$ and $-1 \ge \log(\mathrm{[SIII]/[SII]}) \le -0.7$ that correspond to the values obtained in our observations (see, e.g., Figures~\ref{fig:chaos_fion_a} and \ref{fig:fion_relation}). The prominent `tails' towards higher $T_e$ for the empirically-calibrated values correspond only to those regions where $\log(\mathrm{[SIII]/[SII]}) > 0.5$ -- only a few such regions were used for calibration of Equation~\ref{eqn:fion}, thus the calibration is  less certain for \hii\, regions with the highest ionization parameters.}}
    \label{fig:cloudy_te_test}
\end{figure}

\revone{The fact that the `true' values of $f_{ion}$ do not agree with those obtained from  Equation~\ref{eqn} directly follows from the limitation of this equation when applied to unresolved \hii\, regions. Indeed, we can invert our use of Equation \ref{eqn}, and use it to compute the expected \oi/\ha\ based on our Cloudy models. The values of [O~{\sc i}]/H$\alpha$ that we compute from Equation~\ref{eqn} assuming the `true' $f_{ion}$ derived from Equation~\ref{eq:cloudy_fion} are significantly overestimated in comparison with the `true' line ratios from the Cloudy models (left-hand panel of Figure~\ref{fig:cloudy_eq6_test}). At the same time, Equation~\ref{eqn} reproduces well the [O~{\sc i}]/H$\alpha$ ratio when it is applied to the individual local zones of the Cloudy models (central panel of Figure~\ref{fig:cloudy_eq6_test}).}\footnote{\revone{We believe the small departures from the 1-to-1 relation might be caused by the differences in the atomic data used in the Cloudy models compared to our calculations (see Appendix~\ref{app:derivation}), and by the violation of our assumption of a uniform distribution of $T_e$.}}
\revone{We show in Appendix~\ref{app:derivation} that a revised version of Equation~\ref{eqn} that considers the volumetric differences in the \oi\, and \ha\, emitting zones is more complicated (Equation~\ref{eq:app_o1ha_obs}) and cannot be solved for unresolved \hii\, regions. However, we can test the impact introducing a volumetric correction using the  Cloudy models. Integrating the term $\xi'$, dependent on the local $f_{ion}$ over all Cloudy zones, we obtain values for [O~{\sc i}]/H$\alpha$ that are in very good agreement with those in the output of the Cloudy models (right-hand panel of Figure~\ref{fig:cloudy_eq6_test}).}  

\revone{From this analysis we conclude that Equation~\ref{eqn}, which is fundamental to our method, is valid for both resolved and unresolved nebulae, but for latter it is necessary to weigh $f_{ion}$ according to the volumes occupied by the \oi\, and \ha\, emitting zones instead of using the `true' $f_{ion}$. By definition, the values of $f_{ion}$ derived from [O~{\sc i}]/H$\alpha$ and used in the right-hand plot of Figure~\ref{fig:cloudy_s3s2_fion} already include these weights, and that is why they disagree with the `true' $f_{ion}$ (Figure \ref{fig:cloudy_eq6_test}, left). The empirically defined values of $f_{ion}$ obtained by Equation~\ref{eqn:fion} are weighted as the ionization structure in sulphur encodes some of these volumetric effects, and as a result our relation intrinsically corrects for this potential bias.} 

\revone{As the goal of the presented method is to obtain the measurements of electron temperature, we carry out a final test by comparing the values of $T_e$([N~{\sc ii}] derived from Equation~(\ref{eqn}) with the output of our Cloudy models. As follows from Figure~\ref{fig:cloudy_te_test}, in general, both values are indeed in agreement if we use the observationally-defined calibration of $f_{ion}$ (Equation~\ref{eqn:fion}), or solve Equation~\ref{eq:app_o1ha_obs} on a zone-by-zone basis. Moreover, the empirical calibration shows even better agreement with the 1-to-1 relation in the high $T_e$ regime. Note, however, that the method significantly underestimates $T_e$ for the regions ionized by older star clusters.}
In summary, while we have shown the theoretical limitations of this approach, based on the underlying physical mechanisms governing the charge-exchange process, we further demonstrate that our empirical calibration based on integrated nebulae sufficiently accounts for the unresolved volumetric effects. 

\subsubsection{Age as a secondary parameter}

\begin{figure*}
    \centering
    \includegraphics[width=\linewidth]{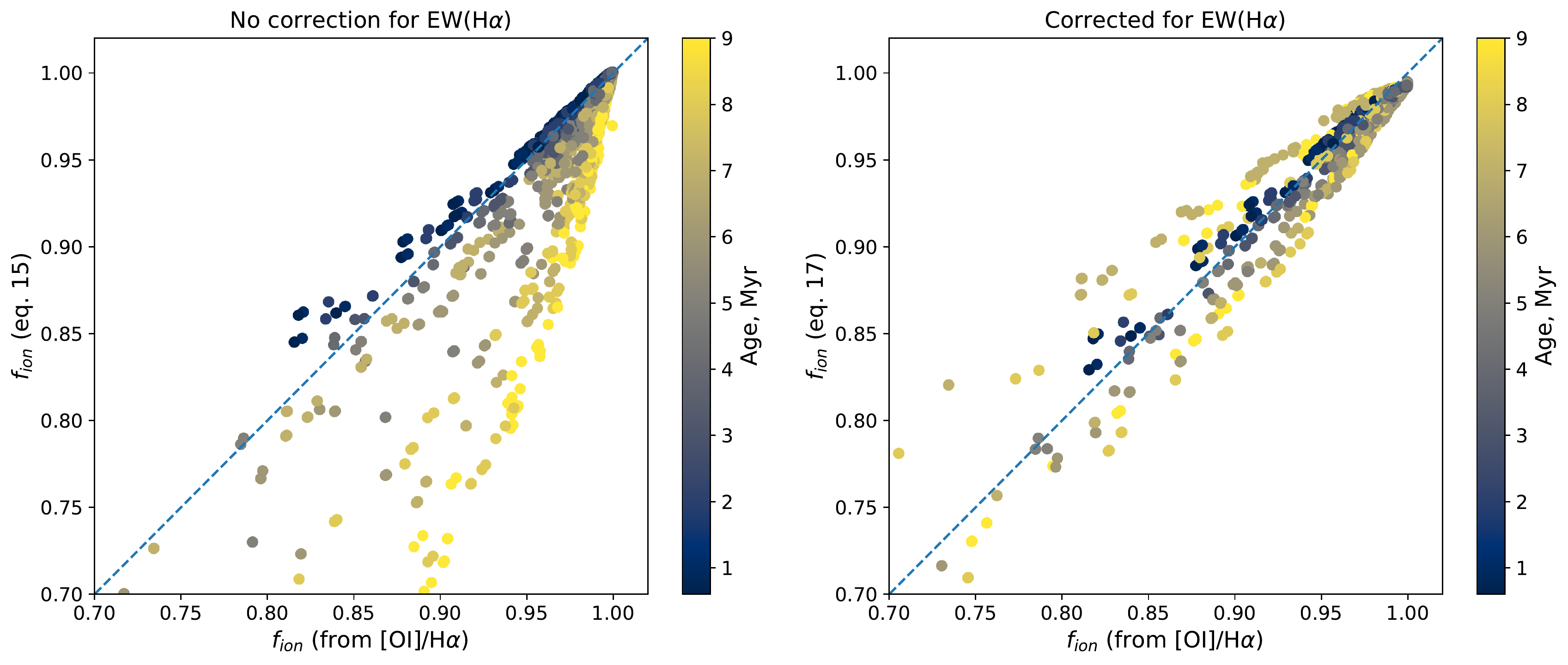}
    \caption{A comparison of the cloudy modeled \fion\ measured from \oi/\ha\ with what is derived using a fit based on \siii/\sii\ (as in Figure \ref{fig:cloudy_s3s2_fion}, right). The left-hand panel depends only on x=log$_{10}$(\siii/\sii) while the right-hand panel includes a secondary dependence on age as traced by the equivalent width of \ha, as y = log$_{10}$(EW(\ha)).  Both plots are colored by the age of the ionizing stellar cluster, and correcting for this secondary dependence on age significantly improves the agreements between the fit and the value for \fion\ in the cloudy models. }
    \label{fig:cloudy_fion_fit}
\end{figure*}

Based on the remarkable visual agreement between the Cloudy models and observed data when using \fion\ inferred from \oi/\ha\ in the models, and the matched biases regarding spatial averaging, we choose to use this modeled value of \fion\ to independently derive a relation between \fion\ and \siii/\sii. We apply our empirical calibration from Equation \ref{eqn:fion} to the Cloudy models. 
As shown in Figure \ref{fig:cloudy_fion_fit} (left), this relation shows relatively good agreement at high \fion\ $>$ 0.9, but a clear secondary dependence on age.  While the age of an \hii\ region is difficult to constrain observationally, strong correlations are seen in models with the equivalent width (EW) of \ha\ \citep{Leitherer2014}.  We use the EW output from the Cloudy models, and include this secondary dependence to perform a third order polynomial fit to Figure \ref{fig:cloudy_s3s2_fion} (right) in order to derive the following functional form:

\begin{equation}
\label{eqn:cloudy_fion_fit}
    f_{\rm ion} = 0.992 + 0.0076 \cdot (x + 0.8295 - 0.731y + 0.148xy)^3,
\end{equation}
where again x = log$_{10}$(\siii/\sii) and y = log$_{10}$(EW(\ha)). In Figure \ref{fig:cloudy_fion_fit} (right), we see that this successfully removes the secondary dependence on age. 

There is one remaining complication in our comparison of the modeled and observed empirical relations. According to the Cloudy models, we find that EW(\ha) $\approx$ 1815 \AA\ at an age of 1 Myr, however based on our observations we find EW(\ha) $\approx$ 126 \AA\ for the same aged nebulae that show a 1-to-1 correspondence with the HST stellar associations (Larsen et al. in prep, Scheuermann et al. in prep). This offset between modeled and observed EW has been previously reported \citep{Morisset2016}, and it can be a consequence of the large escape fraction of ionizing radiation from \hii\ regions, or of the significant contribution of underlying old stellar population -- both factors are not considered in the Cloudy models. 
Likely we are seeing a combination of both effects, as typical escape fractions are expected to be $\sim$50\% \citep{Oey1997, Doran2013, Belfiore2022}, which alone is insufficient to account for the difference. 
To first order a simple scaling is possible whereby we `convert' the observed EW(\ha) to the modeled EW(\ha) at this fixed reference age of 1 Myr, such that the variable y in Equation \ref{eqn:cloudy_fion_fit} is replaced with y = log$_{10}$(EW(\ha)$_{\rm obs}\cdot$ 1815/126). 

\begin{figure}
    \centering
    \includegraphics[width=3.5in]{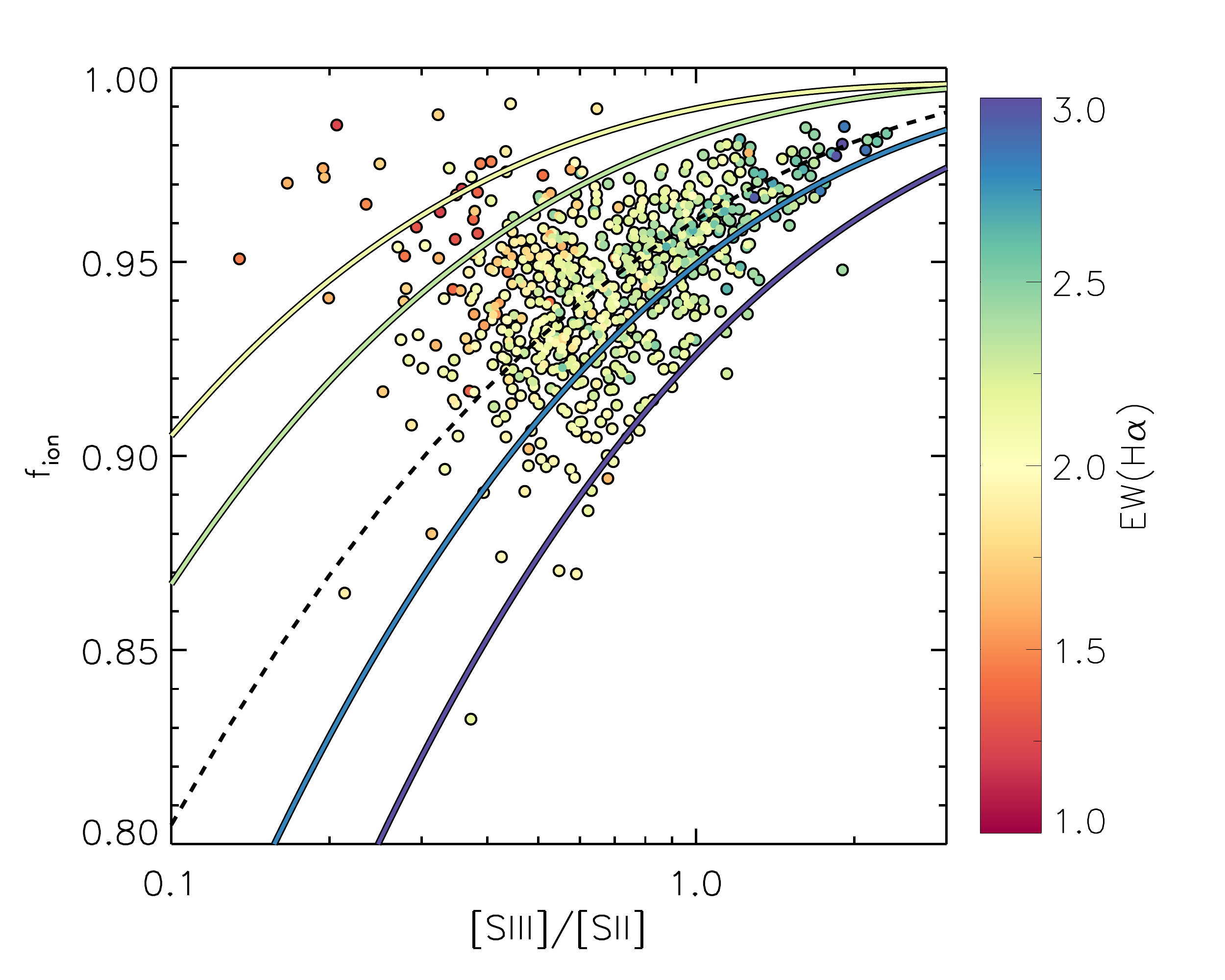}
    \caption{A direct comparison of the observed \fion\ as a function of \siii/\sii\ within the PHANGS-MUSE \hii\ regions with the age-dependent fit resulting from the cloudy models (colored lines, Equation \ref{eqn:cloudy_fion_fit}). The colors correspond to EW(\ha), and for the cloudy model fits they have been scaled as described in the text in order to account for the large quantitative discrepancy between model and observation. The observed data shows a clear gradient, indicating that age trends do play an important role in driving the scatter in this relation. The empirical fit (dashed line) shows good qualitative agreement with the age-dependent cloudy model fit, supporting our choice of this empirical calibration.  However, while the trends are consistent, a significant offset is seen between the EW(\ha) measurements and the predicted values, preventing a straightforward application of our model-driven calibrations to the data. 
    }
    \label{fig:fion_cloudy_data}
\end{figure}

We apply this prescription to the PHANGS-MUSE \hii\ region catalog and show the resulting values in Appendix \ref{fig:app_fion}.   Directly comparing the EW(\ha) measurements with the Cloudy derived model for \fion\ (Figure \ref{fig:fion_cloudy_data}), it is apparent that while the general trends are in agreement, the exact values of EW(\ha) between the models and observations are not in good enough agreement that we can rely on the calibration. However, it demonstrates the potential in future work for improving our \te\ measurements by including age sensitive indicators (such as EW(\ha)) in the prescription, which is not generally done in the literature.

\section{Conclusions}
\label{sec:conclusions}
We develop a new method that exploits the charge-exchange between oxygen and hydrogen to infer the electron temperature, \te, within \hii\ regions using only strong emission lines. This single temperature corresponds to the \oi\ and \ha\ emitting zone.  This method requires three parameters: the \oi/\ha\ line ratio, the gas-phase metallicity 12+log(O/H) and the ionization fraction \fion. Using observations from the CHAOS survey \citep{Berg2020}, where auroral line observations of \te\ are available for $\sim$150 \hii\ regions, we demonstrate that \fion\ correlates with changes in various strong line ratios. We show that the strongest correlation is with \siii/\sii, tracing changes in ionization parameter.  However, due to aperture biases, we cannot derive an empirical relation from this data set alone.

We then use 840 \hii\ regions from the PHANGS-MUSE survey that have auroral line detections of \nii\ $\lambda$5755 to develop an empirical relation between \fion\ and \siii/\sii\ for line fluxes measured across integrated \hii\ regions.  Due to limitations of our calibration sample and concerns about the contribution from diffuse ionized gas, where strong-lines like \sii\ and \oi\ are strongly emitted, we conservatively apply this method only to \hii\ regions that have \siii/\sii\ $>$ 0.5 and a contrast of more than 50\% against the local DIG background. With these restrictions, we recover \te(\nii) to within $\sim$600 K. This uncertainty only increases slightly when we assume a fixed metallicity for each galaxy, demonstrating that this charge-exchange method is not strongly sensitive to the exact determination of metallicity. 

Of the  $\sim$24,000 \hii\ regions in the PHANGS-MUSE nebular catalog \citep{Santoro2022}, we are then able to apply this technique to model \te\ for a total of 4,129 \hii\ regions, $\sim$4 times more than have direct auroral line detections.  We recover positive radial temperature gradients that reflect the expected negative metallicity gradients.  Four galaxies show a scatter in their radial temperature gradient of $\sim$1000 K, well beyond the uncertainties of the method and likely tracing azimuthal variations in the metallicity at fixed radius.  

While some of the assumptions that go into this method would not appear to be met, particularly the assumption that the \oi\ and \ha\ emitting regions have matched electron temperatures, and ideally are co-spatial, we carry out a set of Cloudy models to explore the robustness of our method and in particular to test the empirically derived relation between \fion\ and \siii/\sii. We find that by integrating over the full \hii\ region, our models actually recover the tight observed correlation between \fion\ and \siii/\sii, with a significant secondary dependence on the age of the ionizing stellar cluster. We parameterize this age dependence by changes in the equivalent width of \ha, which shows qualitatively similar agreement to the observed data but significant quantitative differences in the measured values, particularly for EW(\ha). For this reason, we prefer the empirical calibration for \fion, but note that age/EW(\ha) represents a promising additional parameter to consider when deriving accurate \hii\ region temperatures and metallicities.  

This novel method for determining \te\ demonstrates the remarkable potential arising from these new catalogs containing 100--1000s of \hii\ regions with uniform data, as well as future planed surveys (e.g. SIGNALS, \citealt{Rousseau-Nepton2019}; SDSS-V/LVM, \citealt{Kollmeier2017}; AMASE, \citealt{Yan2020}). While this method is physically motivated, there is clearly an exciting potential for the application of machine learning approaches to these data sets.

\begin{acknowledgements}
We thank the referee for their comments, which improved our analysis of the subtleties of this method. We thank Jose Eduardo Mendez Delgado for his comments and input. This work was carried out as part of the PHANGS collaboration. Based on observations collected at the European Southern Observatory under ESO programmes 1100.B-0651, 095.C-0473, and 094.C-0623 (PHANGS--MUSE; PI Schinnerer), as well as 094.B-0321 (MAGNUM; PI Marconi), 099.B-0242, 0100.B-0116, 098.B-0551 (MAD; PI Carollo) and 097.B-0640 (TIMER; PI Gadotti). 
KK, OE and FS gratefully acknowledge funding from the German Research Foundation (DFG) in the form of an Emmy Noether Research Group (grant number KR4598/2-1, PI Kreckel). 
SCOG and RSK acknowledge support from the DFG via SFB 881 ``The Milky Way System'' (Project-ID 138713538; sub-projects B1, B2 and B8) and from the Heidelberg cluster of excellence EXC 2181-390900948 ``STRUCTURES: A unifying approach to emergent phenomena in the physical world, mathematics, and complex data'', funded by the German Excellence Strategy. RSK furthermore thanks for funding from the European Research Council via the ERC Synergy Grant ECOGAL (grant 855130). 
TGW acknowledges funding from the European Research Council (ERC) under the European Union’s Horizon 2020 research and innovation programme (grant agreement No. 694343).

\end{acknowledgements}

%

\begin{thebibliography}{74}
\expandafter\ifx\csname natexlab\endcsname\relax\def\natexlab#1{#1}\fi

\bibitem[{{Allen} {et~al.}(2008){Allen}, {Groves}, {Dopita}, {Sutherland}, \&
  {Kewley}}]{Allen2008}
{Allen}, M.~G., {Groves}, B.~A., {Dopita}, M.~A., {Sutherland}, R.~S., \&
  {Kewley}, L.~J. 2008, \apjs, 178, 20

\bibitem[{{Anand} {et~al.}(2021){Anand}, {Lee}, {Van Dyk}, {Leroy},
  {Rosolowsky}, {Schinnerer}, {Larson}, {Kourkchi}, {Kreckel}, {Scheuermann},
  {Rizzi}, {Thilker}, {Tully}, {Bigiel}, {Blanc}, {Boquien}, {Chandar}, {Dale},
  {Emsellem}, {Deger}, {Glover}, {Grasha}, {Groves}, {Klessen}, {Kruijssen},
  {Querejeta}, {S{\'a}nchez-Bl{\'a}zquez}, {Schruba}, {Turner}, {Ubeda},
  {Williams}, \& {Whitmore}}]{Anand2021}
{Anand}, G.~S., {Lee}, J.~C., {Van Dyk}, S.~D., {et~al.} 2021, \mnras, 501,
  3621

\bibitem[{{Azimlu} {et~al.}(2011){Azimlu}, {Marciniak}, \&
  {Barmby}}]{Azimlu+2011}
{Azimlu}, M., {Marciniak}, R., \& {Barmby}, P. 2011, \aj, 142, 139

\bibitem[{{Bacon} {et~al.}(2010)}]{Bacon2010}
{Bacon}, R. {et~al.} 2010, in \procspie, Vol. 7735, Ground-based and Airborne
  Instrumentation for Astronomy III, 773508

\bibitem[{{Baldwin} {et~al.}(1981){Baldwin}, {Phillips}, \&
  {Terlevich}}]{Baldwin1981}
{Baldwin}, J.~A., {Phillips}, M.~M., \& {Terlevich}, R. 1981, \pasp, 93, 5

\bibitem[{{Balser} {et~al.}(2015){Balser}, {Wenger}, {Anderson}, \&
  {Bania}}]{Balser2015}
{Balser}, D.~S., {Wenger}, T.~V., {Anderson}, L.~D., \& {Bania}, T.~M. 2015,
  \apj, 806, 199

\bibitem[{{Barklem}(2007)}]{Barklem2007}
{Barklem}, P.~S. 2007, \aap, 462, 781

\bibitem[{{Barnes} {et~al.}(2021){Barnes}, {Glover}, {Kreckel}, {Ostriker},
  {Bigiel}, {Belfiore}, {Be{\v{s}}li{\'c}}, {Blanc}, {Chevance}, {Dale},
  {Egorov}, {Eibensteiner}, {Emsellem}, {Grasha}, {Groves}, {Klessen},
  {Kruijssen}, {Leroy}, {Longmore}, {Lopez}, {McElroy}, {Meidt}, {Murphy},
  {Rosolowsky}, {Saito}, {Santoro}, {Schinnerer}, {Schruba}, {Sun}, {Watkins},
  \& {Williams}}]{Barnes2021}
{Barnes}, A.~T., {Glover}, S.~C.~O., {Kreckel}, K., {et~al.} 2021, \mnras, 508,
  5362

\bibitem[{{Belfiore} {et~al.}(2022){Belfiore}, {Santoro}, {Groves},
  {Schinnerer}, {Kreckel}, {Glover}, {Klessen}, {Emsellem}, {Blanc}, {Congiu},
  {Barnes}, {Boquien}, {Chevance}, {Dale}, {Diederik Kruijssen}, {Leroy},
  {Pan}, {Pessa}, {Schruba}, \& {Williams}}]{Belfiore2022}
{Belfiore}, F., {Santoro}, F., {Groves}, B., {et~al.} 2022, \aap, 659, A26

\bibitem[{{Berg} {et~al.}(2020){Berg}, {Pogge}, {Skillman}, {Croxall},
  {Moustakas}, {Rogers}, \& {Sun}}]{Berg2020}
{Berg}, D.~A., {Pogge}, R.~W., {Skillman}, E.~D., {et~al.} 2020, \apj, 893, 96

\bibitem[{{Berg} {et~al.}(2015){Berg}, {Skillman}, {Croxall}, {Pogge},
  {Moustakas}, \& {Johnson-Groh}}]{Berg2015}
{Berg}, D.~A., {Skillman}, E.~D., {Croxall}, K.~V., {et~al.} 2015, \apj, 806,
  16

\bibitem[{{Croxall} {et~al.}(2015){Croxall}, {Pogge}, {Berg}, {Skillman}, \&
  {Moustakas}}]{Croxall2015}
{Croxall}, K.~V., {Pogge}, R.~W., {Berg}, D.~A., {Skillman}, E.~D., \&
  {Moustakas}, J. 2015, \apj, 808, 42

\bibitem[{{Croxall} {et~al.}(2016){Croxall}, {Pogge}, {Berg}, {Skillman}, \&
  {Moustakas}}]{Croxall2016}
{Croxall}, K.~V., {Pogge}, R.~W., {Berg}, D.~A., {Skillman}, E.~D., \&
  {Moustakas}, J. 2016, \apj, 830, 4

\bibitem[{{Dong} \& {Draine}(2011)}]{Dong2011}
{Dong}, R. \& {Draine}, B.~T. 2011, \apj, 727, 35

\bibitem[{{Dopita} {et~al.}(2016){Dopita}, {Kewley}, {Sutherland}, \&
  {Nicholls}}]{Dopita2016}
{Dopita}, M.~A., {Kewley}, L.~J., {Sutherland}, R.~S., \& {Nicholls}, D.~C.
  2016, \apss, 361, 61

\bibitem[{{Doran} {et~al.}(2013){Doran}, {Crowther}, {de Koter}, {Evans},
  {McEvoy}, {Walborn}, {Bastian}, {Bestenlehner}, {Gr{\"a}fener}, {Herrero},
  {K{\"o}hler}, {Ma{\'\i}z Apell{\'a}niz}, {Najarro}, {Puls}, {Sana},
  {Schneider}, {Taylor}, {van Loon}, \& {Vink}}]{Doran2013}
{Doran}, E.~I., {Crowther}, P.~A., {de Koter}, A., {et~al.} 2013, \aap, 558,
  A134

\bibitem[{{Dors} {et~al.}(2011){Dors}, {Krabbe}, {H{\"a}gele}, \&
  {P{\'e}rez-Montero}}]{Dors2011}
{Dors}, Jr., O.~L., {Krabbe}, A., {H{\"a}gele}, G.~F., \& {P{\'e}rez-Montero},
  E. 2011, \mnras, 415, 3616

\bibitem[{{Draine}(2011)}]{Draine2011}
{Draine}, B.~T. 2011, {Physics of the Interstellar and Intergalactic Medium}
  (Princeton University Press)

\bibitem[{{Ekstr{\"o}m} {et~al.}(2012){Ekstr{\"o}m}, {Georgy}, {Eggenberger},
  {Meynet}, {Mowlavi}, {Wyttenbach}, {Granada}, {Decressin}, {Hirschi},
  {Frischknecht}, {Charbonnel}, \& {Maeder}}]{Ekstrom2012}
{Ekstr{\"o}m}, S., {Georgy}, C., {Eggenberger}, P., {et~al.} 2012, \aap, 537,
  A146

\bibitem[{{Emsellem} {et~al.}(2022){Emsellem}, {Schinnerer}, {Santoro},
  {Belfiore}, {Pessa}, {McElroy}, {Blanc}, {Congiu}, {Groves}, {Ho}, {Kreckel},
  {Razza}, {Sanchez-Blazquez}, {Egorov}, {Faesi}, {Klessen}, {Leroy}, {Meidt},
  {Querejeta}, {Rosolowsky}, {Scheuermann}, {Anand}, {Barnes},
  {Be{\v{s}}li{\'c}}, {Bigiel}, {Boquien}, {Cao}, {Chevance}, {Dale},
  {Eibensteiner}, {Glover}, {Grasha}, {Henshaw}, {Hughes}, {Koch}, {Kruijssen},
  {Lee}, {Liu}, {Pan}, {Pety}, {Saito}, {Sandstrom}, {Schruba}, {Sun},
  {Thilker}, {Usero}, {Watkins}, \& {Williams}}]{Emsellem2022}
{Emsellem}, E., {Schinnerer}, E., {Santoro}, F., {et~al.} 2022, \aap, 659, A191

\bibitem[{{Ferland} {et~al.}(2017){Ferland}, {Chatzikos}, {Guzm{\'a}n},
  {Lykins}, {van Hoof}, {Williams}, {Abel}, {Badnell}, {Keenan}, {Porter}, \&
  {Stancil}}]{Ferland2017}
{Ferland}, G.~J., {Chatzikos}, M., {Guzm{\'a}n}, F., {et~al.} 2017, \rmxaa, 53,
  385

\bibitem[{{Georgy} {et~al.}(2013){Georgy}, {Ekstr{\"o}m}, {Eggenberger},
  {Meynet}, {Haemmerl{\'e}}, {Maeder}, {Granada}, {Groh}, {Hirschi}, {Mowlavi},
  {Yusof}, {Charbonnel}, {Decressin}, \& {Barblan}}]{Georgy2013}
{Georgy}, C., {Ekstr{\"o}m}, S., {Eggenberger}, P., {et~al.} 2013, \aap, 558,
  A103

\bibitem[{{Grevesse} {et~al.}(2010){Grevesse}, {Asplund}, {Sauval}, \&
  {Scott}}]{Grevesse2010}
{Grevesse}, N., {Asplund}, M., {Sauval}, A.~J., \& {Scott}, P. 2010, \apss,
  328, 179

\bibitem[{{Hausen} {et~al.}(2002){Hausen}, {Reynolds}, \&
  {Haffner}}]{Hausen2002}
{Hausen}, N.~R., {Reynolds}, R.~J., \& {Haffner}, L.~M. 2002, \aj, 124, 3336

\bibitem[{{Ho}(2019)}]{Ho2019_machine}
{Ho}, I.~T. 2019, \mnras, 485, 3569

\bibitem[{{Ho} {et~al.}(2019){Ho}, {Kreckel}, {Meidt}, {Groves}, {Blanc},
  {Bigiel}, {Dale}, {Emsellem}, {Glover}, {Grasha}, {Kewley}, {Kruijssen},
  {Lang}, {McElroy}, {Kudritzki}, {Sanchez-Blazquez}, {Sand strom}, {Santoro},
  {Schinnerer}, \& {Schruba}}]{Ho+2019}
{Ho}, I.~T., {Kreckel}, K., {Meidt}, S.~E., {et~al.} 2019, \apjl, 885, L31

\bibitem[{{Ho} {et~al.}(2017){Ho}, {Seibert}, {Meidt}, {Kudritzki},
  {Kobayashi}, {Groves}, {Kewley}, {Madore}, {Rich}, \& {Schinnerer}}]{Ho2017}
{Ho}, I.~T., {Seibert}, M., {Meidt}, S.~E., {et~al.} 2017, \apj, 846, 39

\bibitem[{{Hummer} \& {Storey}(1987)}]{hs87}
{Hummer}, D.~G. \& {Storey}, P.~J. 1987, \mnras, 224, 801

\bibitem[{{Kauffmann} {et~al.}(2003){Kauffmann}, {Heckman}, {Tremonti},
  {Brinchmann}, {Charlot}, {White}, {Ridgway}, {Brinkmann}, {Fukugita}, {Hall},
  {Ivezi{\'c}}, {Richards}, \& {Schneider}}]{Kauffmann2003}
{Kauffmann}, G., {Heckman}, T.~M., {Tremonti}, C., {et~al.} 2003, \mnras, 346,
  1055

\bibitem[{{Kepley} {et~al.}(2011){Kepley}, {Chomiuk}, {Johnson}, {Goss},
  {Balser}, \& {Pisano}}]{Kepley2011}
{Kepley}, A.~A., {Chomiuk}, L., {Johnson}, K.~E., {et~al.} 2011, \apjl, 739,
  L24

\bibitem[{{Kewley} \& {Dopita}(2002)}]{Kewley2002}
{Kewley}, L.~J. \& {Dopita}, M.~A. 2002, \apjs, 142, 35

\bibitem[{{Kewley} \& {Ellison}(2008)}]{Kewley2008}
{Kewley}, L.~J. \& {Ellison}, S.~L. 2008, \apj, 681, 1183

\bibitem[{{Kewley} {et~al.}(2001){Kewley}, {Heisler}, {Dopita}, \&
  {Lumsden}}]{Kewley2001}
{Kewley}, L.~J., {Heisler}, C.~A., {Dopita}, M.~A., \& {Lumsden}, S. 2001,
  \apjs, 132, 37

\bibitem[{{Kewley} {et~al.}(2019){Kewley}, {Nicholls}, \&
  {Sutherland}}]{Kewley2019}
{Kewley}, L.~J., {Nicholls}, D.~C., \& {Sutherland}, R.~S. 2019, \araa, 57, 511

\bibitem[{{Klessen} \& {Glover}(2016)}]{Klessen2016}
{Klessen}, R.~S. \& {Glover}, S. C.~O. 2016, Saas-Fee Advanced Course, 43, 85

\bibitem[{{Kollmeier} {et~al.}(2017){Kollmeier}, {Zasowski}, {Rix}, {Johns},
  {Anderson}, {Drory}, {Johnson}, {Pogge}, {Bird}, {Blanc}, {Brownstein},
  {Crane}, {De Lee}, {Klaene}, {Kreckel}, {MacDonald}, {Merloni}, {Ness},
  {O'Brien}, {Sanchez-Gallego}, {Sayres}, {Shen}, {Thakar}, {Tkachenko},
  {Aerts}, {Blanton}, {Eisenstein}, {Holtzman}, {Maoz}, {Nandra}, {Rockosi},
  {Weinberg}, {Bovy}, {Casey}, {Chaname}, {Clerc}, {Conroy}, {Eracleous},
  {G{\"a}nsicke}, {Hekker}, {Horne}, {Kauffmann}, {McQuinn}, {Pellegrini},
  {Schinnerer}, {Schlafly}, {Schwope}, {Seibert}, {Teske}, \& {van
  Saders}}]{Kollmeier2017}
{Kollmeier}, J.~A., {Zasowski}, G., {Rix}, H.-W., {et~al.} 2017, arXiv
  e-prints, arXiv:1711.03234

\bibitem[{{Kopsacheili} {et~al.}(2020){Kopsacheili}, {Zezas}, \&
  {Leonidaki}}]{Kopsacheili2020}
{Kopsacheili}, M., {Zezas}, A., \& {Leonidaki}, I. 2020, \mnras, 491, 889

\bibitem[{{Kreckel} {et~al.}(2020){Kreckel}, {Ho}, {Blanc}, {Glover}, {Groves},
  {Rosolowsky}, {Bigiel}, {Boqu{\'\i}en}, {Chevance}, {Dale}, {Deger},
  {Emsellem}, {Grasha}, {Kim}, {Klessen}, {Kruijssen}, {Lee}, {Leroy}, {Liu},
  {McElroy}, {Meidt}, {Pessa}, {Sanchez-Blazquez}, {Sandstrom}, {Santoro},
  {Scheuermann}, {Schinnerer}, {Schruba}, {Utomo}, {Watkins}, \&
  {Williams}}]{Kreckel2020}
{Kreckel}, K., {Ho}, I.~T., {Blanc}, G.~A., {et~al.} 2020, \mnras, 499, 193

\bibitem[{{Kreckel} {et~al.}(2019){Kreckel}, {Ho}, {Blanc}, {Groves},
  {Santoro}, {Schinnerer}, {Bigiel}, {Chevance}, {Congiu}, {Emsellem}, {Faesi},
  {Glover}, {Grasha}, {Kruijssen}, {Lang}, {Leroy}, {Meidt}, {McElroy}, {Pety},
  {Rosolowsky}, {Saito}, {Sandstrom}, {Sanchez-Blazquez}, \&
  {Schruba}}]{Kreckel2019}
{Kreckel}, K., {Ho}, I.~T., {Blanc}, G.~A., {et~al.} 2019, \apj, 887, 80

\bibitem[{{Kroupa}(2001)}]{Kroupa2001}
{Kroupa}, P. 2001, \mnras, 322, 231

\bibitem[{{Lang} {et~al.}(2020){Lang}, {Meidt}, {Rosolowsky}, {Nofech},
  {Schinnerer}, {Leroy}, {Emsellem}, {Pessa}, {Glover}, {Groves}, {Hughes},
  {Kruijssen}, {Querejeta}, {Schruba}, {Bigiel}, {Blanc}, {Chevance},
  {Colombo}, {Faesi}, {Henshaw}, {Herrera}, {Liu}, {Pety}, {Puschnig}, {Saito},
  {Sun}, \& {Usero}}]{Lang2020}
{Lang}, P., {Meidt}, S.~E., {Rosolowsky}, E., {et~al.} 2020, \apj, 897, 122

\bibitem[{{Law} {et~al.}(2021){Law}, {Ji}, {Belfiore}, {Bershady},
  {Cappellari}, {Westfall}, {Yan}, {Bizyaev}, {Brownstein}, {Drory}, \&
  {Andrews}}]{Law2021}
{Law}, D.~R., {Ji}, X., {Belfiore}, F., {et~al.} 2021, \apj, 915, 35

\bibitem[{{Lee} {et~al.}(2022){Lee}, {Whitmore}, {Thilker}, {Deger}, {Larson},
  {Ubeda}, {Anand}, {Boquien}, {Chandar}, {Dale}, {Emsellem}, {Leroy},
  {Rosolowsky}, {Schinnerer}, {Schmidt}, {Lilly}, {Turner}, {Van Dyk}, {White},
  {Barnes}, {Belfiore}, {Bigiel}, {Blanc}, {Cao}, {Chevance}, {Congiu},
  {Egorov}, {Glover}, {Grasha}, {Groves}, {Henshaw}, {Hughes}, {Klessen},
  {Koch}, {Kreckel}, {Kruijssen}, {Liu}, {Lopez}, {Mayker}, {Meidt}, {Murphy},
  {Pan}, {Pety}, {Querejeta}, {Razza}, {Saito}, {S{\'a}nchez-Bl{\'a}zquez},
  {Santoro}, {Sardone}, {Scheuermann}, {Schruba}, {Sun}, {Usero}, {Watkins}, \&
  {Williams}}]{Lee2022}
{Lee}, J.~C., {Whitmore}, B.~C., {Thilker}, D.~A., {et~al.} 2022, \apjs, 258,
  10

\bibitem[{{Leitherer} {et~al.}(2014){Leitherer}, {Ekstr{\"o}m}, {Meynet},
  {Schaerer}, {Agienko}, \& {Levesque}}]{Leitherer2014}
{Leitherer}, C., {Ekstr{\"o}m}, S., {Meynet}, G., {et~al.} 2014, \apjs, 212, 14

\bibitem[{{Leroy} {et~al.}(2021){Leroy}, {Schinnerer}, {Hughes}, {Rosolowsky},
  {Pety}, {Schruba}, {Usero}, {Blanc}, {Chevance}, {Emsellem}, {Faesi},
  {Herrera}, {Liu}, {Meidt}, {Querejeta}, {Saito}, {Sandstrom}, {Sun},
  {Williams}, {Anand}, {Barnes}, {Behrens}, {Belfiore}, {Benincasa},
  {Be{\v{s}}li{\'c}}, {Bigiel}, {Bolatto}, {den Brok}, {Cao}, {Chandar},
  {Chastenet}, {Chiang}, {Congiu}, {Dale}, {Deger}, {Eibensteiner}, {Egorov},
  {Garc{\'\i}a-Rodr{\'\i}guez}, {Glover}, {Grasha}, {Henshaw}, {Ho}, {Kepley},
  {Kim}, {Klessen}, {Kreckel}, {Koch}, {Kruijssen}, {Larson}, {Lee}, {Lopez},
  {Machado}, {Mayker}, {McElroy}, {Murphy}, {Ostriker}, {Pan}, {Pessa},
  {Puschnig}, {Razza}, {S{\'a}nchez-Bl{\'a}zquez}, {Santoro}, {Sardone},
  {Scheuermann}, {Sliwa}, {Sormani}, {Stuber}, {Thilker}, {Turner}, {Utomo},
  {Watkins}, \& {Whitmore}}]{Leroy2021}
{Leroy}, A.~K., {Schinnerer}, E., {Hughes}, A., {et~al.} 2021, \apjs, 257, 43

\bibitem[{{Li} {et~al.}(2021){Li}, {Krumholz}, {Wisnioski}, {Mendel}, {Kewley},
  {S{\'a}nchez}, \& {Galbany}}]{Li2021}
{Li}, Z., {Krumholz}, M.~R., {Wisnioski}, E., {et~al.} 2021, arXiv e-prints,
  arXiv:2104.14807

\bibitem[{{Luisi} {et~al.}(2018){Luisi}, {Anderson}, {Bania}, {Balser},
  {Wenger}, \& {Kepley}}]{Luisi2018}
{Luisi}, M., {Anderson}, L.~D., {Bania}, T.~M., {et~al.} 2018, \pasp, 130,
  084101

\bibitem[{{Luridiana} {et~al.}(2015){Luridiana}, {Morisset}, \&
  {Shaw}}]{Luridiana2015}
{Luridiana}, V., {Morisset}, C., \& {Shaw}, R.~A. 2015, \aap, 573, A42

\bibitem[{{Maiolino} \& {Mannucci}(2019)}]{Maiolino2019}
{Maiolino}, R. \& {Mannucci}, F. 2019, \aapr, 27, 3

\bibitem[{{Mannucci} {et~al.}(2021){Mannucci}, {Belfiore}, {Curti}, {Cresci},
  {Maiolino}, {Marasco}, {Marconi}, {Mingozzi}, {Tozzi}, \&
  {Amiri}}]{Mannucci2021}
{Mannucci}, F., {Belfiore}, F., {Curti}, M., {et~al.} 2021, \mnras, 508, 1582

\bibitem[{{McKee} \& {Ostriker}(2007)}]{McKee2007}
{McKee}, C.~F. \& {Ostriker}, E.~C. 2007, \araa, 45, 565

\bibitem[{{Metha} {et~al.}(2021){Metha}, {Trenti}, \& {Chu}}]{Metha2021}
{Metha}, B., {Trenti}, M., \& {Chu}, T. 2021, \mnras
  [\eprint[arXiv]{2109.03390}]

\bibitem[{{Mingozzi} {et~al.}(2020){Mingozzi}, {Belfiore}, {Cresci}, {Bundy},
  {Bershady}, {Bizyaev}, {Blanc}, {Boquien}, {Drory}, {Fu}, {Maiolino},
  {Riffel}, {Schaefer}, {Storchi-Bergmann}, {Telles}, {Tremonti}, {Zakamska},
  \& {Zhang}}]{Mingozzi2020}
{Mingozzi}, M., {Belfiore}, F., {Cresci}, G., {et~al.} 2020, \aap, 636, A42

\bibitem[{{Morisset}(2013)}]{Morisset2013}
{Morisset}, C. 2013, {pyCloudy: Tools to manage astronomical Cloudy
  photoionization code}

\bibitem[{{Morisset} {et~al.}(2016){Morisset}, {Delgado-Inglada},
  {S{\'a}nchez}, {Galbany}, {Garc{\'\i}a-Benito}, {Husemann}, {Marino}, {Mast},
  \& {Roth}}]{Morisset2016}
{Morisset}, C., {Delgado-Inglada}, G., {S{\'a}nchez}, S.~F., {et~al.} 2016,
  \aap, 594, A37

\bibitem[{{Oey} \& {Kennicutt}(1997)}]{Oey1997}
{Oey}, M.~S. \& {Kennicutt}, R.~C., J. 1997, \mnras, 291, 827

\bibitem[{{Osterbrock} \& {Ferland}(2006)}]{Osterbrock2006}
{Osterbrock}, D.~E. \& {Ferland}, G.~J. 2006, {Astrophysics of gaseous nebulae
  and active galactic nuclei} (University Science Books)

\bibitem[{{Pilyugin} \& {Grebel}(2016)}]{Pilyugin2016}
{Pilyugin}, L.~S. \& {Grebel}, E.~K. 2016, \mnras, 457, 3678

\bibitem[{{Pilyugin} {et~al.}(2014){Pilyugin}, {Grebel}, \&
  {Kniazev}}]{Pilyugin2014}
{Pilyugin}, L.~S., {Grebel}, E.~K., \& {Kniazev}, A.~Y. 2014, \aj, 147, 131

\bibitem[{{Pineda} {et~al.}(2019){Pineda}, {Horiuchi}, {Anderson}, {Luisi},
  {Langer}, {Goldsmith}, {Kuiper}, {Bryden}, {Soriano}, \&
  {Lazio}}]{Pineda2019}
{Pineda}, J.~L., {Horiuchi}, S., {Anderson}, L.~D., {et~al.} 2019, \apj, 886, 1

\bibitem[{{Pogge} {et~al.}(2010){Pogge}, {Atwood}, {Brewer}, {Byard},
  {Derwent}, {Gonzalez}, {Martini}, {Mason}, {O'Brien}, {Osmer}, {Pappalardo},
  {Steinbrecher}, {Teiga}, \& {Zhelem}}]{Pogge2010}
{Pogge}, R.~W., {Atwood}, B., {Brewer}, D.~F., {et~al.} 2010, in Society of
  Photo-Optical Instrumentation Engineers (SPIE) Conference Series, Vol. 7735,
  Ground-based and Airborne Instrumentation for Astronomy III, ed. I.~S.
  {McLean}, S.~K. {Ramsay}, \& H.~{Takami}, 77350A

\bibitem[{{Reynolds} {et~al.}(1998){Reynolds}, {Hausen}, {Tufte}, \&
  {Haffner}}]{Reynolds1998}
{Reynolds}, R.~J., {Hausen}, N.~R., {Tufte}, S.~L., \& {Haffner}, L.~M. 1998,
  \apjl, 494, L99

\bibitem[{{Rosolowsky} \& {Simon}(2008)}]{Rosolowsky2008}
{Rosolowsky}, E. \& {Simon}, J.~D. 2008, \apj, 675, 1213

\bibitem[{{Rousseau-Nepton} {et~al.}(2019){Rousseau-Nepton}, {Martin},
  {Robert}, {Drissen}, {Amram}, {Prunet}, {Martin}, {Moumen}, {Adamo},
  {Alarie}, {Barmby}, {Boselli}, {Bresolin}, {Bureau}, {Chemin}, {Fernandes},
  {Combes}, {Crowder}, {Della Bruna}, {Duarte Puertas}, {Egusa}, {Epinat},
  {Ksoll}, {Girard}, {G{\'o}mez Llanos}, {Gouliermis}, {Grasha}, {Higgs},
  {Hlavacek-Larrondo}, {Ho}, {Iglesias-P{\'a}ramo}, {Joncas}, {Kam}, {Karera},
  {Kennicutt}, {Klessen}, {Lianou}, {Liu}, {Liu}, {de Amorim}, {Lyman},
  {Martel}, {Mazzilli-Ciraulo}, {McLeod}, {Melchior}, {Millan}, {Moll{\'a}},
  {Momose}, {Morisset}, {Pan}, {Pati}, {Pellerin}, {Pellegrini}, {P{\'e}rez},
  {Petric}, {Plana}, {Rahner}, {Ruiz Lara}, {S{\'a}nchez-Menguiano},
  {Spekkens}, {Stasi{\'n}ska}, {Takamiya}, {Vale Asari}, \&
  {V{\'\i}lchez}}]{Rousseau-Nepton2019}
{Rousseau-Nepton}, L., {Martin}, R.~P., {Robert}, C., {et~al.} 2019, \mnras,
  489, 5530

\bibitem[{{S{\'a}nchez-Menguiano} {et~al.}(2019){S{\'a}nchez-Menguiano},
  {S{\'a}nchez Almeida}, {Mu{\~n}oz-Tu{\~n}{\'o}n}, {S{\'a}nchez}, {Filho},
  {Hwang}, \& {Drory}}]{Sanchez-Menguiano2019}
{S{\'a}nchez-Menguiano}, L., {S{\'a}nchez Almeida}, J.,
  {Mu{\~n}oz-Tu{\~n}{\'o}n}, C., {et~al.} 2019, arXiv e-prints,
  arXiv:1904.03930

\bibitem[{{Santoro} {et~al.}(2022){Santoro}, {Kreckel}, {Belfiore}, {Groves},
  {Congiu}, {Thilker}, {Blanc}, {Schinnerer}, {Ho}, {Diederik Kruijssen},
  {Meidt}, {Klessen}, {Schruba}, {Querejeta}, {Pessa}, {Chevance}, {Kim},
  {Emsellem}, {McElroy}, {Barnes}, {Bigiel}, {Boquien}, {Dale}, {Glover},
  {Grasha}, {Lee}, {Leroy}, {Pan}, {Rosolowsky}, {Saito}, {Sanchez-Blazquez},
  {Watkins}, \& {Williams}}]{Santoro2022}
{Santoro}, F., {Kreckel}, K., {Belfiore}, F., {et~al.} 2022, \aap, 658, A188

\bibitem[{{Schlafly} \& {Finkbeiner}(2011)}]{Schlafly2011}
{Schlafly}, E.~F. \& {Finkbeiner}, D.~P. 2011, \apj, 737, 103

\bibitem[{{Tayal} {et~al.}(2019){Tayal}, {Zatsarinny}, \& {Sossah}}]{Tayal2019}
{Tayal}, S.~S., {Zatsarinny}, O., \& {Sossah}, A.~M. 2019, \apjs, 242, 9

\bibitem[{{Thilker} {et~al.}(2000){Thilker}, {Braun}, \&
  {Walterbos}}]{Thilker2000}
{Thilker}, D.~A., {Braun}, R., \& {Walterbos}, R.~A.~M. 2000, \aj, 120, 3070

\bibitem[{{Tielens}(2010)}]{Tielens2010}
{Tielens}, A.~G.~G.~M. 2010, {The Physics and Chemistry of the Interstellar
  Medium} (Cambridge University Press)

\bibitem[{{Wenger} {et~al.}(2019){Wenger}, {Balser}, {Anderson}, \&
  {Bania}}]{Wenger2019}
{Wenger}, T.~V., {Balser}, D.~S., {Anderson}, L.~D., \& {Bania}, T.~M. 2019,
  \apj, 887, 114

\bibitem[{{Williams} {et~al.}(2022){Williams}, {Kreckel}, {Belfiore}, {Groves},
  {Sandstrom}, {Santoro}, {Blanc}, {Bigiel}, {Boquien}, {Chevance}, {Congiu},
  {Emsellem}, {Glover}, {Grasha}, {Klessen}, {Koch}, {Kruijssen}, {Leroy},
  {Liu}, {Meidt}, {Pan}, {Querejeta}, {Rosolowsky}, {Saito},
  {S{\'a}nchez-Bl{\'a}zquez}, {Schinnerer}, {Schruba}, \&
  {Watkins}}]{Williams2022}
{Williams}, T.~G., {Kreckel}, K., {Belfiore}, F., {et~al.} 2022, \mnras, 509,
  1303

\bibitem[{{Yan} {et~al.}(2020){Yan}, {Bershady}, {Smith}, {MacDonald},
  {Bizyaev}, {Bundy}, {Chattopadhyay}, {Gunn}, {Westfall}, \& {Wolf}}]{Yan2020}
{Yan}, R., {Bershady}, M.~A., {Smith}, M.~P., {et~al.} 2020, in Society of
  Photo-Optical Instrumentation Engineers (SPIE) Conference Series, Vol. 11447,
  Society of Photo-Optical Instrumentation Engineers (SPIE) Conference Series,
  114478Y

\bibitem[{{Zhao} {et~al.}(1996){Zhao}, {Anantharamaiah}, {Goss}, \&
  {Viallefond}}]{Zhao1996}
{Zhao}, J.-H., {Anantharamaiah}, K.~R., {Goss}, W.~M., \& {Viallefond}, F.
  1996, \apj, 472, 54

\end{thebibliography}
%

\bibliographystyle{aa}

\clearpage

\appendix

\section{Derivation of line emissivitiy relation}
\label{app:derivation}

The emissivity of \oi\ $\lambda$6300 transition can be written as:
\begin{equation}
\epsilon_{\rm OI} = \Delta E_{\rm OI} \, R_{\rm OI} \, n({\rm O}^{0}) n({\rm e}^{-}),
\end{equation}
where $R_{\rm OI}$ is the excitation rate coefficient for the $^{3}$P--$^{1}$D transition and $\Delta E_{\rm OI}$ is the energy of the transition. If we assume that the electrons come primarily from ionized hydrogen, then $n({\rm e^{-}}) \simeq n({\rm H^{+}})$ and we can instead write this as
\begin{equation}
\epsilon_{\rm OI} = \Delta E_{\rm OI} \, R_{\rm OI} \, n({\rm O}^{0}) n({\rm H}^{+}).
\end{equation}
The H$\alpha$ emissivity due to H$^{+}$ recombination can be written as
\begin{equation}
\epsilon_{\rm H\alpha} = \Delta E_{\rm H\alpha} R_{\alpha} \, n({\rm e}^{-}) n({\rm H}^{+}) \simeq \Delta E_{\rm H\alpha} R_{\alpha}  \, n({\rm H}^{+})^{2},
\end{equation}
where $R_{\alpha}$ is the rate coefficient for the production of H$\alpha$ photons during recombination (i.e.\ the product of the recombination rate and the fraction of recombinations resulting in H$\alpha$) and $\Delta E_{\rm H\alpha}$ is the energy of H$\alpha$. Taking the ratio of the emissivities yields:
\begin{equation}
\frac{\epsilon_{\rm OI}}{\epsilon_{\rm H\alpha}} = \frac{\Delta E_{\rm OI}}{\Delta E_{\rm H\alpha}} \frac{R_{\rm OI}}{R_{\alpha}} \frac{n({\rm O}^{0})}{n({\rm H}^{+})}.
\end{equation}
We now want to eliminate the dependence on the O$^{0}$ fraction in favour of a dependence on the H$^{+}$/H ratio and the total elemental abundance of oxygen. To do this, we use the fact that the balance between O$^{0}$ and O$^{+}$ is set by charge transfer with hydrogen:
\begin{eqnarray}
{\rm O^{0} + H^{+}} & \rightarrow & {\rm O^{+} + H^{0}}, \\
{\rm O^{+} + H^{0}} & \rightarrow & {\rm O^{0} + H^{+}}.
\end{eqnarray}
If we denote the rate coefficient for the forward reaction (producing O$^{+}$) as $k_{\rm f}$ and the rate coefficient for the reverse reaction as $k_{\rm r}$, then chemical equilibrium between these two reactions implies that
\begin{equation}
k_{\rm f} n({\rm O}^{0}) n({\rm H}^{+}) = k_{\rm r} n({\rm O}^{+}) n({\rm H^{0}}).
\end{equation}
Denoting the total elemental abundance of oxygen as $n({\rm O})$, we can use the relation $n({\rm O^{+}}) = n({\rm O}) - n({\rm O^{0}})$ to write this as:
\begin{equation}
k_{\rm f} n({\rm O}^{0}) n({\rm H}^{+}) = k_{\rm r} \left[n({\rm O}) - n({\rm O^{0}}) \right] n({\rm H^{0}}).
\end{equation}
Rearranging for $n({\rm O^{0}})$ then yields
\begin{equation}
n({\rm O}^{0}) = n({\rm O}) n({\rm H^{0}}) \left[\frac{k_{\rm f}}{k_{\rm r}}  n({\rm H}^{+}) +  n({\rm H^{0}}) \right]^{-1}.
\end{equation}
The difference between the ionization potentials of hydrogen and oxygen is very small and at high temperatures we can ignore it and assume that the energy change in the forward and reverse reactions is zero. In this case, the ratio of the forward and reverse rate coefficients is simply the ratio of the statistical weights of the products, i.e.
\begin{equation}
\frac{k_{\rm f}}{k_{\rm r}} = \frac{g_{\rm O^{+}} g_{\rm H^{0}}}{g_{\rm O^{0}} g_{\rm H^{+}}}.
\end{equation}
With $g_{\rm O^{0}} = 9$, $g_{\rm O^{+}} = 4$, $g_{\rm H^{0}} = 2$ and $g_{\rm H^{+}} = 1$, this yields $k_{\rm f} / k_{\rm r} = 8/9$.
Therefore,
\begin{eqnarray}
n({\rm O}^{0}) & = & n({\rm O}) n({\rm H^{0}}) \left[\frac{8}{9}  n({\rm H}^{+}) +  n({\rm H^{0}}) \right]^{-1}, \\
 & = & n({\rm O}) r \left[\frac{8}{9}  + r \right]^{-1},
\end{eqnarray}
where $r = n({\rm H^{0}}) / n({\rm H}^{+})$. Substituting this into the ratio of the emissivities then yields
\begin{equation}
\frac{\epsilon_{\rm OI}}{\epsilon_{\rm H\alpha}} = \frac{\Delta E_{\rm OI}}{\Delta E_{\rm H\alpha}} \frac{R_{\rm OI}}{R_{\alpha}}  \frac{n({\rm O})}{n({\rm H}^{+})}  \frac{r}{8/9 + r}.
\end{equation}
Finally, we can use the fact that 
\begin{equation}
 n({\rm H}) = n({\rm H^{+}}) + n({\rm H^{0}}) = n({\rm H^{+}}) \left[1 + r\right]
\end{equation}
to write this as
\begin{equation}
\frac{\epsilon_{\rm OI}}{\epsilon_{\rm H\alpha}} = \frac{\Delta E_{\rm OI}}{\Delta E_{\rm H\alpha}} \frac{R_{\rm OI}}{R_{\alpha}}  r \xi \frac{n({\rm O})}{n({\rm H})},
\end{equation}
where $\xi = (1+r) / (8/9 + r)$. Comparing this with Equation~1 in \citet{Reynolds1998} shows us that they use the following expression for $\Delta E_{\rm OI} R_{\rm OI} / \Delta E_{\rm H\alpha} R_{\alpha}$:
\begin{equation}
\frac{\Delta E_{\rm OI}}{\Delta E_{\rm H\alpha}} \frac{R_{\rm OI}}{R_{\alpha}} = 2.63 \times 10^{4} \frac{T_{4}^{1.85}}{1 + 0.605 T_{4}^{1.105}} \exp \left(-\frac{2.284}{T_{4}} \right),  \label{rey98}
\end{equation}
where $T_{4} = T / 10000$~K.

For our updated calculation, we have computed $R_{\rm OI}$ for large number of different temperatures by numerically integrating the cross-section data\footnote{Available at \tt{https://github.com/barklem/public-data/tree/master/inelastic-O+e}} from \citet{Barklem2007} and have then fit a function to the resulting rate. The best fit that we have found is the following:
\begin{equation}
R_{\rm OI} = 9.54 \times 10^{-10} f_{\rm OI}(T) \exp \left(-\frac{2.284}{T_{4}} \right),
\end{equation}
where 
\begin{equation}
f_{\rm OI}(T) = \sum_{i=0}^{6} a_{i} T_{4}^{i}.
\end{equation}
This fit is accurate to within 3\% for temperatures in the range $3000 < T < 30000$~K, but is not guaranteed to give sensible results outside of this temperature range.

\begin{table}[h]
\begin{tabular}{cl}
Coefficient & Value \\
\hline
$a_{0}$ & $-1.00$ \\
$a_{1}$ & $+6.4854011$ \\
$a_{2}$ & $-4.17358515$ \\
$a_{3}$ & $+1.81446389$ \\
$a_{4}$ & $-0.514022051$ \\
$a_{5}$ & $+8.39069326 \times 10^{-2}$ \\
$a_{6}$ & $- 5.93343677 \times 10^{-3}$ \\
\hline
\end{tabular}
\end{table}

Since $\Delta E_{\rm OI} = 3.155 \times 10^{-12}$, it then follows that:
\begin{equation}
\Delta E_{\rm OI} R_{\rm OI} = 3.01 \times 10^{-21} f_{\rm OI}(T) \exp \left(-\frac{2.284}{T_{4}} \right).
\end{equation}
For H$\alpha$, we can use the accurate fit given in \citet{Draine2011} to the data of \citet{hs87}:
\begin{equation}
\Delta E_{\rm H\alpha} R_{\alpha} = 4\pi j_{\rm H\alpha} = 3.54 \times 10^{-25} T_{4}^{-0.942 - 0.031 \ln T_{4}}.
\end{equation}
This fit is also accurate to within a few percent in the temperature range of interest. Combining these then yields 
\begin{equation}
\frac{\Delta E_{\rm OI}}{\Delta E_{\rm H\alpha}} \frac{R_{\rm OI}}{R_{\alpha}} = 8492 \frac{f_{\rm OI}(T)}{T_{4}^{-0.942 - 0.031 \ln T_{4}}} \exp \left(-\frac{2.284}{T_{4}} \right),
\end{equation}
and hence
\begin{equation}
\frac{\epsilon_{\rm OI}}{\epsilon_{\rm H\alpha}} = 8492 \frac{f_{\rm OI}(T)}{T_{4}^{-0.942 - 0.031 \ln T_{4}}} \exp \left(-\frac{2.284}{T_{4}} \right) r \xi \frac{n({\rm O})}{n({\rm H})}.
\label{eq:app_o1ha_emis}
\end{equation}

\begin{figure}
    \centering
    \includegraphics[width=3.5in]{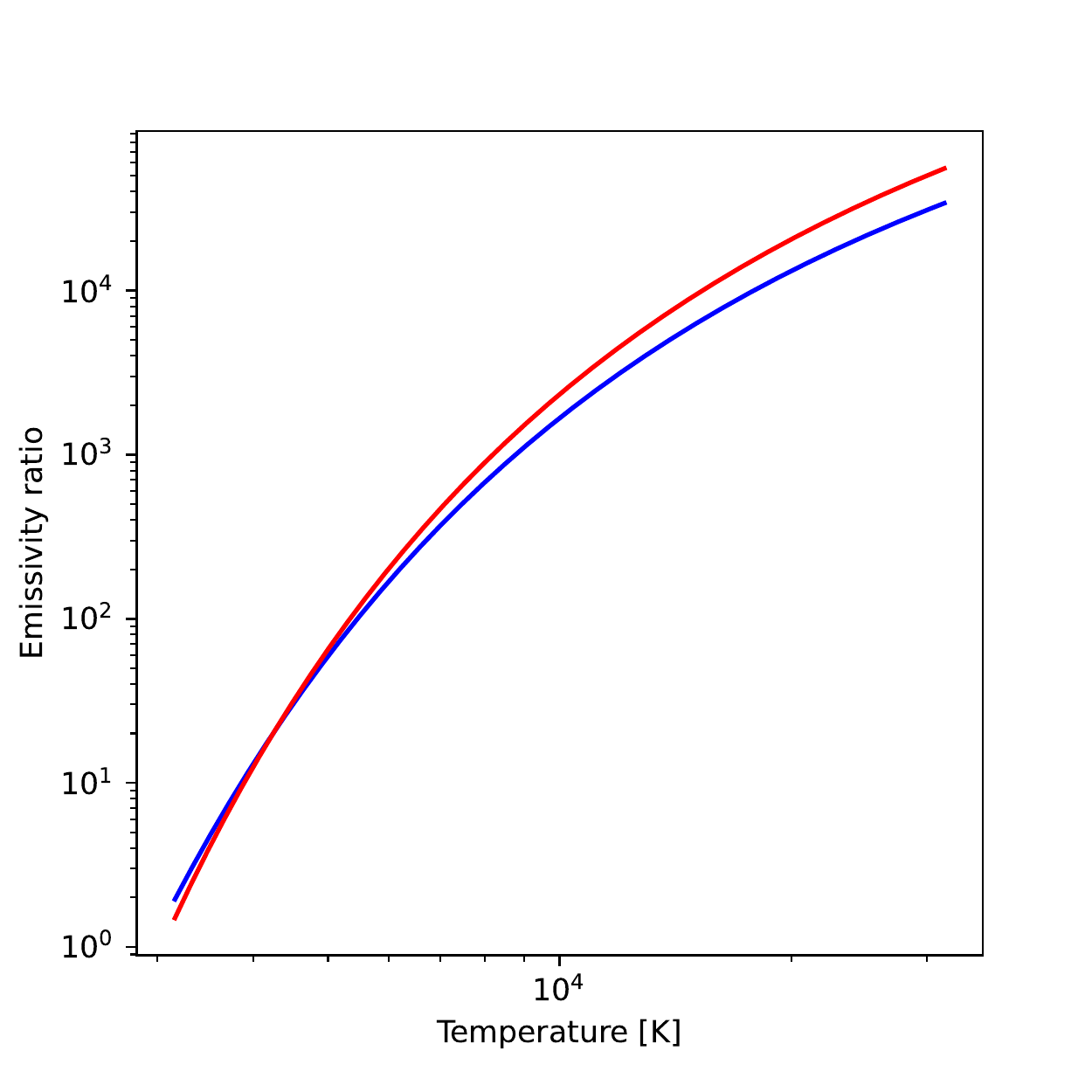}
    \caption{A comparison of the emissivity coefficients computed in Section \ref{app:derivation} (red) with the value provided in \citet{Reynolds1998} (blue). For \hii\ region temperatures in the normal range for PHANGS galaxies, the difference is around 20-40\%.
    \label{fig:app_coeffs1}}
\end{figure}

A comparison of the relations for $\frac{\Delta E_{\rm OI}}{\Delta E_{\rm H\alpha}} \frac{R_{\rm OI}}{R_{\alpha}}$ provided by \citet{Reynolds1998} and derived here is shown in Figure \ref{fig:app_coeffs1}. 

\revone{
Note that in general, the ratio of the total observed fluxes in \oi\, and \ha\, is not equal to the ratio of their emissivities, but can be expressed from the equations above as follows (assuming a uniform $T_e$ across the \oi\, and \ha\, emitting zones):
\begin{equation}
\begin{split}
    I(\mathrm{[OI]}/I(\mathrm{H\alpha}) & = \frac{\int_{V_\mathrm{[OI]}}\epsilon_{\rm OI} dV}{\int_{V_\mathrm{H\alpha}}\epsilon_{\rm H\alpha} dV} = \\ & = 8492 \frac{f_{\rm OI}(T)}{T_{4}^{-0.942 - 0.031 \ln T_{4}}} \exp \left(-\frac{2.284}{T_{4}} \right) \xi' \frac{n({\rm O})}{n({\rm H})}
    \label{eq:app_o1ha_obs}
\end{split}
\end{equation}
where $\xi'$ depends on the distribution of the hydrogen ionization fraction $f_{ion} = n({\rm H^+})/n({\rm H}) = 1/(1+r)$ in the \oi\, and \ha\, emitting zones as
\begin{equation}
    \xi' = \frac{\int_{V_\mathrm{[OI]}}\frac{r}{(1+r)(8/9+r)}dV}{\int_{V_\mathrm{H\alpha}}1/(1+r)^2 dV}.
\end{equation}
Equation (\ref{eq:app_o1ha_obs}) is equal to Equation~\ref{eq:app_o1ha_emis} (and thus - to Equation~\ref{eqn}) if one assumes that \oi\, and \ha\, come from the same zones. We discuss the limitation of this assumption in Section~\ref{sec:cloudy_cospatial}.
}

\section{Application of the model-based prescription for \fion}

As described in Section \ref{sec:cloudy}, we use a series of Cloudy models to derive \fion\ as a function of \siii/\sii\ across integrated \hii\ regions. Given the similarities between the model results and our observations, we explore using the Cloudy models to derive a new prescription for \fion, and improve upon them by adding an age dependence based on variations in EW(\ha) (Equation \ref{eqn:cloudy_fion_fit}).

In this section, we explore how well this model-driven parameterization of \fion\ applies to our observed PHANGS-MUSE \hii\ regions. For each \hii\ region, we calculate EW(H$\alpha$) to use as an age tracer. Here, we assume that the stellar continuum is dominated by light from the young stellar population, which may not be the case for low EW(H$\alpha$) regions. Given the added dependence on age (via inclusion of EW(\ha)), we can remove our requirement that \siii/\sii $>$ 0.5. We find that this doubles the number of \hii\ regions for which we can apply the charge exchange method to a total of $\sim$9000 \hii\ regions,  nearly half of all \hii\ regions in the PHANGS-MUSE sample. 

\begin{figure*}
    \centering
    \includegraphics[width=6in]{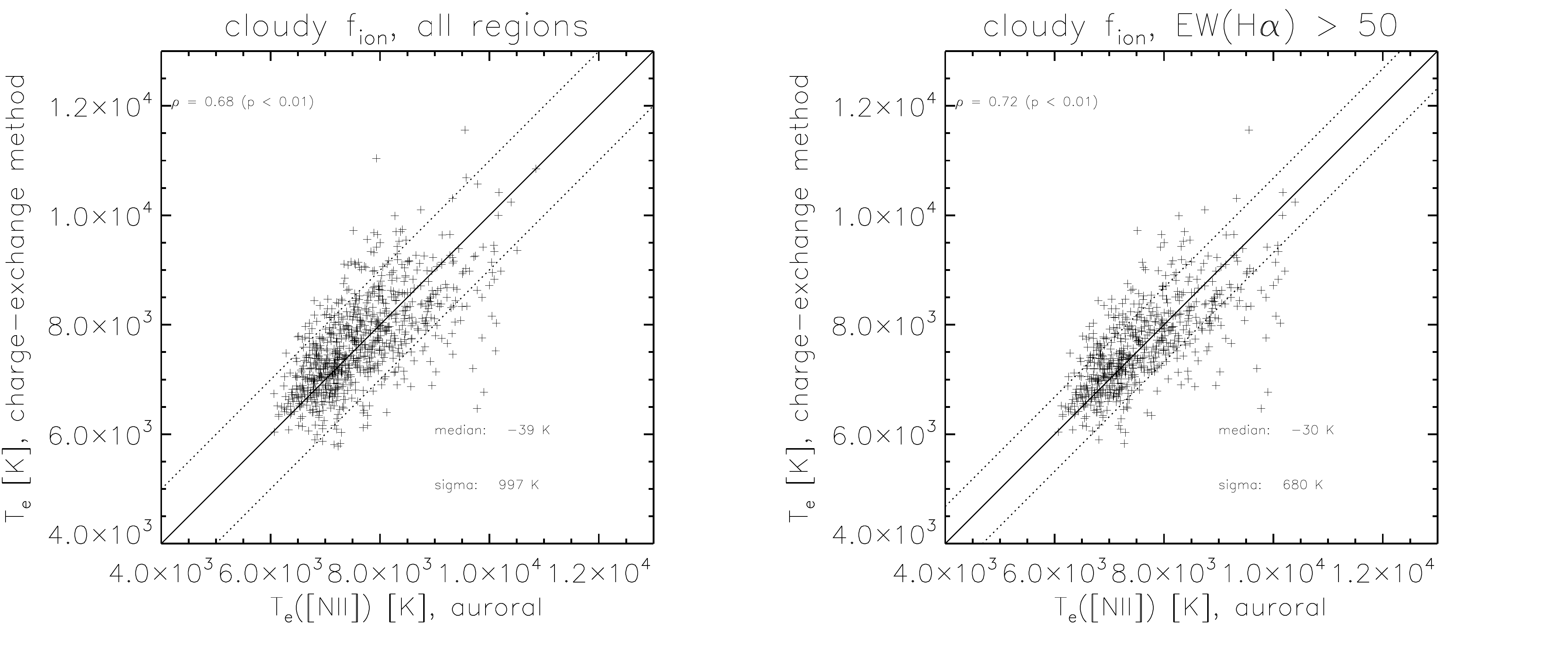}
    \caption{A comparison of \te\, derived from the charge-exchange method, with \te(\nii), derived from auroral line detections, for \hii\ regions in the PHANGS-MUSE sample. For the charge-exchange method, we use the parametization of \fion\ derived from Cloudy models (Equation \ref{eqn:cloudy_fion_fit}). We find systematically larger scatter ($\sim$1000~K) compared to the relation in Figure \ref{fig:predict_obs}, where \fion\ was empirically calibrated. The scatter is still larger, even when we only consider younger (higher EW(\ha)) \hii\ regions. }
    \label{fig:app_fion}
\end{figure*}

In Figure \ref{fig:app_fion} we compare the charge-exchange method \te\ with the \te\ derived from auroral line detections. We note that the scatter is significantly increased ($\sim$1000 K instead of 600 K).  Considering only \hii\ regions with EW(\ha) $>$ 50 \AA\ significantly reduces this scatter but also limits the total number of \hii\ regions. 

\begin{figure*}
    \centering
    \includegraphics[width=6in]{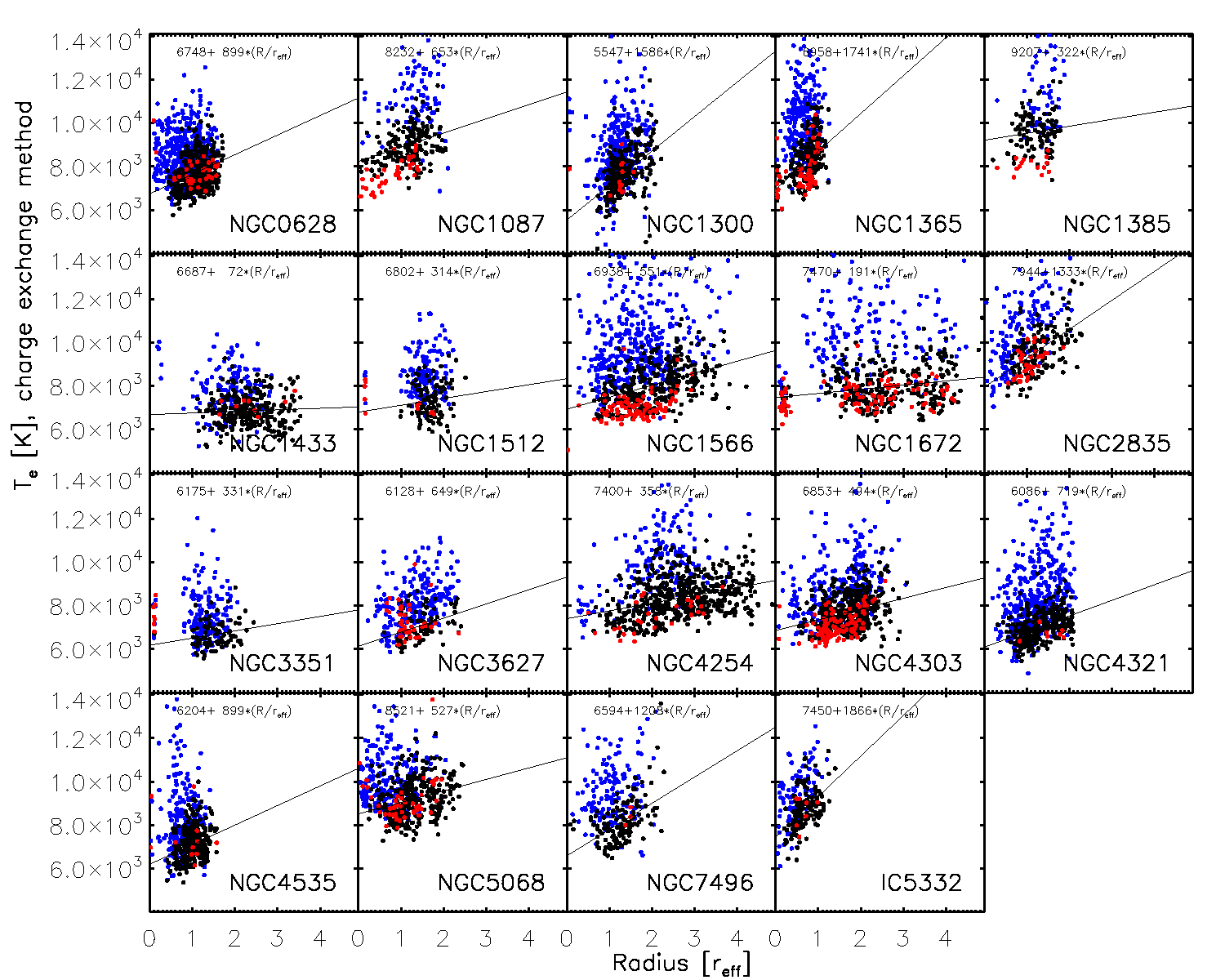}
    \caption{Radial gradients in \te\ for each galaxy, using \fion\ parameterized from Cloudy models (Equation \ref{eqn:cloudy_fion_fit}). \hii\ regions marked in blue have EW(\ha)$<$50 \AA\, a regime poorly sampled by our calibrations and at the extremes of the Cloudy models. These all have systematically higher \te, suggesting problems with the calibration. \hii\ regions with younger ages (black) show better agreement with the auroral line measurements (red), similar to Figure \ref{fig:te_grad0}.}
    \label{fig:app_fion2}
\end{figure*}

In Figure \ref{fig:app_fion2} we plot the radial gradient in \te\ for each galaxy. Again, we see that for many of the \hii\ regions at low EW(\ha) this prescription for \fion\ results in particularly large \te\ values (10,000--12,000 K).

As this revised parameterization for \fion\ does not result in improved agreement with \te(\nii) and results in unphysically high \te\ values along the radial gradient, combined with the systematic offsets between modeled and measured EW(\ha), we choose in this paper to adopt the empirically calibrated relation for \fion (Equation \ref{eqn:fion}). 

\end{document}